\documentclass[lettersize,journal]{IEEEtran}

\usepackage{caption}
\usepackage{amsmath,amsfonts}
\usepackage{algorithmic}
\usepackage{algorithm}
\usepackage{array}
\usepackage[caption=false,font=normalsize,labelfont=sf,textfont=sf]{subfig}
\usepackage{textcomp}
\usepackage{stfloats}
\usepackage{url}
\usepackage{verbatim}
\usepackage{graphicx}
\usepackage{cite}
\UseRawInputEncoding

\usepackage{multirow}
\usepackage{graphicx}
\usepackage{longtable}
 \usepackage{lscape}
\usepackage[table,xcdraw,dvipsnames]{xcolor}

\begin{document}

\title{Precoding for High Throughput Satellite Communication Systems: A Survey}
\author{Malek~Khammassi,
        Abla~Kammoun,~\IEEEmembership{Member,~IEEE,}
        and~Mohamed-Slim~Alouini,~\IEEEmembership{Fellow,~IEEE}
}
\markboth{}%
\IEEEpubid{}

\maketitle

\begin{abstract}
With the expanding demand for high data rates and extensive coverage, high throughput satellite (HTS) communication systems are emerging as a key technology for future communication generations. However, current frequency bands are increasingly congested. Until the maturity of communication systems to operate on higher bands, the solution is to exploit the already existing frequency bands more efficiently. In this context, precoding emerges as one of the prolific approaches to increasing spectral efficiency. This survey presents an overview and a classification of the recent precoding techniques for HTS communication systems from two main perspectives: 1) a problem formulation perspective and 2) a system design perspective. 
From a problem formulation point of view, precoding techniques are classified according to the precoding \textcolor{black}{optimization problem}, group, and level. From a system design standpoint, precoding is categorized based on the system architecture, the precoding implementation, and the type of the provided service. Further, practical system impairments are discussed, and robust precoding techniques are presented. Finally, future trends in precoding for satellites are addressed to spur further research. 

\end{abstract}


\section{Introduction}

\IEEEPARstart{S}{atellite} communication systems have been attracting intensified attention during the last few decades, extending their application from conventional media broadcasting to data services, namely broadband SatComs. This shift has been motivated by two main factors: 1) the increasing demand for higher capacity in the already covered areas, and 2) the need to extend the broadband coverage to under-served areas. More specifically, a vast majority of the highly populated urban areas are now suffering scarcity in services and coverage due to the limited capacity of the conventional telecommunication infrastructure. Furthermore, vast exurban regions lacking a reliable Internet and communication technology infrastructure are still disseminated worldwide. Consequently, the research community and industries have been investigating more innovative broadband solutions that can provide more coverage while seamlessly integrating the diverse wireless and wired technologies. In this context, satellite communication systems offer a plethora of ways to increase coverage and data rates while matching the objectives of future communication generations.

Satellites have been witnessing various technological breakthroughs revealing their potential in future communication systems, namely, new constellations and on-board capabilities \cite{46}. Traditionally, SatComs have been dominated by geostationary (GEO) satellites since they allow for high coverage and bypass mobility between terminals and the satellite transceivers. However, the last few decades have been marked by new ambitious constellations driven by low manufacturing and launch cost. In this context, low earth orbit (LEO) constellations have been receiving fast-growing attention after Teledesic introduced it 25 years ago \cite{LEO0}. Unlike GEO satellites, LEO constellations have the advantage of low latency at the expense of reduced coverage. They have been attracting various big companies like OneWeb, Amazon, and SpaceX, planning prominent LEO constellations and launching demo satellites. In addition to LEO constellations,  medium earth orbit (MEO) satellites are emerging as a trade-off between the low latency of LEO constellations and the high coverage of GEO constellations. In fact, O3B is a constellation of 20 MEO satellites launched by SES at an altitude of 8063 km, with the first four satellites reaching orbit in 2013 and the last four satellites reaching orbit in 2019. These new constellations have also motivated hybrid constellations exploiting advantages of various orbits such as hybrid MEO and GEO constellations \cite{geo_meo} and a combination of LEO constellations with higher orbits \cite{leo}.

In addition to the new constellation types, new on-board capabilities have been emerging for satellite systems \cite{46}. Traditionally, satellite communication technology advances have been majorly constrained by the on-board processing capacity. This is mainly due to the limited power supply and flexibility. In fact, much of the on-board power consumption is due to the large mass and path loss of traditional satellites. Additionally, due to the limited maintenance of on-board components, satellite technologies' flexibility is constrained by ultra-reliability and robustness. Therefore, a great number of satellites operate as relays limited to performing frequency conversion, amplification, and forwarding. Nevertheless, more ambitious on-board processing has recently been possible thanks to developments in the efficiency of power generation, analog, and digital processing \cite{30}, as well as the cheap manufacturing and launching costs. Advanced processing techniques include beamforming, signal regeneration, flexible routing, and channelization \cite{46}.

Another significant milestone for satellite communication systems is the shift from single-beam to multi-beam transmission \cite{singleVSmultipleBeams}, which came with various advantages \cite{Tronc2014}. In contrast to single-beam architectures, multi-beam architectures allow spatially multiplexed communication by concurrently transmitting different data streams to geographically separated areas. Further, the available radio resources can be reused in the different beams, increasing the user bandwidth. However, adjacent beams face increased interference because of multi-beam antenna side-lobes. In-orbit multi-beam satellites usually handle this interference by partial polarization and/or partial frequency reuse across the beams. In this case, radio resources are distinct among the adjacent beams and shared among spatially separated beams. Thus, this fractional frequency reuse approach guarantees orthogonality while reusing the radio resources to a certain extent. Nevertheless, the partial spectrum reuse strategy limits the throughput of the system. Higher efficiency can be attained when all the available radio resources are reused in every beam. While such aggressive spectrum reuse enhances the user bandwidth, it sacrifices the interference experienced across the beams. Some current multi-beam high throughput satellite (HTS) systems in-orbit include \cite{7}: Anik F2 \cite{51}, Wildblack-1 \cite{50}, Sky Muster \cite{53}, and KA-SAT \cite{52}. Multi-user multiple-input multiple-output (MU-MIMO) techniques are often exploited to cope with the high interference in such systems. For example, one can readily employ multi-user detection and precoding techniques by modeling the return link as a MIMO multiple access channel and the forward link as a MIMO broadcast channel. This approach is also called joint multi-beam processing in the literature \cite{Christopoulos2012}, and it has been featured in the latest extensions of broadband satellite communication standards DVB-S2X \cite{54}. In this survey, we focus on precoding as a prolific interference mitigation technique in the forward link of multi-beam satellite communication systems. 

MIMO techniques, such as precoding, have more than two decades of astounding results in terrestrial applications \cite{47}. However, precoding designs for satellite communication systems require different considerations than terrestrial systems. Their design accounts for different interfering sources, frequency bands, and wave propagation conditions. It also considers the particular antenna structures, onboard power, and processing limitations, as well as transceiver impairments of satellite systems \cite{Sharma2021}. During the last decade, there have been various works surveying precoding and beamforming techniques. The work in \cite{Fatema2018} surveys linear precoding schemes for massive MIMO scenarios, specifically single-cell and multi-cell. Reference \cite{Alodeh2018} presents a comprehensive overview and categorizes precoding schemes according to the precoding switching rate and group. In \cite{LiAng2020}, a tutorial on interference exploitation using precoding in wireless communication systems is provided. It focuses on symbol-level precoding and addresses constructive and destructive interference classification. However, the aforementioned works are not specific to precoding for satellite systems and focus more on terrestrial communication systems. In the context of satellite communications, \cite{46} provides a comprehensive overview of the state-of-the-art in SatComs, covering everything from the system aspects to the networking, medium access, air interface, and testbeds. However, this survey gives a more general overview of satellite communication and all the technical aspects involved without a deep focus on precoding techniques in particular. On the other hand, \cite{47} focuses on precoding for satellites by reviewing MIMO techniques for satellite communications until 2010. Although the aforementioned survey covers the works that laid the ground for precoding to flourish in satellite communication, a tremendous amount of work has been proposed since 2010, and a more comprehensive survey, including the recent contributions, has to be carried out. 

Therefore, this survey offers an in-depth overview and a classification of the up-to-date precoding techniques for HTS communication systems. The precoding techniques are classified according to two main perspectives: 1) a problem formulation perspective and 2) a system design perspective. From a problem formulation perspective, precoding can be classified according to its group, \textcolor{black}{optimization problem}, and level. The precoding group refers to the scheduling strategy, which can be unicasting or multicasting. The precoding \textcolor{black}{optimization problem} involves the optimized objective function, and the surveyed objectives are spectral efficiency, energy efficiency, secrecy, and mixed objectives. Finally, the precoding level refers to the switching rate of precoding, including block-level precoding and symbol-level precoding. From a system design perspective, precoding can be categorized according to the architecture, implementation, and type of service. The system architecture includes single-gateway (single-GW) systems and multi-gateway (multi-GW) systems. Precoding implementation can be classified into on-board, on-ground, and hybrid on-ground/on-board implementations. Finally, the service type refers to fixed and mobile satellite services. Further, this survey covers the practical payload and channel impairments for satellite communication systems and presents robust precoding techniques in this context. This classification of precoding techniques is illustrated in Fig.~ \ref{fig:classification}. \textcolor{black}{Finally, we present technologies driving the current research trends in precoding for HTS. }

The rest of this paper is organized as follows. Section II describes the key aspects of satellite communication systems, namely, the system and the modeling aspects. Section III presents a brief summary of signal processing techniques for satellite systems, from the medium access control layer to the physical layer. Section IV presents the precoding techniques classification according to six aspects: the objective, group, level, architecture, implementation, and the type of the provided service. Section V presents practical satellite communication systems impairments and surveys robust precoding techniques. Open topics in precoding for satellites are covered in section VI, and the paper is concluded in section VII. \textcolor{black}{Further, the lists of acronyms and notations are respectively  given in Table I and Table II}.

\section{Preliminaries}

\subsection{\textcolor{black}{ Key Requirements of Satellite Communication}}\textcolor{black}{Precoding exploits multiple antennas at the transmitter and receiver to improve wireless communication systems' capacity and data rate. It is widely adopted in terrestrial communication systems, and its incorporation into satellite communication systems can provide similar benefits. However, it requires additional considerations due to the unique characteristics of satellite communication.}

\textcolor{black}{The frequency bands allocated for satellite communication (typically the Ku and Ka bands) systems differ from their terrestrial counterparts. As a result, satellite communication systems benefit from more bandwidth since they operate on higher frequency bands. However, they are also more susceptible to interference from other sources. For example, interference can originate from other communication systems operating on the same frequency, which can be other satellites or terrestrial systems. Interference can also come from natural causes such as solar flares. In fact, the radiation emitted by the sun can degrade the satellite's signal and cause a short communication outage when the sun is aligned with the satellite. These types of interference can significantly degrade the precoding algorithm's performance.}

\textcolor{black}{In addition to interference, satellite communication systems vary from their terrestrial counterpart by their propagation conditions. In fact, a satellite's transmitted signal faces atmospheric attenuation, rain, and multipath fading when it travels through the atmosphere. Atmospheric attenuation arises when atmospheric gases absorb the transmitted signal.
Rain fading results from raindrops scattering the signal, which can cause a significant attenuation. Multipath fading occurs when neighboring objects reflect or diffuse the signal, especially in mobile satellite services. In this case, the receiver captures the transmitted signal from different paths, each with a different delay and phase. These different propagation conditions can considerably impact the performance of the precoding algorithm and must be accounted for in the design.}

\textcolor{black}{Satellite systems also have specific antenna structures that account for the size and weight constraints on-board the satellite. As a result, satellite communication systems have a limited number of compact and lightweight antennas. Furthermore, the antenna size influences its gain and radiation pattern, further affecting precoding performance. Another constraining fact of precoding for satellite communication systems is the limited on-board power and processing capabilities. Therefore, processing time and power consumption must be considered in the design of precoding schemes.}

\textcolor{black}{Finally, the precoding design for satellite communication systems must also account for transceiver impairments unique to satellite systems. Specifically, transceiver impairments such as phase noise, non-linear amplification, and carrier frequency offset can degrade the performance of the precoding algorithm. In fact, these impairments can cause significant signal distortion, which can impact the channel estimation's accuracy and the precoding algorithm's performance.}

\textcolor{black}{In the rest of this section, we discuss in more detail some preliminaries about the system and modeling aspects of satellite communication systems.} The system aspects cover the main components of a multi-beam satellite system, namely, the ground, space, user segments, and the utilized spectrum. On the other hand, the modeling aspects represent the necessary considerations in channel modeling and a mathematical formulation of the precoding problem. 

\subsection{System Aspects}

Satellite communication systems have three main components: an on-ground segment, including gateway (GW) stations and control facilities, an on-board segment consisting of the satellite constellation, and a user segment consisting of mobile or fixed user terminals (UT). These three components transmit and receive signals within a certain allocated spectrum, namely the extremely high-frequency band. The end-to-end link from the GW through the satellite to the UTs is the forward link, while the end-to-end link in the opposite direction is the return link, as it is shown in Fig.~ \ref{fig:SatCom_architecture}. Finally, the user link denotes the link between the satellite and the UTs, while the feeder link denotes the link between the satellite and the GW. Commonly, both the user link and the feeder link operate on radio frequencies. Although the alternative of using free space optics for the feeder link has been attracting a lot of attention recently \cite{11}, this work focuses on feeder links operating on the radio frequency band. Further, frequency division duplex is employed since time division duplex can be highly inefficient due to propagation delays. 

\subsubsection{On-ground Segment} 

\begin{figure}[t!]
    \includegraphics[width=\linewidth]{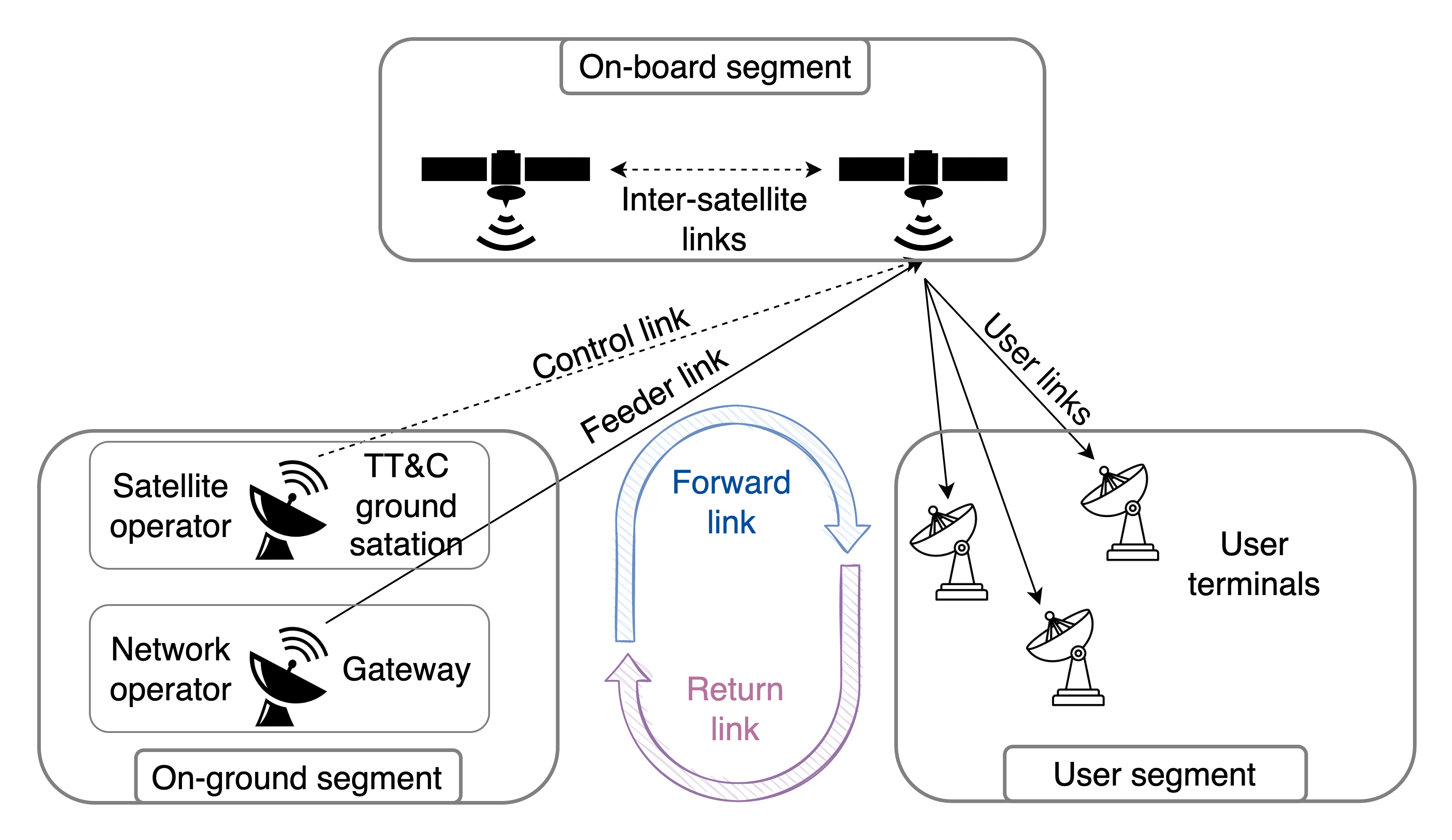}
    \caption{SatCom system architecture\textcolor{black}{\cite{46}}.}
    \label{fig:SatCom_architecture}
\end{figure}

The on-ground segment of a satellite communication system is operated and maintained by two parties, namely, the satellite operator and the network operator, as it is illustrated in Fig.~ \ref{fig:SatCom_architecture}. Satellite operators supervise the satellite subsystems through telemetry, tracking, and control (TT\&C) stations to maintain the correct orbits and configurations and handle failures and updates. On the other hand, the network operator runs and maintains the GW stations to handle network access and backhauling. Reference \cite{Elbert2000} presents more details about the on-ground segment.

\subsubsection{On-board Segment}  

The on-board segment represents the satellite constellation. Depending on the orbit’s altitude, there are three prominent satellite constellations: GEO, MEO, and LEO \cite{46}. GEO refers to geosynchronous equatorial orbit and has an altitude of 36,000 km, MEO refers to medium earth orbit and has an altitude ranging from 5,000 to 25,000 km, and LEO refers to low earth orbit, and its altitude range is 500 to 900 km. The constellation type is a crucial aspect of satellite communications since it influences latency, attenuation, and coverage. In fact, the higher the altitude, the higher the latency and attenuation and the broader the coverage. Further, the versatile nature of satellite constellations has given rise to hybrid constellations combining trade-offs from the different orbits \cite{geo_meo,leo}.

\subsubsection{User Segment} 
The user segment varies depending on the various use cases of satellite communications. The authors in \cite{46} describe the role of satellites in 5G use cases as three-fold: enhanced mobile broadband, ultra-reliable and low latency communications, and massive machine-type communication. Thus, satellites have a wide range of use cases, including fixed network backhaul, IoT connectivity, broadcasting, and direct connectivity in remote areas. Furthermore, satellite systems provide connectivity solutions for on-board moving stations, such as trains and aircrafts, either directly or complementary. They are also used in aeronautical and maritime tracking applications.

\subsubsection{Spectrum}
\begin{figure}[t!]
    \includegraphics[width=\linewidth]{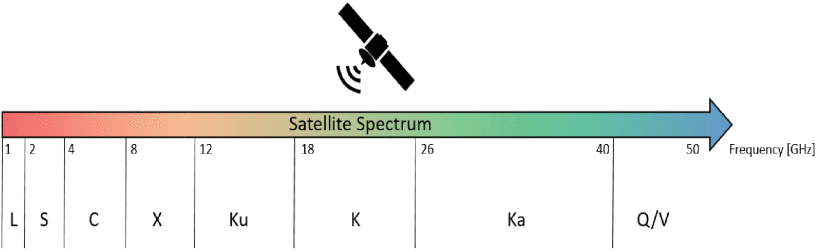}
    \caption{Satellite spectrum \cite{46}}\label{fig:2}
  \end{figure}

Satellite communication systems operate between 1-50 GHz, namely, in the extremely high frequency (EHF) band. Frequency bands are assigned considering the type of delivered service, the type of users, and the climate conditions. Further, they can be split into lower frequencies: L-band, S-band, X-band, and C-bands and higher frequencies: Ku-band, K-band, Ka-band, and Q/V band, as shown in Fig.~ \ref{fig:2}. The L-band is used for mobile satellite services such as Iridium and Inmarsat, as well as global positioning system (GPS) applications. NASA satellites communicating with the international space station operate on the S-band. The S-band is also used for surface ships and weather radars. Further, TT\&C operates on the L and S bands. The C-band is allocated for raw satellite feeds or full-time satellite TV networks. The X-band is used for military, weather monitoring, and terrestrial earth exploration. Finally, the Ku-band, K-band, and Ka-band are used for fixed and broadcasting satellite services. The C-band and Ku-band are now suffering from spectrum scarcity, pushing satellite operators to move to the Ka-band, which provides larger bandwidth at the expense of increased susceptibility to weather conditions.

\subsection{Modeling Aspects}

The most relevant modeling aspect of satellite communication systems in the context of precoding is channel modeling. Various channel models can be considered according to satellite systems’ architectures, constellations, configurations, UTs, and services. Nevertheless, there are three common aspects to the different models \cite{46}: 
\begin{enumerate}
    \item Long-term components: define the first-order statistics of the channel. 
    \item Dynamic components: define the second-order statistics of the channel.
    \item Scatterer-free transmission environment near the satellite antennas. 
\end{enumerate}

Modeling the satellite channel is essentially done by characterizing the propagation conditions. The main two categories of satellites influencing channel modeling and their propagation properties are presented in what follows \cite{47}.  

\begin{table}[t!]
\centering
\caption{List of acronyms}
\begin{tabular}{|c|c|}
\hline
Acronym                                   & Definition \\ \hline
Geosynchronous equatorial orbit           & GEO        \\ \hline
Low earth orbit                           & LEO        \\ \hline
Medium earth orbit                        & MEO        \\ \hline
High throughput satellite                 & HTS        \\ \hline
Medium access control                     & MAC        \\ \hline
Physical                                  & PHY        \\ \hline
Extremely high frequency                  & EHF        \\ \hline
Gateway                                   & GW         \\ \hline
Single-gateway                            & single-GW  \\ \hline
Multi-gateway                             & multi-GW   \\ \hline
User terminals                            & UT         \\ \hline
Telemetry, tracking, and control          & TT\&C      \\ \hline
Additive white Gaussian noise             & AWGN       \\ \hline
Line-of-sight                             & LOS        \\ \hline
Non-line-of-sight                         & NLOS       \\ \hline
Multiple-input multiple-output            & MIMO       \\ \hline
Multi-user multiple-input multiple-output & MU-MIMO    \\ \hline
Channel state information                 & CSI        \\ \hline
Data information                          & DI         \\ \hline
Time division multiplexing                & TDM        \\ \hline
Single feed per beam                      & SFPB       \\ \hline
Multiple feeds per beam                   & MFPB       \\ \hline
Multi-beam antennas                       & MBA        \\ \hline
Radio frequency                           & RF         \\ \hline
Beamforming network                       & BFN        \\ \hline
Signal-to-interference-plus-noise         & SINR       \\ \hline
Minimum mean squared error                & MMSE       \\ \hline
Linear minimum mean squared error         & LMMSE       \\ \hline
Tomlinson-Harashima precoding             & THP        \\ \hline
Dirty paper coding                        & DPC        \\ \hline
Zero-forcing                              & ZF        \\ \hline
Regularized zero-forcing                  & RZF        \\ \hline
Regularized channel inversion             & RCI        \\ \hline
Spectral efficiency                       & SE        \\ \hline
Energy efficiency                         & EE        \\ \hline
Fixed satellite service                   & FSS        \\ \hline
Mobile satellite service                  & MSS        \\ \hline
High power amplifier                      & HPA        \\ \hline
Traveling-wave tube amplifier             & TWTA       \\ \hline
\end{tabular}
\end{table}

\subsubsection{Fixed Satellites}

Fixed satellites operate above 10 GHz to serve fixed satellite terminals. Due to the scatterer-free environment around the user terminals, they are characterized by an additive white Gaussian noise (AWGN) channel and a line-of-sight (LOS) communication. Nevertheless, the Ku-band and especially the Ka-band undergo several atmospheric fading phenomena \cite{Cottis2004}. According to \cite{47}, these mechanisms can be divided into dynamic and long-term channel effects.

\paragraph{Long Term Channel Effect}
These effects typically involve the first-order statistics of the channel. They include attenuation due to precipitation which results in a flat fading process, cloud attenuation,  tropospheric scintillations, Gaseous absorption, signal depolarization, and sky noise increase. A model that considers all of the aforementioned attenuation effects was developed in \cite{48} and is described by 

\begin{equation}
    A_{tot}(p) = A_g(p) + \sqrt{A_s(p)^2 + [A_r(p) + A_c(p)]^2 },
\end{equation}

\noindent \textcolor{black}{where} $A_{tot}(p)$ is the total attenuation, $A_g(p)$, $A_s(p)$, $A_r(p)$, and $A_c(p)$  are gaseous, scintillation, rain, and cloud attenuation respectively. They can be determined empirically for p\% of annual time.

\paragraph{Dynamic Channel Effects}
These effects are related to the temporal behavior of the AWGN channel in the face of rain fading. They are represented using stochastic models that provide second-order statistics of the satellite channel, like the fade duration and slope. For more details about dynamic channel effects modeling, kindly refer to \cite{47}.

\subsubsection{Mobile Satellites} 

Mobile satellites operate below 10 GHz to serve mobile user terminals. They are characterized by LOS and non-line-of-sight (NLOS) components due to the obstructed environment near the user terminals and the non-static propagation property. On the other hand, unlike fixed satellite systems, propagation in the lower frequency bands hinders tropospheric phenomena for mobile satellites \cite{47}. 

Multiple aspects are taken into consideration in the modeling depending on the properties of the mobile terminal environment (rural, urban, suburban). These aspects include time dispersion (wide-band and narrow-band channels), the Doppler power spectrum, and spatial correlation. Further, since a single distribution can be insufficient to model the narrow band mobile satellite channel, various combinations of statistical distributions have been proposed in the literature, including single-state and multi-state models. For more details about mobile satellite channel modeling, kindly refer to \cite{47}.

\begin{table}[t!]
\centering
\caption{List of notations}
\begin{tabular}{ll}
\hline
Notation & Definition\\
\hline
$a, A$ & Scalar or set\\
$\textbf{a}$ & Vector\\
$\textbf{A}$ & Matrix\\
$\textbf{A}^*$ & Conjugate of $\textbf{A}$ \\
$\textbf{A}^T$ & Transpose of $\textbf{A}$\\
$\textbf{A}^H$ & Transpose conjugate of $\textbf{A}$\\
$\textbf{a}_{k}$ & Column $k$ of matrix $\textbf{A}$\\
$a_{i,j}$ & Element $(i,j)$ of matrix $\textbf{A}$\\
$|a|$ & Magnitude of $a$ \\
$\mathrm{E}(.)$ & Statistical expectation\\
$\mathrm{P}(.)$ & Probability measure\\
\hline
\end{tabular}
\end{table}

\subsubsection{General Mathematical Framework} 

The shift from broadcast to broadband satellite services has led to a transition from single-beam to multi-beam architectures. Thus, instead of a global beam covering a fixed geographical region, served regions are now served using multiple narrower beams enabling higher frequency reuse and spectral efficiency. However, the increased throughput comes at the price of inter-beam interference. In this context, joint multi-beam processing techniques (\textit{i.e.} MU-MIMO techniques) allow interference mitigation between users in different beams using multi-user detection and linear or non-linear precoding in the return and forward links, respectively. Unlike multi-user detection and the non-linear dirty paper coding (DPC) precoding \cite{49}, linear precoding is more suitable for multi-beam satellite communication systems thanks to its near-optimal performance and low computational cost. Further, satellite communication standards \cite{54} are designed to cope with severe fading phenomena and long propagation delays characterizing the satellite channel. Thus, long forward error correction codes (FEC) and adaptive coding and modulation (ACM) are used. In this case, multiple users can be accommodated in the same physical frame (codeword) to avoid wasting dummy frames. Therefore, the precoding cannot be designed on a user level, and the precoding problem has to be formulated in the multicast scenario. In this context, we present a general mathematical framework for linear precoding in multi-beam satellite systems, as it is the focus of this work.

Consider the forward link of a multi-beam HTS system employing full frequency reuse. The system is composed of a single GW (multiple GWs can also be considered), a satellite with $N_f$ antenna feeds, generating $N_b$ beams, and a set $U$ of single-antenna UTs with $|U| = N_u$. The UTs are distributed along the $N_b$ beams, such that the $i$-th beam contains a set $U_i$ of users with $|U_i|=N_{u_i}$ verifying $U = \cup_{i=1}^{N_b} U_i$. 

In each time slot, sets $G_i \subset U_i, \forall i \in \{1, \ldots, N_b\}$ of users, with $|G_i| = N_{g_i} $, are selected to be served by multiplexing the data of users of the same beam in the same physical frame. Therefore, in each time slot, a total set of $G = \cup_{i=1}^{N_b} G_i$ of users with cardinal $N_g = \sum_{i=1}^{N_b} N_{g_i}$ is served across all the beams. 

The received signal $y_{i,j}$ of the $j^{th}$ user at the $i^{th}$ beam with $i\in (1..N_b)$ and $j\in (1..N_{g_i})$ can be written as\textcolor{black}{\cite{Christopoulos2015paper}} 

\begin{align}
    y_{i,j} &=  \sum_{i=1}^{N_b} \textbf{h}_{i,j}^H \textbf{w}_{i} s_{i} + n_{i,j}\\
            &= \textbf{h}_{i,j}^H \textbf{w}_{i} s_{i} + \sum_{l\neq i} \textbf{h}_{i,j}^H \textbf{w}_{l} s_{l} + n_{i,j},
\label{eq:linear}
\end{align}
where $s_{i}\in \mathbb{C}^{1\times 1}$ is the symbol/frame destined to be transmitted in beam $i \in (1..N_b)$, $\textbf{w}_{i}\in \mathbb{C}^{N_f\times 1}$ is the precoding vector transforming the beam signal $s_{i}$ to the feed transmissions, $\textbf{h}_{i,j}\in \mathbb{C}^{N_f\times 1}$ is the channel vector that includes all the budget link between the $N_f$ feeds and of the $j^{th}$ user at the $i^{th}$ beam, and $n_{i,j}$ is an AWGN with variance $\sigma_i^2$. 

\textcolor{black}{If we collect all channel vectors $\textbf{h}_{i,j}$ for $i\in (1..N_b)$ and $j\in (1..N_{g_i})$, we will have a total of $N_g$ vectors, each of size $N_f \times 1$ that can be stacked into the multi-beam channel matrix $\textbf{H}$ of size $N_g \times N_f$ \cite{Christopoulos2015paper}, such that}
\begin{equation}
    \textbf{H} = \boldsymbol{\Phi} \textbf{B},
\end{equation}
\textcolor{black}{with $\boldsymbol{\Phi}$, a diagonal $N_g \times N_g$ matrix representing the phase variations due to propagation effects, and $\textbf{B}$ an $N_g \times N_f$ matrix representing the multi-beam antenna pattern, which depends on numerous parameters, including the gain between a particular feed and the user of interest, receive antenna gain, the distance between the satellite and the user, frequency of operation and bandwidth \cite{Sharma2021, Qi2018}.}
 
In this context, the signal-to-interference-plus-noise (SINR) ratio for the $j^{th}$ user at the $i^{th}$ beam with $i\in (1..N_b)$ and $j\in (1..N_{g_i})$ can be written as \textcolor{black}{\cite{Christopoulos2015paper}}

\begin{equation}
     \mathrm{SINR}_{i,j} = \frac{|\textbf{h}_{i,j}^H \textbf{w}_{i}|^2}{\sum_{l\neq i} |\textbf{h}_{i,j}^H \textbf{w}_{l}|^2 + \sigma_i^2}.
\end{equation}

According to \cite{1}, precoding can be defined as ``the design of the transmitted signal to efficiently deliver the desired data stream at each user exploiting the multiantenna spatial degrees of freedom, data, and channel state information (CSI) while limiting the inter-stream interference". Formally, precoding is generally designed by solving a constrained optimization problem where the objective is a performance metric, function of the precoding weights, and involving the CSI and/or data information (DI). The constraints include quality of service (QoS) requirements, which are measured in terms of the SINR, and/or transmitted power \cite{Alodeh2018}. A generic precoding optimization problem can be written as

\begin{align}
\begin{split}
    &\max_{\textbf{w}_1, \ldots,\textbf{w}_{N_b}} \Psi(\textbf{w}_1, \ldots,\textbf{w}_{N_b})\\
    & \mathrm{s.t} ~ \mathrm{SINR}_{i,j} \geq \gamma_i, ~ i \in \{1, \ldots N_b\} \\ 
    & \mathrm{s.t} ~ \rho_l (\textbf{w}_1, \ldots,\textbf{w}_{N_b}) \leq q_l,  ~ l \in \{1, \ldots L\},
\end{split}
\end{align}
where $\Psi(.)$ is a performance metric to optimize, $\gamma_i$ is the SINR threshold for the $i$-th beam (QoS constraints), $\rho_l(.)$ is a function of the precoding weights corresponding to the $l$-th power threshold $q_l$ (power constraints).

\begin{table}[t!]
\centering
\caption{Summary of System Model Parameters}
\begin{tabular}{|c|l|}
\hline
Parameter & \multicolumn{1}{c|}{Definition}                         \\ \hline
$N_f$     & Number of satellite feeds                               \\ \hline
$N_b$     & Number of generated beams                               \\ \hline
$U$       & Total set of user                                       \\ \hline
$N_u$     & Total number of users                                   \\ \hline
$U_i$     & Set of users in the $i$-th beam                         \\ \hline
$N_{u_i}$   & Number of users in the $i$-th beam                      \\ \hline
$G$       & Total set of served users                               \\ \hline
$N_g$     & Total number of served users                            \\ \hline
$G_i$     & Set of served users in the $i$-th beam                  \\ \hline
$N_{g_i}$   & Number of served users in the $i$-th beam               \\ \hline
$y_{i,j}$ & Signal received by the $j$-th user in the $i$-th beam   \\ \hline
$h_{i,j}$ & Channel of the $j$-th user in the $i$-th beam           \\ \hline
$w_{i,j}$ & Precoding vector for the $j$-th user in the $i$-th beam \\ \hline
$s_i$     & Symbol/frame destined to the $i$-th beam                \\ \hline
$n_{i,j}$ & AWGN for the $j$-th user in the $i$-th beam             \\ \hline
$L$       & Number of linear power constraints                      \\ \hline
$N$       & Number of non-linear power constraints                  \\ \hline
$M$       & Number of gateways                                      \\ \hline
\end{tabular}
\end{table}

\subsection{\textcolor{black}{Summary \& Learnt Lessons}}
\textcolor{black}{This section lays the necessary foundations and key concepts for this survey. It starts by specifying the key requirements of satellite communication systems compared to their terrestrial counterpart. In fact, satellite communication systems are characterized by different propagation conditions, interfering sources, antennas, and transceiver impairments. This imposes different considerations on MIMO techniques for satellites compared to terrestrial communication systems. Then, this section proceeds to dive deeper into the aspects of satellite communication systems, namely,  the system and modeling aspects. The system aspects cover the main component of a satellite communication system, namely, the on-ground, on-board, and user segments, in addition to the spectrum allocation of satellite communication systems. On the other hand, the modeling aspects present the main considerations in the channel modeling of satellite communication systems, which separate channel models into two categories: fixed satellite channels and mobile satellite channels. It also presents a general mathematical framework of the precoding problem that will be used through out this survey.}

\section{Precoding: a Cross-layer Design }

In this section, we cover the signal processing techniques for the forward link of multi-beam satellites that interact with the design of precoding from two perspectives: a medium access control (MAC) layer perspective and a physical (PHY) layer perspective. Due to the limited scope of our work, this section is not meant to survey all the MAC and PHY layers concepts. Instead, it provides brief descriptions that will serve as a reference in the analysis of precoding as a cross-layer optimization framework. We refer interested readers  to \cite{46} for more details about the MAC protocols and air interface for satellite communication systems.  

\subsection{Multiple Access Controls Layer}

The MAC layer is at the core of satellite communication systems as it coordinates the transmitted frames, retransmits damaged packets, and resolves data collisions. \textcolor{black}{ The signal processing techniques in this layer address problems such as scheduling, resource allocation, and carrier aggregation. Scheduling protocols in this context can be looked at from a forward or return link perspective. At the forward link, the data streams corresponding to multiple UTs are multiplexed at the satellite and transmitted over a shared channel to the UTs. Thus, appropriate scheduling protocols are employed considering the nature of the satellite communication channel, as well as the satellite communication standards. On the other hand, at the return link, the multiple UTs send their data to the satellite through a shared channel. Therefore, multiple access protocols are used to manage the transmitted frames. These protocols can be broadly classified into fixed and random multiple-access protocols. } 

\textcolor{black}{Fixed multiple access protocols allocate distinct resources to each user. In other words, the network assigns a separate time slot, frequency band, or spreading code to each UTs, which guarantees a  collision-free transmission. Protocols under this category include time division multiple access (TDMA), frequency division multiple access (FDMA), code division multiple access (CDMA), and their variants. However, fixed multiple access protocols can become non-optimal in situations of bursty traffic. In fact,  the resource allocation is typically based on a pre-determined pattern, which makes it more efficient when the number of users is limited, and the traffic pattern is predictable.}

\textcolor{black}{On the other hand, random multiple-access protocols allow users to send their data over the shared channel without prior coordination, inducing possible packet collisions. Protocols under this category include ALOHA, and they are quite compatible with Internet of Things (IoT) communication over satellites. However, despite being implemented in IoT technologies such as NB-IoT\cite{Cluzel2018}, protocols based on ALOHA can have limited performance in modern IoT satellite systems due to the propagation delay in satellite channels. This has driven the design of more advanced random multiple access schemes for satellite IoT \cite{46}, such as enhanced spread-Spectrum ALOHA \cite{6324672}, asynchronous contention resolution diversity ALOHA \cite{6847724}, and contention resolution diversity ALOHA \cite{4155680}, which are shown to have promising spectral and energy efficiency \cite{Mengali2018}. In the rest of this section, we discuss the above-mentioned key components of the MAC layer for HTS systems in more detail.}

\subsubsection{Scheduling} 

In general, scheduling is the action of distributing the satellite resources, in time, among the different UTs. Scheduling is implemented for both the forward link (\textit{i.e.} forward link scheduling) and the return link ( \textit{i.e.} return link scheduling). 

\paragraph{Return link scheduling} \textcolor{black}{It} is needed when the multiple UTs require access to the satellite channel to transmit their data. The return link access for satellite systems is done using multi-frequency time division multiple access (MF-TDMA) \cite{46}, which assigns different carriers to UTs during assigned time slots. Random access can also be implemented for the return link scheduling of satellite systems \cite{46}. In these protocols, each node decides when to access the channel independently from the activity of the other nodes. Therefore, random access has no coordination and may result in collisions. On the other hand, it has the advantage of achieving more efficient use of resources when traffic is bursty compared to fixed assignment protocols. 

\paragraph{Forward link scheduling} \textcolor{black}{It} is needed to enable the transmission of independent data streams from the satellite to different UTs or spot beams. This type of scheduling is closely related to precoding because both techniques are applied for the forward link of a multi-beam satellite. Therefore, we focus on forward link scheduling in what follows, and we refer to \cite{46} for more details about the return link scheduling. Ever since the release of DVB-S2 \cite{540} standard, the satellite communication community has been working on enhancing the performance of forward packet schedulers to guarantee an efficient time distribution of resources among the UTs while satisfying QoS requirements. Typically, the scheduling design takes multiple factors into consideration, namely, channel status, packet priority, QoS requirements, and buffer congestion. It is typically done prior to precoding in the full frequency reuse systems, which closely relate the performance of precoding to the scheduling strategy. Two types of scheduling strategies can be differentiated \cite{46} as it is detailed in what follows.

\begin{figure}[t!]
    \includegraphics[width=0.5\textwidth]{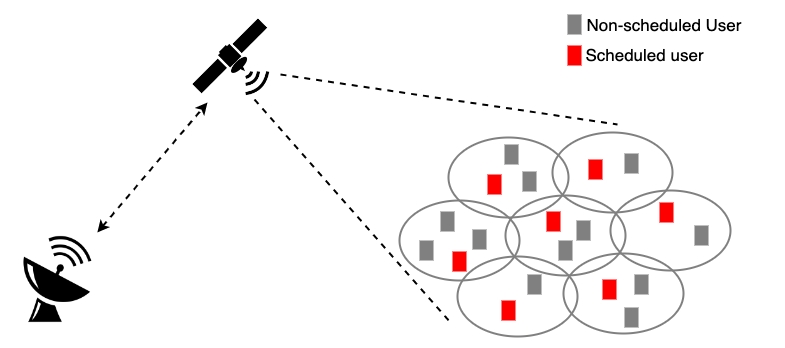}
    \caption{Unicast scheduling for a multi-beam satellite \textcolor{black}{\cite{Vazquez2016}}.}
    \label{fig:unicast}
\end{figure}

\textbf{Unicast} means that the scheduling strategy selects a single user per beam in each frame. Consequently, a set of $N_b$ users, each from a distinct beam, are assigned to synchronous frames in each time slot. This is illustrated in Fig. \ref{fig:unicast}, where the scheduled users are colored in red. In this case, the different users across the beams are served during different assigned time slots in a time division multiplexing (TDM) fashion. The design of unicast scheduling has to take two aspects into account: (i) demand satisfaction and (ii) interference mitigation. Demand satisfaction prioritizes users with large pending data. However, this also depends on the service-level agreement signed by users with the satellite operator, which specifies the minimum, maximum, and average rate over time, as well as the tolerated latency. Interference mitigation imposes on the scheduled users within a set of synchronous frames to have almost orthogonal channel vectors \cite{Qian2019}. 

\begin{figure}[t!]
    \centering
    \includegraphics[width=0.5\textwidth]{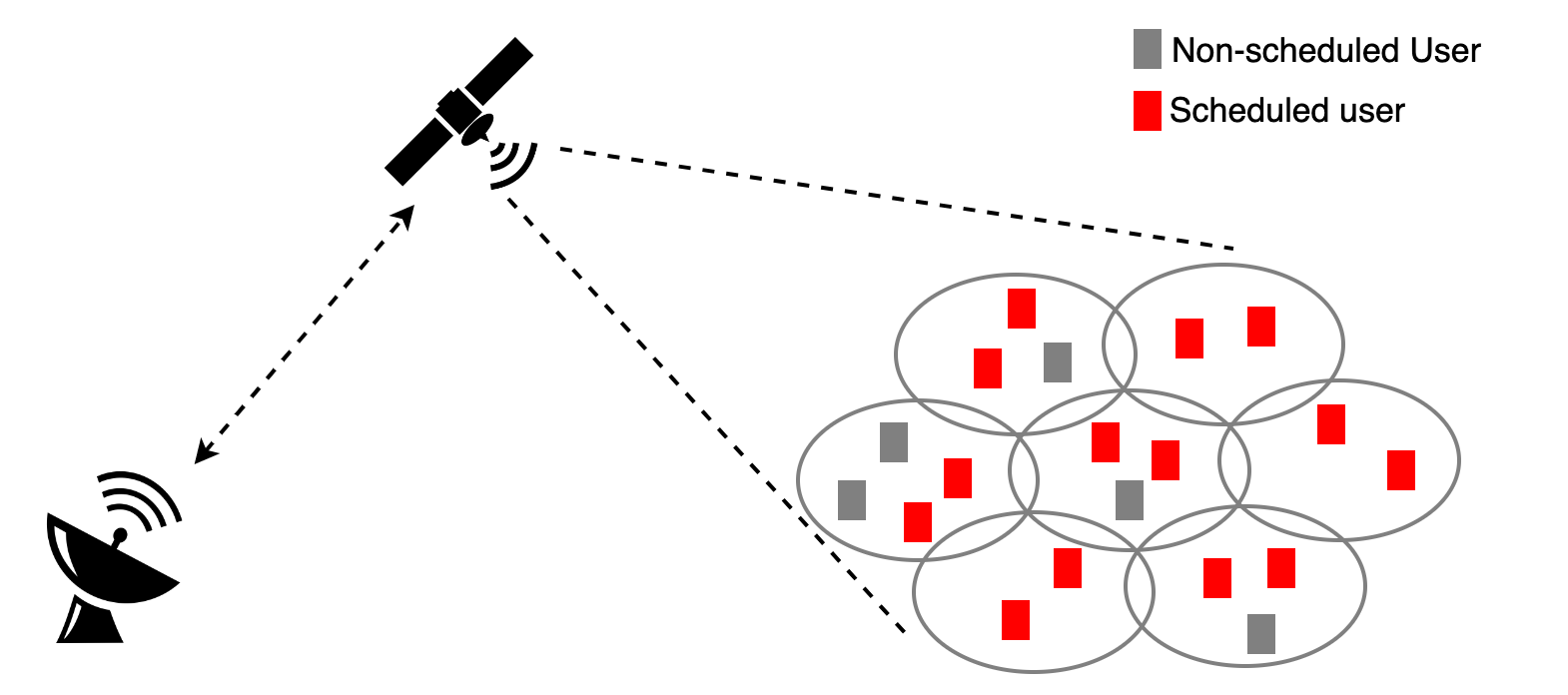}
    \caption{Multicast scheduling for a multi-beam satellite \textcolor{black}{\cite{Vazquez2016}}.}
    \label{fig:multicast}
\end{figure}

\textbf{Multicast} consists of scheduling multiple users per beam in each frame, as it is illustrated in Fig.~ \ref{fig:multicast}. It stems from satellite communication standards that rely on long correction codes making multicasting multiple users in one frame more efficient than a unicasting approach. Therefore, serving only one user per frame is not a practical assumption for a real system. However, the multicasting approach comes with additional design constraints. In fact, instead of scheduling one user in each beam, a group of users has to be scheduled per beam. Then, since all the users within one group (frame) are served via the same modulation and coding (MODCOD) scheme, the scheduler has to group users according to the similarity in their propagation conditions. 

Scheduling for satellite communication systems has been heavily studied in the literature. Works such as \cite{Tropea2011, Castro2007, Rendon2011, Neely2003} do not consider aggressive frequency reuse in the design of the scheduling strategy. On the contrary, employing full frequency reuse entails the need for scheduling and precoding design since they become closely dependent and they greatly influence each other's performance. In this context, multiple scheduling strategies have been proposed in the literature. Some works assume a constant number of users to be scheduled in each beam across all beams \cite{Christopoulos2015paper, Joroughi2017, Taricco2014}. The scheduling in the multicast scenario can be done through various approaches. One approach consists of randomly choosing the first user in each group, then constructing the groups associated with each of the randomly chosen users \cite{Taricco2014,Joroughi2017}. The latter approach can be further improved by choosing the first user in each group based on some orthogonality criteria \cite{Christopoulos2015paper}. Another approach consists of choosing the users per group all at once based on specific geographical strategies \cite{Guidotti2018, 8510717}.

\subsubsection{Power Assignment}

Power is an expensive resource for satellites and has to be efficiently distributed to guarantee the QoS requirements. Traditional single-beam satellites are designed considering long intervals of traffic and are not optimized for the bursty nature of data traffic. However, with the increasing demands for higher data rates and multi-beam architecture, flexibility in power assignment becomes essential to dynamically adapt to the time-varying nature of demands. Traffic demands and propagation conditions are considered, in \cite{ratematching}, to propose a power allocation and packet scheduling strategy. However, this work assumes that adjacent beams are inactive to bypass inter-beam interference, which highly depends on the assigned power per beam. Therefore, inter-beam interference has to be accounted for in the power allocation process to enhance flexibility \cite{6112304}, which closely relates power allocation and precoding.

\subsubsection{Spectrum Assignment}

In addition to flexibility in power allocation, an additional degree of flexibility can be added with spectrum assignment, namely, bandwidth and carrier allocation. Dynamic spectrum assignment techniques can be divided into three categories according to how much spectrum is reused \cite{46}: (i) full frequency reuse, in which all the radio resources are reused in each beam, (ii) semi-orthogonal asymmetric carrier assignment \cite{ratematching, 6112304, 7039249, 5586902}, in which a fraction of the spectrum is reused across the beams, and (iii) orthogonal asymmetric carrier assignment, where the beams do not share any of the spectrum resources. Although (iii) eliminates inter-beam interference, (i) and (ii) are more promising thanks to their increased capacity. Precoding is essential in (iii), while in (ii), certain carriers can be precoded according to the demand.

\subsubsection{Beamhopping}

Typically, all the beams of a multi-beam HTS system are constantly illuminated despite the bursty nature of demand. In fact, traffic demand fluctuates across the beams daily or seasonally \cite{46}. Thus, the heterogeneous nature of data demand over the satellite beams calls for a more dynamic resource allocation strategy. This has motivated a new beam-illumination technique, namely beam hopping \cite{558686}, which allows to flexibly activate beams depending on the demand over time. Specifically, beam hopping exploits all available satellite resources to serve a subset of beams for an allocated time slot. The subset of served beams in each time slot is the illuminated set of beams, which changes in each time frame depending on designed repetition patterns. Thus, beam hopping offers different capacities for the beams to match the uneven demand. However, it comes with a set of new challenges for satellite communication, such as the acquisition and synchronization of bursty transmitted data \cite{Giraud2018} as well as the design of a perfect beam-illumination pattern to match the demand \cite{patent}. Furthermore, beam hopping with full frequency reuse can significantly deteriorate the system's performance due to inter-beam interference \cite{28}.

\subsection{Physical Layer}

The PHY layer handles the data transmission service through channel selection, physical transceivers, energy, and signal management. 
Multiple functionalities are performed at the PHY layer level: modulation, multiplexing, precoding, and beamforming. Here, we focus on the enablers of precoding from a PHY layer perspective, namely, satellite transceivers and MU-MIMO techniques.

\subsubsection{Payloads and Antennas}

\paragraph{Satellite Payload}
 Satellite payloads can be generally categorized into two categories \cite{Sharma2021}: (i) repeaters and (ii) regenerative transponders. Repeaters can further be classified into bent-pipe transponders and digital transparent transponders. Transponders that operate on the bent-pipe principle are limited to tasks such as amplification, filtering, frequency translation, and routing only with analog components. Bent-pipe transponders are widely employed in the current in-orbit satellites, thanks to their simplistic configuration. Nevertheless, there has been a growing interest in incorporating more digital processing into satellite transponders \cite{AngelettiP2008, AngelettiP2009}, such as digital transparent processing and regenerative processing. Digital transparent transponders sample the received signal to process it in the digital domain. They can perform digital beamforming, channelization, routing, digital filtering, and programmable gain amplification. However, neither demodulation nor decoding is carried out \cite{AngelettiP2008}. Compared with analog bent-pipe transponders, digital transparent transponders allow for higher flexibility and power efficiency. Missions employing digital transponders include INMARSAT-4 and SES 12 \cite{SES12}. 
 
 In addition to repeaters, there is the regenerative processing paradigm. Regenerative transponders operate on the digital base-band signal obtained after sampling, demodulation, and decoding. Then, the processed signal is encoded, re-modulated, and retransmitted. Missions employing regenerative transponders include  HISPASAT-AG1, Iridium, and Spaceway3. Therein, regenerative processing is used essentially for routing, switching, and multiplexing. Although this paradigm presents a generalization of digital transparent processing that decouples the user link from the feeder link, it has an increased complexity with a full transmitter and receiver chain for each transponder.

 \paragraph{Satellite Antennas}
 
 Satellite antenna models can be categorized into two groups \cite{46}: (i) passive antennas and (ii) active antennas. While active antenna arrays are more of a recent concept, passive reflector antennas dominate the state-of-the-art with highly efficient models optimized over the years \cite{Sharma2021}. The transition from broadcast to broadband missions gave rise to multi-beam antennas (MBAs) which forged reflector antennas towards multiple feeds per beam (MFPB) and single feed per beam (SFPB) configurations \cite{Rao2015, Palacin2017}.

 In SFPB antennas, each beam is illuminated from a single feed. SFPB MBAs have the advantage of low side-lobe levels and high gain. However, their architecture usually requires 3 or 4 reflectors to attain contiguous coverage \cite{Rao2015, Palacin2017}, which limits additional satellite missions. On the other hand, in MFPB MBAs, each beam is generated by a cluster of feeds. Thus, MFPB has the advantage of ensuring contiguous coverage with fewer reflectors \cite{Rao2015, Palacin2017}, at the cost of a more complex beamforming network (BFN). Hence, MFPB is not always the preferred choice for passive antenna architectures \cite{Fenech2016}. GEO HTS systems are dominated by SFPB and MFPB, which are driven towards more flexibility using multiport amplifiers \cite{Egami1987}, and larger reflectors \cite{46}. However, the fast-growing need for flexibility in power allocation and directivity is attracting efforts to design active antenna array solutions. 
 
 An active antenna is based on integrated amplifiers within the radiating elements. Therefore, compared with a passive antenna, an active antenna allows for a spatially distributed amplification of the signal. Further, active antenna arrays have more reliability thanks to their limited peak RF power levels and power distribution. Active antennas can be implemented using an array-fed reflector (AFR) or a direct radiating array (DRA). Each of these architectures comes with a set of trade-offs that can be assessed depending on the system requirements \cite{46}.

\subsubsection{Multiple Antenna Techniques}

 Two multiple antenna techniques that fall within the scope of this survey are precoding and beamforming. Here, a clarification of the commonly adopted terminology in literature is necessary \cite{Christopoulos2015}. 

\paragraph{Beamforming}

 In terrestrial communication systems, beamforming refers to the process that steers the antenna's radiation beam to follow the users' position. This terminology stems from uniform linear arrays and their ability to direct their main lobe towards any angular direction. Nevertheless, beamforming in satellite communications implies a fixed beam pattern generation based on geographic zones. The beam radiation patterns are generated using an on-board BFN that combines the feed signals. Usually, the generated beams provide coverage considering traffic requirements. Such beam patterns have a quasi-static nature and are not adapted to temporal variations of traffic. In this context, user requirements and propagation conditions are exploited in the beam pattern design rather than the instantaneous CSI of the users. This traffic-aware, user-agnostic, and quasi-static technique refers to beamforming. It is typically implemented on-board the satellite using analog processing due to limited flexibility \cite{Shankar2021}.

\paragraph{Precoding}

 Unlike beamforming, precoding exploits the instantaneous CSI to optimize the transmission to the different users. Therefore, precoding provides traffic-agnostic, and user-aware transmission \cite{Shankar2021}. On the other hand, it requires an adaptive BFN to handle the time-varying nature of CSI. Precoding is carried out on-ground using base-band digital processing, which guarantees large computational resources but necessitates a larger feeder link requirement because the feed signals must be transmitted to the satellite \cite{Shankar2021}. 
 Recently, with the growing need for the efficient exploitation of satellite resources, there have been various precoding approaches involving cross-layer optimization designs. Power allocation is typically done jointly with carrier assignment \cite{Barcelo2009, Lei2010, Schubert2004}. However, there have been many works combining precoding with power allocation or carrier assignment in joint designs. Another common approach combines precoding with scheduling in a cross-layer optimization design, where the networking layer provides the traffic class, the MAC layer provides the destination of the packets, and the PHY layer handles the CSI and applies precoding \cite{Zhang2019, Shankar2021}. 

\subsection{\textcolor{black}{Summary \& Learnt Lessons}}
\textcolor{black}{This section discusses signal processing techniques for HTS from a MAC layer and PHY layer perspective. Specifically, this section focuses on the signal processing techniques that interact with precoding at these two layers. At the MAC layer, techniques such as forward and return link scheduling, power assignment, spectrum assignment, and beamhopping are discussed. On the other hand, aspects like the satellite payload and antennas are described in the PHY layer, and multiple antenna techniques, such as beamforming and precoding, are defined. Therefore, this section emphasizes the cross-layer design of precoding for HTS. In fact, although precoding and beamforming, which are the focus of this survey, are PHY layer techniques, their performance is directly influenced by higher layer techniques such as MAC protocols. Therefore, optimal techniques result from combining these signal processing techniques in a cross-layer design, which will be further discussed in the next section by using this section as a point of reference. }

\begin{figure*}[t!]
    \centering
    \includegraphics[width=\linewidth]{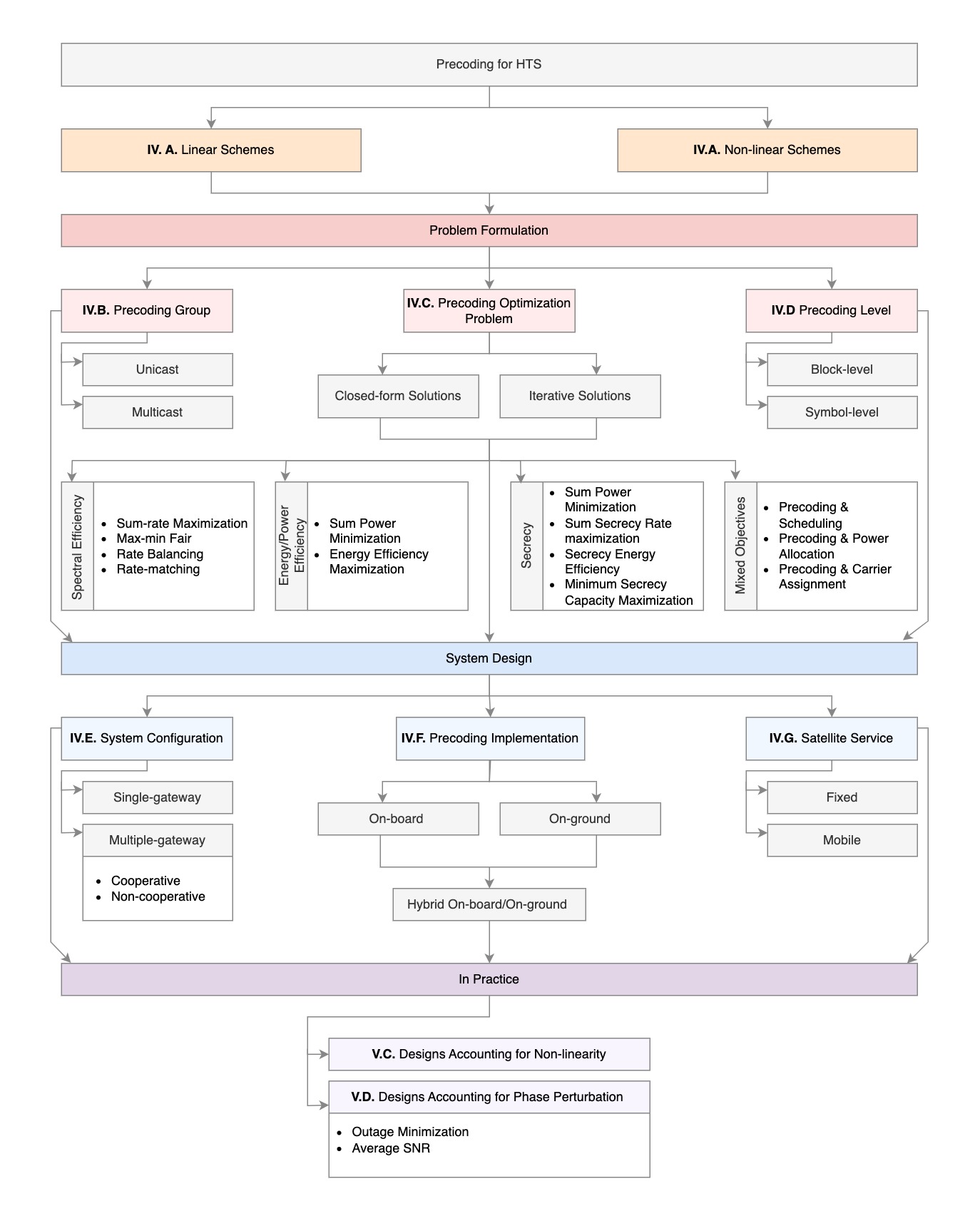}
    \caption{\textcolor{black}{Classification of precoding techniques for HTS.}}
    \label{fig:classification}
\end{figure*}

\section{Precoding: a Classification}

 \subsection{\textcolor{black}{Classification Outline}}

 \textcolor{black}{Precoding is a signal processing technique that exploits information about the channel and the transmitted data to mitigate interference, and it can be broadly categorized into linear and non-linear techniques.}
 
\textcolor{black}{ Linear precoding consists of multiplexing the data symbols using a linear transformation. It is commonly done by multiplying the vector of data symbols by a matrix of complex weights called the precoding matrix. Then, the resulting precoded signal is transmitted over the wireless channel. These linear techniques include zero-forcing (ZF) precoding \cite{Taesang2005, TaesangYoo2006}, minimum mean square error (MMSE) precoding \cite{Peel2005, Ruly2003}, and singular value decomposition (SVD) precoding \cite{SVD}. ZF precoding is derived by designing the precoding matrix orthogonal to the interference subspace, nullifying the interference. However, it may amplify the noise at the receiver, leading to poor performance. MMSE precoding minimizes the mean square error between the transmitted and received signals and takes both the interference and the noise into consideration in the precoding matrix design. Finally, SVD-based precoding consists of decomposing the precoding matrix using SVD decomposition and exploiting the orthogonality of the resulting matrices to use the singular values of the channel matrix to perform precoding.
}

\textcolor{black}{Non-linear precoding, on the other hand, applies non-linear processing on the data symbols before transmission. Non-linear precoding techniques include dirty paper coding (DPC) \cite{DPC} and Tomlinson-Harashima precoding (THP) \cite{THP}. DPC achieves the channel capacity without a power constraint. However, it assumes that the transmitter has knowledge of the interference. A suboptimal approximation of DCP is THP precoding, which relies on feedback loops between the transmitter and receiver to design the precoding scheme. These techniques can enhance the performance of the system compared to linear precoding. However, they are generally highly costly and impractical to implement.}

\textcolor{black}{In the context of multi-beam satellite communication, interference mitigation between users in different beams can be performed using linear or non-linear precoding in the forward link. Unlike the non-linear dirty paper coding (DPC) precoding \cite{49}, linear precoding is more suitable for multi-beam satellite communication systems thanks to its near-optimal performance and low computational cost. Therefore, it will be the focus of our survey.}

\textcolor{black}{In addition to categorizing precoding techniques according to the linearity aspect, the classification of precoding techniques can be approached from two main perspectives: problem formulation and system design. }

\textcolor{black}{From a problem formulation perspective, precoding is categorized according to its group, optimization problem, and level. The precoding group is related to the scheduling strategy incorporated into the precoding scheme, which may be unicasting or multicasting. Unicasting is when only one user per beam is scheduled in each time slot, while multicasting schedules more than one user per beam to be served in each time slot. The precoding optimization problem is related to the optimized objective function, the considered constraints, and the solution. Specifically, objective functions include spectral efficiency, energy efficiency, secrecy maximization, or mixed objectives, and they are generally optimized under power or quality of service constraints. The solution to the precoding optimization problem can be closed-form or can be determined by an iterative optimization algorithm, which we refer to as an iterative solution. Finally, the precoding level is determined by the switching rate of precoding. When the precoding scheme is updated with every block of symbols, we refer to it as block-level precoding. On the other hand, when the precoding needs to be updated with every new symbol, we refer to it as symbol-level precoding. }

\textcolor{black}{From a system design perspective, precoding can be categorized according to the architecture, implementation, and type of service. The system architecture can be based on a single-gateway (single-GW) system or a multi-gateway (multi-GW) system. The multi-GW systems can be cooperative or non-cooperative depending on whether or not CSI and data information are exchanged. Precoding implementation can be done either on-board the satellite, on-ground at the GW, or hybrid on-ground/on-board. Lastly, the service type of a satellite communication system can be either fixed or mobile depending on the mobility of the UTs. Moreover, this survey covers the practical payload and channel impairments for satellite communication systems and presents robust precoding techniques in this context. A summary of this classification is given in Fig. \ref{fig:classification}.}

In this section, we provide a detailed classification of precoding techniques for multi-beam satellite systems according to six aspects discussed above. The first three aspects, namely, precoding group, objective, and level, are from a problem formulation point of view. The three last aspects are from a system design perspective, which include system configuration, precoding implementation, and satellite services.  

\subsection{Precoding Group}

 In practical satellite systems with multi-beam joint processing, each beam covers a large geographical area with multiple users to serve. Therefore, it is common for users to outnumber the beam antennas. This deteriorates the performance of broadcast channel MU-MIMO techniques \cite{47}. In the literature, the problem of serving multiple users per beam has been addressed in two main fashions. The first formulates the precoding problem in the context of unicasting, and the second considers a multicasting context.

\subsubsection{Unicast} 

 In general, unicast is a service type where a transmitter communicates with multiple receivers by sending an individual message to each receiver. Thanks to MIMO technologies, multiple streams corresponding to multiple receivers can be sent simultaneously, and precoding can be used to mitigate multi-user interference. Unicasting with full-frequency reuse in the multi-beam satellite forward link consists of considering only a single user per beam to be served in each time frame.
 Thus, users within the same beam are served in TDM fashion, and the user selection for each time slot is made through a prior scheduling process. This guarantees that the number of users scheduled per time slot is equal to the number of beams $N_b$. In this case, the SINR ratio of the user at the $k^{th}$ beam, with $i\in (1..N_b)$, at a given time slot, can be written as \textcolor{black}{\cite{Shankar2021}}

\begin{equation}
     \mathrm{SINR}_{i} = \frac{|\textbf{h}_{i}^H \textbf{w}_{i}|^2}{\sum_{l\neq i} |\textbf{h}_{i}^H \textbf{w}_{l}|^2 + \sigma_i^2},
\end{equation}
 where $\textbf{h}_{i}\in \mathbb{C}^{N_f\times 1}$ is the channel vector that includes all the budget link between the $N_f$ feeds and the selected user at the $i^{th}$ beam for the considered time slot.

 In the context of precoding, assuming one user per beam per transmission leads to user level precoding scheme. Among the first studies of MU-MIMO precoding for broadband satellite services was done in \cite{EURECOM1, EURECOM2} and it summarized an activity carried out by ESA \cite{esa}. Both \cite{EURECOM1} and \cite{EURECOM2} investigate the performance of the MMSE linear precoder considering a unicast scenario for transparent multi-beam satellite systems in the forward link as per the  DVB-S2 standard \cite{540}. Moreover, \cite{EURECOM2} addresses the issue of fairness among users by considering the UpConst algorithm to maximize system throughput while maintaining fairness between UT regardless of location. Regarding user selection, \cite{EURECOM1} assumes one user per beam, which is not common in practical systems. On the other hand, \cite{EURECOM2} considers multiple users per beam and randomly selects one UT per beam per transmission before applying the precoder. This random selection does not fully exploit the system capabilities because the randomly scheduled user is not necessarily the best. This problem has been addressed in \cite{Zorba2008} where opportunistic beamforming was investigated to increase the sum rate, including a simple UT selection process based on the corresponding SNIRs fed back to the GW. These initial works highlight the potential of linear precoding to enhance the capacity of multi-beam satellite systems and feature the practical impairments, such as channel estimation and on-board non-linearity, that may limit its performance. 

 Later on, the non-linear Tomlinson-Harashima precoding (THP) was studied in \cite{4409391} and \cite{5198782} under the same assumption of one user per beam. \cite{4409391} shows the potential performance boost of applying THP for broadband satellites compared to linear precoding and traditional frequency division among beams. However, it also identifies multiple implementation issues such as synchronization, channel estimation, and non-linear effects due to satellite high-power amplifiers. \cite{5198782} provides a bit error rate analysis of THP for multi-beam satellite systems and various insights on its capacity improvement. Further, it highlights the impact of the positions of simultaneously served users (located in the different beams) on THP performance. More work under the same assumption of user scheduling followed up in \cite{6190179}, \cite{Christopoulos2012} and \cite{Zheng2012}. In \cite{6190179}, linear MMSE precoding and power optimization have been studied for satellite forward link channels. In \cite{Christopoulos2012}, the analysis is further extended to include the return and forward link channels of multi-beam satellites. In the forward link, linear precoding is studied compared to the non-linear DPC as the upper limit for the performance. In the return link, the linear MMSE receiver is studied compared to successive interference cancelation as the upper bound of the performance.

 Although precoding schemes serving only one user per beam offer considerable capacity gain, satellite precoding is essentially different from its terrestrial counterpart by its multicasting nature. In fact, a single DVB-S2 \cite{540} or DVB-S2X \cite{54} codeword (frame) is designed to incorporate data bits addressing multiple UTs \cite{Arapoglou2016}. Therefore, more than one user needs to be served by the same DVB-S2X frame, and consequently, one codeword can be decoded by more than one user. This has motivated a new research direction for satellites called frame-based precoding \cite{Christopoulos2015}, \textcolor{black}{which will be discussed in the following subsection}.

\subsubsection{Multicast}

 In general, multicasting is a service type where a transmitter communicates with multiple users by sending a common message to a subset of the users (the multicast group). The first extreme case, where the addressed subset of users contains only one user, represents unicasting. The second extreme case, where the addressed multicast group contains all the users, represents broadcasting. Thanks to multiple antennas, multiple streams corresponding to different multicast groups can be transmitted simultaneously, and interference management techniques can be used to limit interference between the different subsets of users. \textcolor{black}{In the terrestrial context, multicasting has been applied in two different manners. The first is single-group multicasting which allocates distinct radio resources to each multicast group \cite{730451, 1311613, 1634819, 4289084}. The second is multigroup multicasting which allows subsets of multicast groups to share the same radio resources \cite{1574196, 1574217, 9093950, 4305444}}. In the forward link of multi-beam satellites with full frequency reuse, it is more practical and compatible with satellite standards to formulate the transmission as a multicast problem. More specifically, the framing structure featured in satellite communication standards \cite{Arapoglou2016} imposes multiplexing the packets of multiple users together within one frame by interleaving their bits. Therefore, these users receive and decode the same physical layer frame and thus, can be considered as a multicast group. Further, with one multicast group considered per beam, all the multicast groups share the same radio resources in the full-frequency reuse scenario. In this case, the same precoding weights will be applied to symbols of multiple users belonging to the same frame. Therefore, designing an optimal frame-based precoding scheme has to be done in the context of multigroup multicasting \cite{Christopoulos2015}.

 The multigroup multicast formulation raises the problem of user clustering, which addresses the strategy of grouping users in each beam ( \textit{i.e.} the multicast groups selection). Once the multicast groups are identified, the follow-up question is how to process the channel of the users in each beam (\textit{i.e.} multicast group) to have one representative channel per beam (\textit{i.e.}  per multicast group) \cite{Guidotti2018}. In the context of multigroup multicasting, the SINR of the \textcolor{black}{$j^th$} user at the $i^{th}$ beam, with $i\in (1..N_b)$, $j\in (1..N_{g_i})$ can be written as \textcolor{black}{\cite{Shankar2021}}

\begin{equation}
     \mathrm{SINR}_{i,j} = \frac{|\textbf{h}_{i,j}^H \textbf{w}_{i}|^2}{\sum_{l\neq i} |\textbf{h}_{i,j}^H \textbf{w}_{l}|^2 + \sigma_i^2},
\end{equation}

 The first work modeling multi-beam precoding for satellites as a multigroup multicast problem was carried out in \cite{Taricco2014}, where geographic user clustering (further investigated in \cite{Guidotti2018} and \cite{Guidotti2020}) was proposed to select the users served by the same DVB-S2 frame. In the precoding design, the set of selected users is modeled as one user with a representative channel corresponding to the arithmetic mean of the selected channels. The scheme shows promising spectral efficiency results. Nevertheless, accounting for several users per frame deteriorates the channel averaging process. Later on, the authors in \cite{Christopoulos2014} formally made the connection between frame-based precoding for multi-beam satellites and physical layer multicasting to multiple cochannel groups, namely, multigroup multicast precoding. To this end, \textcolor{black}{they extended the weighted fair multigroup multicast precoding \cite{Christopoulos2014Aug, Christopoulos2014Oct}, originally proposed for general multi-antenna systems,} to a multi-beam satellite transmitter with a random user scheduling and show a considerable throughput gain with respect to four-color frequency reuse and Multicast Aware MMSE \cite{Silva2009}. In \cite{Christopoulos2015paper}, the sum rate maximization problem constrained by per-antenna power consumption is investigated for frame-based precoding. Unlike \cite{Christopoulos2014} that adopts random scheduling, the work in \cite{Christopoulos2015paper} employs a CSI-based user scheduling to achieve a considerable throughput gain over conventional system configurations.

 Later on, a low complex ground precoding scheme was proposed in \cite{Joroughi2017}. It consists of a two-stage linear precoding technique \textcolor{black}{similar to the generalized design of precoding matrices in \cite{Stankovic2008}. Further, it} offers higher spectral efficiency than the average MMSE scheme in \cite{Taricco2014}, the regularized zero-forcing (RZF) in \cite{Silva2009}, and the frame-based precoding scheme in \cite{Christopoulos2015paper}. The work in \cite{Joroughi2017} differs from \cite{Christopoulos2015paper} by decoupling the problems of precoding and user scheduling. It employs a variation of the k-means algorithm for user scheduling. However, rather than accounting only for channel magnitudes for user grouping as in \cite{Taricco2014}, the work in \cite{Joroughi2017} considers both the channel magnitudes and the phase effects. In \cite{8510717}, the MMSE precoder, which coincides with the popular RZF approach \cite{Vazquez2016}, \cite{4599181}, was investigated for a broadband GEO multi-beam satellite system with aggressive frequency reuse. In this work, the precoding at the physical layer is closely coupled with scheduling from the MAC layer and QoS constraints imposed by upper layers. Therefore, unlike \cite{Christopoulos2015paper}, authors in \cite{8510717} consider heterogeneous broadband traffic by giving priority to delay-sensitive packets and ensuring a satisfactory throughput to non-real-time packets.

 Most of the aforementioned works, namely \cite{Taricco2014}, \cite{Christopoulos2014}, \cite{Christopoulos2015paper}, \cite{Joroughi2017} and \cite{8510717} assume perfect CSI at the GW. However, authors in \cite{Wang2018} proposed more robust precoding designs addressing the fairness and power minimization problem with partial CSI. Further, the work proposed user clustering to enhance the precoding robustness and performance. Another important consideration when combining resource allocation techniques with precoding in the multigroup multicast scenario is satellite traffic heterogeneity over time and space. In fact, maintaining the same level of offered capacity across all beams all the time can lead to inefficient utilization of radio resources. This problem of bursty traffic demand in the satellite coverage area has been recognized and addressed in multiple European Space Agency (ESA) projects \cite{ESABH1, ESABH2}. In this context, beam hopping (BH) emerges as a potential solution to adapt the offered capacity to the changes in the traffic demand over space and time \cite{558686}. BH illuminates a subset of the satellite beams for a certain time based on a designed time-space transmission pattern \cite{Angeletti2006,5586860,5586902,6112304,558686,Hu2020, 41}. This concept has been investigated in conjunction with precoding in \cite{ 28, Zhang2021} to provide more flexible precoding designs.

\subsection{Precoding \textcolor{black}{Optimization Problem}}

In principle, precoding for multi-beam satellite communication systems consists of exploiting the available information about the channel to manage interference. Formally, a precoding scheme is designed to optimize a specific objective under several constraints. \textcolor{black}{ Closed-form solutions involve the popular  ZF precoding \cite{Taesang2005, TaesangYoo2006}, RZF \cite{Nguyen2008}, and MMSE precoding \cite{Peel2005, Ruly2003}. In addition to the closed-form solutions, other precoding schemes are designed as the solutions for the numerical optimization of certain objective functions under a set of constraints.} Classically, two main optimization objectives can be targeted by design: power minimization and spectral efficiency, and more recently, mixed objectives that combine the precoding scheme design with radio resource management such as scheduling, bandwidth, or power allocation are proposed. 

\subsubsection{Spectral Efficiency}
  
 The need for high throughput satellite communications is ever-increasing due to limited spectral resources and technological immaturity for higher bands. In this context, precoding allows more efficient exploitation of the already used frequency bands by utilizing the knowledge about the channel to precode the transmitted signals. The design of the precoding scheme is determined via the optimization of certain objective functions, and some of the common objectives related to spectral efficiency include fairness and sum rate maximization problems. Other objective functions that account for the traffic demand can also be considered to maximize spectral efficiency, namely, rate balancing and rate matching. These objectives are generally optimized under the set of power constraints that can be linear, such as sum power constraints and per-antenna power constraints, or non-linear,\textcolor{black}{which we represent in a generalized form as in \cite{Zheng2012} in what follows.} 
  
  \paragraph{General Linear Power Constraints} An arbitrary number of $L$ linear power constraints can be expressed as 
  
  \begin{equation}
     \sum_{k=1}^{N_b} \textbf{w}_k^H \textbf{Q}_l \textbf{w}_k \leq q_l, ~ \forall l \in \{1,\ldots,L\}.
  \end{equation}

 Here $q_l > 0$ is the power threshold, and $\textbf{Q}_l$ is a positive semi-definite shaping matrix. It is worth mentioning that this representation takes into account the sum-power constraint as a special case for $\textbf{Q}_l = \textbf{Q} = \textbf{I} \text{ and } L = 1$. It also includes the per-antenna power constraints as a special case if we take  $L = N_f$ and $\textbf{Q}_l$ to be a matrix whose elements are zero except the $l$-th diagonal element is 1.
 
  \paragraph{General non-linear Power Constraints} An arbitrary number of $N$ non-linear power constraints can be expressed as 
  
  \begin{equation}
     \sum_{k=1}^{N_b} f_{k,n}(\textbf{w}_k^H \textbf{Q}_n \textbf{w}_k) \leq q_n, ~ \forall n \in \{1,\ldots,N\}, 
  \end{equation}
 where $f_{k,n}(.)$ is a non-linear function assumed to be continuously increasing, relating the beam output power to the input power. These non-linear functions of beam powers are needed, for instance, to map the output power of the HPA to the available direct current power.

 Objective functions can be written differently depending on the problem formulation as unicast or multicast. \textcolor{black}{In what follows, we present these problems under the unicasting assumption, for simplicity, as in \cite{Chatzinotas2011, Zheng2012}}. However, these representations can be generalized under the assumption of multi-group multicasting as in \cite{Christopoulos2014Oct, Christopoulos2014, Christopoulos2015paper, Joroughi2017}. Further, we present the optimization problems considering linear power constraints as they are widely adopted. We consider the Shannon rate to determine the throughput. However, it is essential to highlight that, in practical systems, the throughput performance is constrained by modulation. Therefore, modulation-aware precoding schemes consider the spectral efficiency of the MODCOD scheme of the studied standards, such as DVB-S2X \cite{Christopoulos2015paper}.

  \paragraph{Sum-rate Maximization Problem}
 
  \begin{align}
  \begin{split}
      &\max_{\textbf{w}_1, \ldots, \textbf{w}_{N_b}}  \sum_{k=1}^{N_b} log(1+\mathrm{SINR}_{k})\\
      & \mathrm{s.t} \sum_{k=1}^{N_b} \textbf{w}_k^H \textbf{Q}_l \textbf{w}_k \leq q_l, ~ \forall l \in \{1,\ldots,L\},
  \end{split}
  \end{align}
  
  In these types of objectives, a weighted sum can be adapted to indicate delay constraints or the priority of each user or user group.
  
  \paragraph{Max-min Fair Problem}
  
  \begin{align}
  \begin{split}
      &\max_{\textbf{w}_1, \ldots, \textbf{w}_{N_b}} \min_{ 1 \leq k \leq N_b}\mathrm{SINR}_{k}\\
      & \mathrm{s.t} \sum_{k=1}^{N_b} \textbf{w}_k^H \textbf{Q}_l \textbf{w}_k \leq q_l, ~ \forall l \in \{1,\ldots,L\},
  \end{split}
  \end{align}

  \paragraph{Rate Balancing Problem}
  
  \begin{align}
  \begin{split}
      &\max_{\textbf{w}_1, \ldots, \textbf{w}_{N_b}} \min_{ 1 \leq k \leq N_b} \frac{ log(1+\mathrm{SINR_{k}})}{F_k}\\
      & \mathrm{s.t} \sum_{k=1}^{N_b} \textbf{w}_k^H \textbf{Q}_l \textbf{w}_k \leq q_l, ~ \forall l \in \{1,\ldots,L\},
  \end{split}
  \end{align}
  where $F_1, \ldots, F_{N_b}$ are the traffic demands in each of the beams. 
  
  \paragraph{Rate-matching Problem}
  
  \begin{align}
  \begin{split}
      &\min_{\textbf{w}_1, \ldots, \textbf{w}_{N_b}} \sum_{k=1}^{N_b} |log(1+\mathrm{SINR_{k}}) - F_k |^p\\
      & \mathrm{s.t} \sum_{k=1}^{N_b} \textbf{w}_k^H \textbf{Q}_l \textbf{w}_k \leq q_l, ~ \forall l \in \{1,\ldots,L\},
  \end{split}
  \end{align}
  where $p$ is a predefined order. Rate-matching problems are used in the context of satellite communications to tackle the discrepancy often found in traffic demand \cite{ratematching}. Further, we can see that for $p=1$, the rate matching problem becomes equivalent to a sum-rate maximization problem since we have 
  \begin{equation}
  \sum_{k=1}^{N_b} | log(1+\mathrm{SINR_{k}}) - F_k | = \sum_{k=1}^{N_b} F_k - \sum_{k=1}^{N_b}  log(1+\mathrm{SINR_{k}}). 
  \end{equation}
 
 The objective functions listed above have been widely studied in the literature on multi-beam satellites with full frequency reuse. Authors in \cite{Zheng2012} considered a fixed multi-beam satellite system and developed a generic linear precoding optimization framework in the forward link. The algorithm can be applied to optimization problems featuring the data rate in the objective subject to general power constraints, and it is proven to always converge.  \textcolor{black}{ Fairness using the max-min problem under total power consumption constraint was investigated for general multi-antenna systems in \cite{Karipidis2008, 1634819, Silva2009}, and again under per-antenna power consumption constraints in \cite{Christopoulos2014Oct}. Then, building on these works, authors in \cite{Christopoulos2014Aug, Christopoulos2014Oct} provided a consolidated precoding scheme addressing a fairness objective formulated as weighted max-min subject to per-antenna power control for multigroup multicasting. Specifically, Gaussian randomization, semi-definite relaxation, and bisection were exploited to provide accurate solutions. Later on, the authors in \cite{Christopoulos2014} applied the framework and solution proposed in \cite{Christopoulos2014Aug, Christopoulos2014Oct} to frame-based precoding for multi-beam satellites. Numerical results highlight the performance gain of this multigroup multicast solution compared to classical heuristic precoding techniques.}

 \textcolor{black}{The problem of sum rate maximization in the context of multigroup multicast was first studied  for terrestrial systems. Two variations were investigated: one in \cite{Kaliszan2012}, which considered sum-power constraints, and another in \cite{DChristopoulos2014}, which considered per-antenna constraints. Building on these prior works, a frame-based precoding scheme for satellite systems was proposed in \cite{Christopoulos2015paper} to optimize the sum-rate while imposing per-antenna power constraints.} The problem was also extended to include minimum rate constraints. Further, the authors in \cite{Christopoulos2015paper} proposed and heuristically solved a modulation-aware throughput maximization problem that relates the spectral efficiency to the MODCOD schemes of DVB-S2X rather than the Shannon rate. Later on, in \cite{Joroughi2017}, the maximum sum-rate problem was formulated by considering the rates of the worst users over all the beams. It was solved by splitting the precoding matrix into two matrices, one for maximizing the intra-frame rate and the other for interference management. In \cite{7}, a ZF precoding scheme was investigated as a sub-optimal solution for the sum-rate maximization problem. It employed partial CSI to decrease signaling overheads and lower latency. Additionally, to meet power constraints at the GW matrix, vector normalization was considered. 

 Another objective that falls under spectral efficiency accounts for the traffic demand when optimizing the offered capacity instead of maximizing the capacity without considering the demand. For example, the work in \cite{Chatzinotas2011} studied the rate-balancing performance of linear precoding under flexible power constraints for multi-beam satellite systems with full frequency reuse to achieve much higher spectral efficiency than the conventional partial frequency reuse. Spectral efficiency can also be improved by employing techniques such as faster-than-Nyquist signaling \cite{12, 14}, and rate-splitting-multiple access \cite{9145200, 3, 9473795, SiZhi2022} combined with linear precoding.

\subsubsection{Power Minimization/ Energy Efficiency}

 Although most of the precoding literature for multi-beam joint processing addresses spectral efficiency problems, saving power for satellites is a crucial task to extend the lifetime of the on-board services. One approach to address this problem is via solving the power minimization problem under QoS constraints. In this case, the objective can be minimizing the total transmitted power \cite{22} or the per-antenna transmitted power \cite{Christopoulos2014Oct}. Another interesting approach aims to design precoding schemes to maximize energy efficiency under service and power-control constraints. Energy efficiency is the system throughput per power unit consumption. In the literature on multi-beam satellite communications \cite{6190179, 35}, energy efficiency is defined as the sum rate of all the users or the worst users, divided by the total power of the satellite. These optimization problems can be presented differently based on the considered scenario (\textit{i.e.} unicast or multicast). In the following, we present the common problems of sum power minimization and energy efficiency in the case of a unicast scenario \textcolor{black}{as in \cite{Bengtsson1,6190179}}. However, these representations can be extended to the multigroup multicast scenario \cite{Karipidis2008, Christopoulos2014Oct, 35}. 

\paragraph{Sum Power Minimization}
 
 \begin{align}
  \begin{split}
      &\min_{\textbf{w}_1, \ldots, \textbf{w}_{N_b}}  \sum_{k=1}^{N_b} ||\textbf{w}_{k}||\\
      & \mathrm{s.t} ~ \mathrm{SINR}_{k} \geq \gamma_{k}, ~  1 \leq k \leq N_b, 
  \end{split}
  \label{powermin}
  \end{align}
where $\gamma_{k}$ is the SINR target of the user at the $k$-th beam. 

\paragraph{Energy Efficiency Maximization}

\begin{align} 
\label{eq:ee}
    \begin{split}
         &\max_{\textbf{w}_1, \ldots, \textbf{w}_{N_b}} \frac{\sum_{k=1}^{N_b} \log(1+\mathrm{SINR}_{k})}{P_0 + \sum_{k=1}^{N_b} |\textbf{w}_k|^2 } \\
         & \mathrm{s.t} ~ \mathrm{SINR}_{k} \geq \gamma_{k}, ~  1 \leq k \leq N_b \\
         & \mathrm{s.t} \sum_{k=1}^{N_b} \textbf{w}_k^H \textbf{Q}_l \textbf{w}_k \leq q_l, ~ \forall l \in \{1,\ldots,L\},
    \end{split}
\end{align}
 where $P_0$ is is the aggregation of the power consumed by amplifiers for the feeder, user and control links. It also includes the satellite signal processing units power consumption.

 \textcolor{black}{For general multi-antenna systems, the sum power minimization problem under SINR constraints was studied in \cite{Bengtsson1} and a solution was proposed to tackle the intricacies of the problem using the classical semi-definite relaxation. This problem was extended to the multigroup multicast scenario in \cite{Karipidis2008}, where the authors propose a computationally efficient and accurate approximate solution by combining cochannel multicast power control with Lagrangian relaxation strategies. Other precoding problems for general multi-antenna systems} target a per-antenna power minimization as in  \cite{Christopoulos2014Oct}, where a relationship between the latter and the weighted max-min fair problem was established, offering an elegant framework for it to be solved. \textcolor{black}{ For multi-beam satellite systems, \cite{Qi2018} aimed to minimize the sum power of the forward link. However, in order to account for phase perturbations, the SINR constraints are modeled as probability inequalities, which maintain a sufficiently low inter-beam interference statistically. }

 On the contrary, simple linear precoders such as MMSE and ZF strike a balance between complexity and performance \cite{Christopoulos2012, 7, Vazquez2016}. Precoding schemes employing ZF and sequential convex approximation were studied in \cite{Qi2018} for the energy efficiency maximization problem constrained by the QoS levels and the total power consumption. The results imply that reducing the satellite operation power can be effective in improving energy efficiency. However, the authors in \cite{Qi2018} hypothesize a unicast scenario that employs a TDM scheduling in which, for each time slot, only one user is scheduled per beam. Later on, the work in \cite{Qi2018} was extended to the multigroup multicast scenario in \cite{35}. The authors in \cite{35} formulated an optimization problem that maximizes the energy efficiency under power and $\mathrm{SINR}$ constraints per beam. Since the problem is non-convex, fractional programming was applied along with the first-order Taylor low-bound approximation. Then, the Charnes-Cooper transformation was employed to convexify the optimization problem. To this end, the designed precoding scheme demonstrates a higher energy efficiency performance compared to the multi-beam interference mitigation algorithm in \cite{Joroughi2017}, the regularized ZF algorithm in \cite{Stankovic2008}, and MMSE algorithm in \cite{EURECOM2}. Authors in \cite{34} also hypothesize a unicast scenario, where they investigate a robust ZF THP scheme for multi-beam satellite forward link systems to enhance the energy efficiency when CSI at the transmitter is imperfect. Therein, simulation results show the great influence of system parameters, such as the transmit power, the circuit power, and the CSI error, on the energy efficiency performance.
  
\subsubsection{Physical Layer Secrecy Maximization}

 Satellite communication systems suffer from an inherent broadcasting nature and LOS properties. Therefore, privacy and security issues in satellite networks have always received great efforts, especially in space \cite{securitySpace} and military applications.
 Security of the wireless communication medium is a mature research topic that has been addressed using multiple approaches \cite{wang2019}, namely, cryptography \cite{Spinsante2006} in higher layers, and information theory in the physical layer \cite{Kalantari2015, Hamamreh2019}. Regarding satellite communication, it is commonly addressed in higher layer protocols \cite{Slim2009, Liang2009, Spinsante2006}. Recently, with the new satellite standard DVB-S2x \cite{54} laying the ground for precoding schemes implementation, physical layer secrecy using precoding for multi-beam satellites has been receiving increased attention. The problem is formulated considering a satellite transmitting to a number of legitimate users messages that a number of eavesdroppers can intercept. Then, a precoding scheme is commonly formulated to optimize a secrecy performance metric while considering power and QoS requirements. Common secrecy metrics include secrecy capacity, ergodic capacity, rate,  outage probability, and energy efficiency. We refer interested readers to \cite{LiBin2020} for further details about the secrecy metrics.

 One of the first works to address physical layer secrecy using beamforming for multi-beam satellites was done in \cite{Lei2011}. Therein, the problem of transmit power minimization with individual secrecy rate constraints was investigated, assuming fixed beamforming weights and perfect CSI. Further, a joint beamforming and power control problem was solved to provide sub-optimal beamforming weights and power allocation. The performance evaluation studies the impact of CSI, the number of individual secrecy rate requirements, and the number of beams and antenna elements. Further, it favors the joint beamforming scheme over the fixed beamforming scheme. The problem of total transmit power minimization under secrecy rate constraints was also addressed later on, in \cite{Zheng2012sec}. However, Unlike \cite{Lei2011} where only a single common eavesdropper is assumed for all legitimate users, a general eavesdropping scenario was investigated, where each legitimate user can be surrounded by either their own multiple eavesdroppers or common multiple eavesdroppers. A gradient-based method and semi-definite relaxation are investigated to find an approximate optimal solution. Further, the beamforming design was extended to the case when only channel covariance matrices about eavesdroppers were available. 

 The aforementioned works focus on the beamforming design for a unidirectional service of a satellite communication system with fixed users. On the other hand, \cite{Kalantari2015} investigated a scenario of bidirectional satellite communication, which considers two eavesdroppers and a pair of users communicating via a transparent multi-beam satellite. The beamformer and the return and forward link time allocations are optimally designed to maximize the sum secrecy rate. The performance evaluation is then performed with and without the XOR network coding scheme \cite{Ahlswede2000}, and it shows the advantage of network coding on secrecy rate. More recently, \cite{Li2018} proposed a secure beamforming scheme in a cognitive satellite-terrestrial network to solve a total-transmit power minimization problem subject to outage constraints. Further, \cite{Du2018} investigated an optimization framework to design secure cooperative and non-cooperative beamforming schemes via maximizing the secrecy rate constrained by power and transmission quality levels.

 Nevertheless, the majority of the aforementioned literature \cite{Lei2011, Zheng2012sec, Li2018} merely focused on power minimization problems from the aspect of the network design with predefined secrecy rate constraints. The first work to adopt the sum secrecy rate as an objective to maximize, subject to transmitted power constraints, is \cite{Lu2019}. Herein, the beamforming design considers imperfect CSI of eavesdroppers to design a robust scheme, achieving superiority to perfect CSI approaches. Another objective function, namely, the secrecy energy efficiency, was investigated by the authors in \cite{Lin2019}, under secrecy constraints at the eavesdroppers, signal-to-noise ratio (SNR) constraints at the earth station, and per-antenna power constraints at satellite antenna feeds to design a secure beamforming scheme. In addition to the power minimization, the secrecy energy efficiency, and the sum secrecy rate problems, authors in \cite{Lin2019Apr, Schraml2020, Zhang2021sec} addressed the objective of the minimum secrecy capacity maximization to secure the forward links of multiple users against multiple eavesdroppers, under power and QoS constraints. Recently, the emergence of hybrid satellite-terrestrial relay networks, satellite-terrestrial integrated networks, and cognitive satellites triggered an intensified attention to precoding for physical secrecy \cite{Lin2021, Lin2021RSMA, Lin2021unknownEVE, Lu2021, Zhao2021, Guo2021sec, 11}. We refer interested readers to \cite{LiBin2020} for a comprehensive survey about recent developments of physical layer secrecy in satellite communication.

 \subsubsection{Mixed Objectives} \label{mixed}

 Satellite resources are expensive, and efficient resource allocation is crucial to match the service demands. Usually, resource management happens at the higher layers. However, much work in the literature has been addressing cross-layer optimization problems that combine the precoding design with radio resource management to allocate resources while mitigating interference efficiently. More specifically, each of the objectives addressed in the three previous sections can be extended to a mixed objective that is not only a function of the precoding vectors but also other variables representing resource allocation. The common joint problems include combining precoding with scheduling, power allocation, or carrier assignment. However, \textcolor{black}{in the literature of multi-antenna systems, the term joint design can refer to i) non-iterative decoupled approaches, ii) iterative decoupled approaches, and iii) coupled approaches \cite{Bandi2020MISO}}. For the non-iterative decoupled approaches, resource allocation and precoding are approached as two distinct problems where users' scheduling is performed according to certain considerations, then precoding is designed for the scheduled users \cite{Taesang2006, Lee2018, Song2008}. In iterative decoupled approaches, resource allocation and precoding are still considered distinct problems. Nevertheless, the precoding and scheduling variables are tuned iteratively to optimize the objective considering the previous stages' feedback \cite{Yu2013, Matskani2008, Li2016, Kountouris2008}. Unlike the last two approaches, coupled approaches formulate the problem jointly depending on both the precoding and scheduling variables \cite{Ku2015, Yu2012, Douik2016}. \textcolor{black}{The decoupled approaches for multi-beam satellite systems have been covered in the previous three sections since the considered objective functions depend only on the precoding vectors \cite{Christopoulos2015paper, 8510717, Guidotti2018}, and coupled approaches with mixed objective functions are the focus of this section.}

 \textcolor{black}{User scheduling} is the process of selecting the set of users to be addressed in each transmission. Therefore, it depends on the problem formulation (\textit{i.e.} unicast or multicast). For example, unicast scenarios schedule one user per beam in each transmission. Thus, the scheduler has to select $N_b$ users (one user in each beam) among a total number of $N_u$ users for each time slot. The user selection is usually designed considering the users' channel conditions and their level of interference with each other. This closely relates scheduling to precoding as they significantly impact each other's performance. Therefore, the optimal transmit scheme involves the design of a joint precoding and scheduling algorithm, which has been well studied during the last decade \cite{Eduardo2017}. An example of a coupled scheduling and sum-rate maximization problem in the unicast scenario can be represented as\textcolor{black}{\cite{Shankar2021}}
 
 \begin{align}
  \begin{split}
      &\max_{\{\textbf{w}_i, u_i \in U_i\}_{i=1}^{N_b}} \sum_{i=1}^{N_b} log(1+ \mathrm{SINR}_i)\\
      & \text {s.t}~ \mathrm{SINR}_{i} \geq \gamma_i, ~ i \in \{1, \ldots, N_b \} \\
      & \text {   }~ \sum_{k=1}^{N_b} \textbf{w}_k^H \textbf{Q}_l \textbf{w}_k \leq q_l, ~ \forall l \in \{1,\ldots,L\},
  \end{split}
  \end{align}
where $\mathrm{SINR}_i$ denotes the SINR of user $u_i$ considering the set of scheduled users $u_1, \ldots, u_{N_b}$.

 In a multicast scenario such as frame-based precoding, user scheduling refers to selecting and aggregating the information transmitted in each physical frame. In fact, from a practical implementation perspective, DVB-S2 \cite{540} and DVB-S2X \cite{54} standards have two modes of operation, namely, the short FEC frame mode with a frame length of 16Kbit and the normal FEC frame mode with a frame length of 64Kbit \cite{54}. The amount of data to be multiplexed in each FEC frame depend on the code rate. For instance, the lowest code rate in the normal FEC frame mode allows up to four users to be scheduled in the same frame \cite{Christopoulos2015}. In this case, scheduling directly impacts the MODCOD for each frame since all users served by the same physical frame will be transmitted with the same ModCod. Furthermore, scheduling determines the set of users for which the precoding is designed. Therefore, it impacts the precoding design, which in turn impacts the level of interference, which determines the choice of the MODCOD. Thus, naturally, an optimal design will consider a joint objective of scheduling and precoding parameters across the users in all the beams. An example of a coupled scheduling and sum-rate maximization problem in the multicast scenario can be represented as\textcolor{black}{\cite{21}}
 
 \begin{align}
  \begin{split}
      &\max_{\{\textbf{w}_i, S_i \in D_i\}_{i=1}^{N_b}} \sum_{i=1}^{N_b} log(1+ \Omega_i)\\
      & \text{s.t}~ \Omega_i = \min_{j \in S_i} \mathrm{SINR}_{i,j}, ~ i \in \{1, \ldots, N_b \} \\
      & \text {   }~ \mathrm{SINR}_{i,j} \geq \gamma_i, ~ i \in \{1, \ldots, N_b \} \\
      & \text {   }~ \sum_{k=1}^{N_b} \textbf{w}_k^H \textbf{Q}_l \textbf{w}_k \leq q_l, ~ \forall l \in \{1,\ldots,L\},
  \end{split}
  \end{align}
 with $S_i$ being any subset of $U_i$ with cardinality $|G_i| = N_{g_i} $, and $D_i$ being the dictionary of all the subsets of type $S_i$, $i \in \{1, \ldots, N_b\}$. Surely, the cardinality of $D_i$ is $\binom{N_{u_i}}{N_{g_i}}$.
 
 These coupled approaches are generally NP-hard problems, and several ad-hoc optimization frameworks have been investigated to provide approximate solutions\cite{Bandi2020MISO}. Coupled scheduling and precoding variables hinder their joint update in these approaches. \textcolor{black}{ For cellular systems, solving these optimization problems is usually done either by alternatively updating the precoding and scheduling variables \cite{Yu2012}, or by ignoring the scheduling constraints \cite{Ku2015}.}
 
 In the context of multi-beam satellites, the authors in \cite{21} developed a joint precoding and scheduling scheme for multigroup multicast systems. The proposed scheme maximized the sum-rate under per-beam  power control and target SINR constraints. They transform the sum-rate objective with non-convex SINR constraints and discrete scheduling variables into a difference of convex problems that allows for updating the precoding and scheduling jointly. The simulation results show that this coupled approach offers up to 50\% gain in sum-rate with respect to the heuristic decoupled approach considered in \cite{Christopoulos2015paper}. 

 The most common resource management problems aim to optimally assign bandwidth, power, time slots, and precoding weights among the served users. A common approach to address such problems consists of assigning one representative user for all the users to be served in each beam. This reduces complexity by decoupling the precoding and scheduling design from the spectrum and power allocation problem. Consequently, power allocation is typically studied jointly with carrier allocation \cite{40, Barcelo2009, Lei2010, Schubert2004}. However, there have been several works coupling power allocation with precoding \cite{6190179, Christopoulos2012, Gao2021, 2} and bandwidth allocation with precoding \cite{Vu2022}.

\subsection{Precoding Level: Block-level and Symbol-level}

\begin{figure*}
    \centering
    \includegraphics[width=0.8\linewidth]{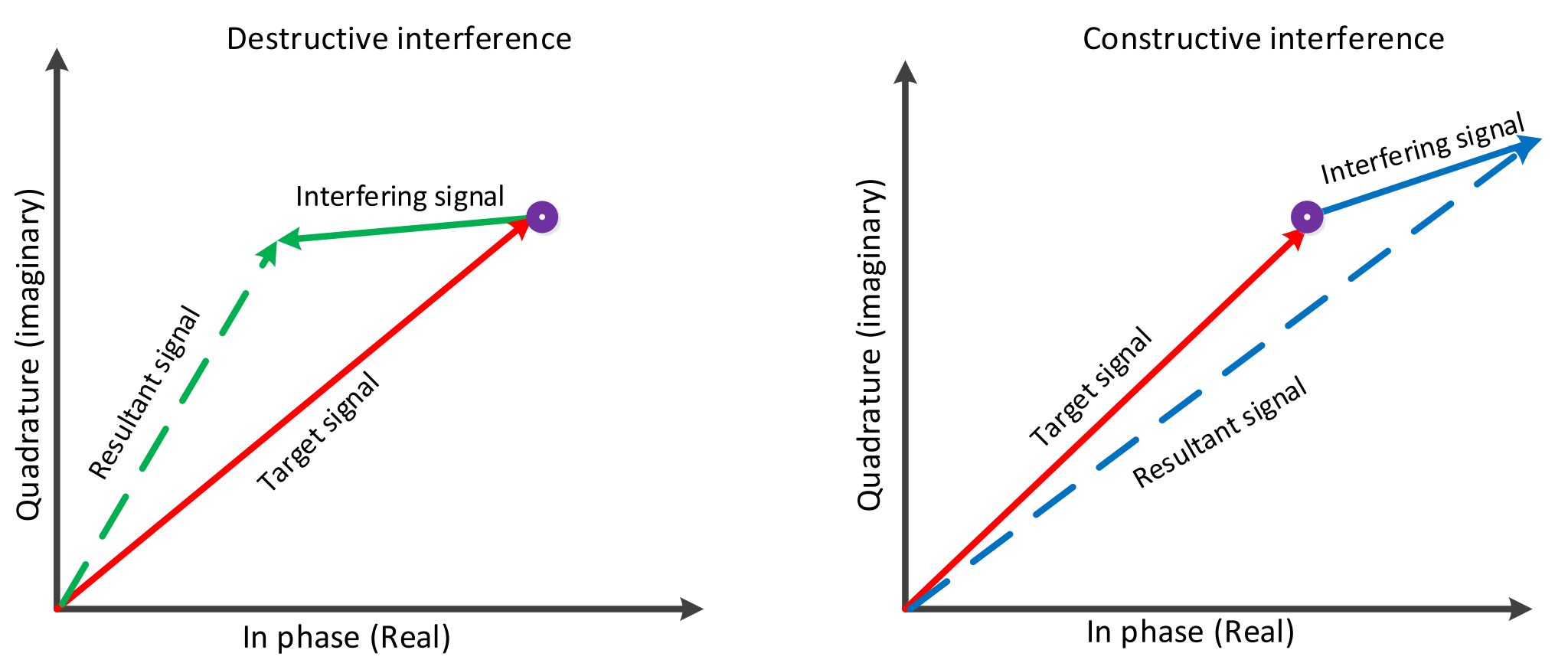}
    \caption{The first quadrant of QPSK with constructive interference and destructive interference \cite{Alodeh2015}.}
    \label{fig:constructiveVSdestructive}
\end{figure*}

 In the precoding literature for forward link transmission, the conventional approach uses the CSI to mitigate multiple user interference. Therefore, the precoding scheme has to be updated whenever the CSI changes. In the context of linear precoding, a new precoding matrix has to be calculated every time coherence of the channel. Since the time coherence of the channel is greater than the symbol duration in slow flat fading, the same CSI will be used for a set/block of consecutively transmitted symbols. Therefore, this conventional precoding technique is referred to as block-level precoding or alternatively channel-level precoding. Closed-form solutions, in this category, include the popular ZF precoding \cite{Taesang2005, TaesangYoo2006}, RZF \cite{Nguyen2008}, and MMSE precoding \cite{Peel2005, Ruly2003}. Apart from the closed-form solutions, other block-precoding schemes are designed as the solutions for the numerical optimization of certain objective functions under a set of constraints. The latter precoding schemes have been covered in the previous section. In this line, \cite{Bengtsson1} aims to minimize of the total power under QoS constraints, \cite{Schubert2004} investigates the max-min fair problem subject to sum-power constraints, \cite{YU2007} accounts for the per-antenna power constraints, \cite{Dartmann2013} and \cite{Zheng2012} consider generalized power constraints, and \cite{Christopoulos2014Oct} tackles precoding in the multi-group multicast scenario.

 In addition to block-level precoding, a more sophisticated technique is symbol-level precoding. This approach uses the symbols' data information (DI) and CSI to design the precoding scheme. Therefore, the precoding scheme has to be updated with every new symbol. Therefore, a new precoding matrix has to be calculated for every symbol duration. The main difference between block-level precoding and symbol-level precoding is that, while block-level precoding aims to eliminate interference completely, symbol-level precoding aims just to control it. Specifically, symbol-level precoding schemes are designed to allow constructive interference and eliminate destructive interference. Therefore, this approach carefully manages and exploits interference to adapt the precoding scheme to each individual symbol at the cost of a high switching rate as opposed to block-level precoding. 

 \textcolor{black}{The literature of multi-antenna systems investigates symbol-level precoding in two contexts: the analog context with directional modulation in the antenna and propagation field and the digital context for constructive interference in the signal processing field.} In a directional modulation-based transceiver, a set of gain-amplifiers and phase shifters are driven by a single RF chain to control the amplitude and phase of the received signal on a symbol level in the analog domain. The single RF chain transceiver has the advantages of simple structure and minimal power consumption, and it has been attracting a growing research interest \cite{6645431, 7045476, 7446283, 6096357, 6544472, 6746064, 4684619}. Nevertheless, this technology of analog symbol-level precoding (directional modulation) comes with a lack of algorithmic foundation and implementation limitations \cite{Alodeh2018}. Nonetheless, digital symbol-level precoding has a stronger algorithmic foundation. It guarantees constructive interference at the receiver by precoding the transmitted signals digitally. However, it requires a full digital transceiver, hindering its use in large antenna arrays \cite{Alodeh2018}.

\textcolor{black}{Herein, we focus on digital symbol-level precoding, which we will refer to as symbol-level precoding for simplicity. Readers seeking an in-depth overview of directional modulation may refer to \cite{Alodeh2018} for further details. As the majority of the current literature on symbol-level precoding pertains to terrestrial systems, we initially introduce the technique within the broader context of multi-antenna systems. We provide a brief overview of the existing works proposed in this area, not as a survey of symbol-level precoding for terrestrial systems but rather as a point of reference for the extension of symbol-level precoding to multi-beam satellite systems, which will be addressed subsequently. For a comprehensive survey of symbol-level precoding techniques for terrestrial systems, readers can refer to \cite{Alodeh2018}. }

\paragraph{\textcolor{black}{Symbol-level Precoding for Terrestrial Systems}}
 Symbol-level precoding aims to correct the phase of each interfering signal, shifting it to the detection region of the target signal. This hinders the detrimental effect of interference because the interfering signals add constructively to the target signal and remain in the correct detection region. In this context, two types of interference are distinguished: constructive interference, which maintains the received symbol in its correct detection region, and destructive interference, which drives the received symbol away from its correct detection region. The classification of interference as constructive or destructive has been carried out in \cite{Alodeh2015} for phase shift keying (M-PSK) and in \cite{Masouros2009} for binary phase shift keying (BPSK) and quadrature phase shift keying (QPSK).

 The interference unit-power of the $k$-th data stream on the $l$-th user is measured by assessing the orthogonality \cite{Taesang2006} of the $k$-th stream precoding vector and the $l$-th user channel, and it can be written as\textcolor{black}{\cite{Alodeh2018}}

\begin{equation}
\psi_{lk} = \frac{\textbf{h}_l^T \textbf{w}_k}{|\textbf{h}_l| |\textbf{w}_k|}.
\end{equation}
We can see that the $l$-th user channel vector and the $k$-th stream precoding vector are orthogonal when $\psi_{lk}$ is zero, and co-linear when $|\psi_{lk}|$ is one. 

 Using the above definition of unit-power interference, the conditions for an M-PSK symbol $s_k$ to interfere constructively with another M-PSK symbol $s_l$ are\textcolor{black}{\cite{Alodeh2018}}

\begin{align}
\begin{split}
     &\mathrm{Arg}(s_l) - \frac{\pi}{M} \leq \mathrm{arctan}\left( \frac{\mathrm{Im}(\psi_{lk} s_k)}{\mathrm{Re}(\psi_{lk} s_k)} \right) \leq \mathrm{Arg}(s_l) + \frac{\pi}{M},\\
    &\mathrm{Re}(\psi_{lk})\mathrm{Re}(s_k) > 0, ~\mathrm{Im}(\psi_{lk})\mathrm{Im}(s_k) > 0.
\end{split}
\end{align}
 The first condition guarantees that the phase of the received symbol lies in the correct detection region of the transmitted symbol, while the last two conditions guarantee that the detected symbol has the same direction as the transmitted symbol. These conditions ensure mutuality of constructive interference. More specifically, if the interference caused by the $k$-th user on the $l$-th user is constructive, then the interference caused by $l$-th user on the $k$-th user is also constructive \cite{Alodeh2015}. An example of destructive interference and constructive interference is illustrated in Fig.~ \ref{fig:constructiveVSdestructive}. 

 The design of symbol-level precoding can be determined based on the optimized objective function and the constructive interference constraints. \textcolor{black}{A common optimization problem, investigated in \cite{Alodeh2015, Alodeh2015MQAM, AlodehMaha2016, Alodeh2017}}, consists of total transmit power minimization with QoS and constructive interference constraints, and it can be written, for M-PSK modulation, as 

\begin{align}
  \begin{split}
      &\min_{\textbf{w}_1, \ldots, \textbf{w}_{N_b}} || \sum_{k=1}^{N_b} \textbf{w}_{k} s_k||^2\\
      & \mathrm{s.t} ~ |\textbf{h}_l^T \sum_{k=1}^{N_b} \textbf{w}_{k} s_k |^2 \geq  \zeta_{l} \sigma_l^2, ~ \forall l \{1, \ldots, N_b\}, \\
      & \mathrm{s.t} ~ \mathrm{Arg} \left(\textbf{h}_l^T \sum_{k=1}^{N_b} \textbf{w}_{k} s_k \right) = \mathrm{Arg}(s_l), ~ \forall l \{1, \ldots, N_b\}, 
  \end{split}
\end{align}
where $\zeta_1, \ldots, \zeta_{N_b}$ are signal-to-noise ratio targets that should be granted by the transmitter. 

 Various symbol-level precoding designs have been studied in the literature. The concept of interference exploitation was first proposed in \cite{Masouros2009} and \cite{Masouros2011}. Therein, an analytical classification of interference was proposed, and a precoding scheme based on channel inversion was investigated. The analysis, in \cite{Masouros2009} and \cite{Masouros2011}, demonstrates that higher receive SNRs can be guaranteed by exploiting constructive interference instead of increasing the per-user transmit power. Later on, \cite{Alodeh2014} and \cite{Alodeh2015} investigated the application of the constructive interference, introduced in \cite{Masouros2009} and \cite{Masouros2011}, for the forward link beamforming. In particular, authors in \cite{Alodeh2015} tackled max-min fairness, transmit power minimization, and sum-rate maximization using symbol-level precoding. The work in \cite{Masouros2015} builds up on the aforementioned works to improve such symbol-level precoders using a more relaxed optimization framework. More specifically, their proposed precoding scheme incorporates constructive interference, as a source of signal power, into the QoS constraints to enhance the gain in the transmit power. Further, the work was extended to account for imperfect CSI with bounded errors for a more robust design. In \cite{AlodehMaha2016}, the maximization of the minimum SINR under total transmit power constraints and minimization of the transmit power under SINR constraints were studied for symbol-level precoding, according to the concept of relaxed detection region and the definition of constructive interference in \cite{AlodehMaha2016}. In more detail, authors in \cite{AlodehMaha2016} generalize the symbol-level precoding design by relaxing the correct detection region. The analysis shows a trade-off between symbol error rate and power, which enhances the energy efficiency of the relaxed scheme compared to existing schemes in \cite{Alodeh2015}.

 Most of the aforementioned works focus on constructive interference for M-PSK modulation. However, in \cite{Alodeh2015MQAM}, the constructive interference precoding was extended for multi-level modulation (MQAM) to study the transmit power minimization under QoS constraints. Further, The problem of per-antenna transmit power minimization under the SINR constraints was first considered in \cite{Spano2016} in the context of symbol-level precoding. The proposed scheme enhances the peak-to-average power ratio and power peaks compared to \cite{Alodeh2015}. 

 \textcolor{black}{The work in \cite{Haqiqatnejad2018} uses the notion of distance-preserving constructive interference regions, firstly proposed in \cite{Haqiqatnejad13}, and fully characterized in \cite{Haqiqatnejad13}, to derive a simplified problem formulation of total transmit power minimization under SINR constraints. The analysis of the latter using the Karush-Kuhn-Tucker conditions provides a closed-form sub-optimal symbol-level precoding solution, which proves to be more computationally efficient than the optimal solution. \cite{LiAng2018} highlights the problem of the computational cost that symbol-level precoding approaches have to pay to outperform block-level precoding approaches. They propose computationally efficient closed-form precoding schemes. Lagrangian and Karush-Kuhn-Tucker conditions are considered to derive the optimal precoding for strict phase rotations. Then, for the non-strict phase rotation, authors in \cite{LiAng2018} prove that the optimization problem is equivalent to a strict phase rotation problem with double the dimensions. Power and hardware efficiency of symbol-level precoding were addressed in \cite{Domouchtsidis2019}. Therein, three efficient transmitter architectures for symbol-level precoding in large-scale antenna systems are proposed. They are designed so that the average Euclidean distance between the target and received signals is minimized while satisfying architecture-dependent constraints. Moreover, Field-programmable gate array (FPGA) implementation of symbol-level precoding has been investigated in \cite{Krivochiza2019, Haqiqatnejad2021} to improve the computational efficiency and enable a real-time operation mode.}

\paragraph{\textcolor{black}{Symbol-level Precoding for Satellite Systems}}
 In the context of multi-beam satellites, an important challenge is the distortion of the transmitted waveforms due to the non-linearity of the satellite traveling-wave-tube amplifiers \cite{Andrenacci2015}. An approach to address this problem is pre-distortion in the single-user links \cite{Casini2004, Karam1991, Piazza2014}. Nevertheless, extending pre-distortion techniques to multi-beam joint processing systems is challenging because precoding correlates the transmitted data streams. In \cite{SpanoDanilo2016}, a symbol-level precoding scheme for the forward link of multi-beam satellites proves effective in mitigating non-linear distortions. More precisely, authors in \cite{SpanoDanilo2016} consider per-antenna power constraints to account for the lack of flexibility in on-board payloads and address the problem of max-min fairness using symbol-level precoding. They show that per-antenna power constraints in a symbol-level precoding design control the instantaneous per-antenna transmit power, thus restricting the power peaks to the linear region of the traveling-wave-tube amplifiers. 
 Later on, authors in \cite{Spano2017} build up on the works in \cite{Spano2016} and \cite{SpanoDanilo2016} to propose a robust symbol-level precoding design against non-linearities of the satellite channel, specifically differential phase distortion. However, the novelty in \cite{Spano2017} is to address non-linear distortions by the minimization of the spatial peak-to-average power ratio instead of considering the per-antenna power constraints as in \cite{Spano2016} and \cite{SpanoDanilo2016}. Their formulation considers QoS constraints, and it has enhanced spatial peak-to-average power ratio compared to the techniques in \cite{Spano2016} and \cite{SpanoDanilo2016}. The works of Spano \textit{et al.} in \cite{Spano2016} and \cite{SpanoDanilo2016} was also extended, more recently, in \cite{SpanoDanilo2019}. Therein, the problem of per-antenna power minimization under QoS constraints is addressed by accounting for multi-level modulation schemes. To this end, the symbol-level precoding in \cite{SpanoDanilo2019} has enhanced robustness by reducing the power peaks for each symbol slot across the antennas. 

 In \cite{Duncan2018}, a hardware testbed is presented to illustrate precoded communication in the forward link of a multi-beam satellite system. Realistic benchmarks of symbol and block-level precoding are performed in unicast and multicast scenarios. It experimentally demonstrates the feasibility of precoding using the framing structure of DVB-S2X. More recently, \cite{Maturo2019} also presented a real-time demonstrator of precoding techniques for full frequency reuse in the forward link of multi-beam satellite systems. Specifically, it shows the feasibility of implementing the computationally efficient symbol-level precoding in \cite{Krivochiza2018, Krivochiza2019}, as well as MMSE precoding in \cite{Peel2005} for satellite communications considering the DVB-S2X standard.

\subsection{Configuration: Single-gateway and Multi-gateway Systems }

\begin{figure}
    \includegraphics[width=\linewidth]{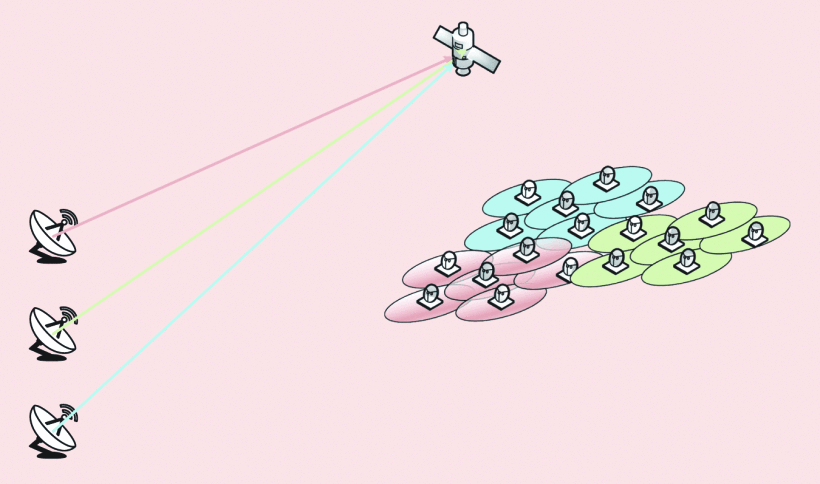}
    \caption{Multi-GW satellite system  \cite{30}.}\label{fig:3}
  \end{figure}

 Most precoding literature for multi-beam satellite communications assumes an ideal scenario where a single GW can handle all the system's users. However, this assumption is unrealistic as it implies a large feeder link bandwidth requirement, unavailable under current frequency allocations (Ka-bands). A general consideration in satellite communication consists of minimizing the payload complexity. Consequently, the majority of computations have to be performed at the GW, and the already precoded signals should be fed to the satellite. In this case, the feeder link must support the whole traffic of the satellite (equal to the product of the total user bandwidth, the number of polarizations, and the number of beams). This results in a large bandwidth requirement that is even larger with the deployment of full frequency reuse since it increases the user link bandwidth. 
 
 One way to address this problem consists of moving the feeder links to the Q/V bands \cite{Gharanjik2013, GharanjikAhmad2013, Gayrard2009}. However, rain attenuation in these bands majorly deteriorates the feed signals compared to the Ka-band. Another solution consists of using optical feeder links \cite{Saathof2017, Poulenard2015} to exploit their large bandwidth. Other approaches address this problem from a system design point of view instead of merely switching to higher frequency bands. The first approach is on-board beamforming, where the satellite payload performs the beam processing to alleviate the bandwidth requirement of the feeder link. In this case, the feeder link transmits only the user signals without precoding. Nevertheless, this scheme is known to reduce the achievable rates \cite{Arnau2012, JoroughiDevillers2013}. Further, for a large number of beams matching the intensive broadband demands, the feeder link still would not be able to support all the user signals \cite{Joroughi2016}, even with on-board beamforming. On the other hand, the combination of on-board and on-ground precoding in a hybrid design has been an interesting alternative \cite{Devillers2011}. The on-board and on-ground precoding and their hybrid designs will be addressed in the next section. 

 The second approach is deploying multiple GWs instead of a single GW \cite{WangJiZhou2019}. In this case, each GW can use the whole bandwidth to serve a subgroup of beams in the satellite coverage. The configuration of a multi-GW system is presented in Fig. \ref{fig:3}. Therefore, considering multiple GWs reduces the bandwidth requirement of each of the feeder links. In fact, the aggregation of all feeder links belonging to the different GWs represents the equivalent feeder link of the system. Additionally, adopting multiple GWs reduces the processing complexity at each GW since each GW handles a small number of beams. Furthermore, multi-GW systems can manage GW fails and adverse fading cases by routing the traffic to the functioning GWs. 
 
 If we consider $M$ GWs, we can associate a block diagonal precoding matrix $W$ to the whole system of the multiple GWs, as\textcolor{black}{\cite{30}}

\begin{equation}
    \textbf{W} = \mathrm{diag} \{ \textbf{W}_1, \ldots, \textbf{W}_m, \ldots, \textbf{W}_M \}, 
\end{equation}
where $\textbf{W}_m \in \mathbb{C}^{N_b^m \times N_f^m}$ is the precoding matrix transmitted by the $m$-th GW for $m \in \{1, \ldots, M \}$. 

 We can see that in this multiple GW system, the total number of feeds is $N_f = \sum_{m=1}^{M} N_f^m$, and the total number of beams is $N_b = \sum_{m=1}^{M} N_b^m$. This means that the $m$-th GW only transmits $N_b^m N_f^m$ precoded signals instead of $N_b N_f$, which reduces the bandwidth requirement of each of the GWs. Nevertheless, multi-GW systems face a number of challenges in real deployment. The first challenge is \textcolor{black}{decreasing} the degrees of freedom for interference mitigation. In fact, if we consider a multi-beam satellite communication system with $N_b$ beams and $N_f$ feeds. Then, unlike the single-GW \textcolor{black}{that} has all the $N_f$ feeds for inter-beam interference mitigation, in a multi-GW system, each GW will have a fraction of the $N_f$ feed signals and, therefore, fewer degrees of freedom to design its precoding matrix. The second challenge is the collective processing of the precoding among the GWs. Centralized approaches avoid signaling overhead between GWs. However, they lack robustness against failure and are far from being realistic because of the large-bandwidth backhaul links required to connect all the GWs \cite{Zheng2012multi}. On the other hand, distributed approaches are more robust in the face of the failure of one of the GWs. However, they require perfect connectivity between the GWs and more signaling overhead. In fact, in multi-GW satellite systems, each GW has to design a precoding matrix that mitigates interference between its beams, on the one hand, and the other GWs' beams, on the other hand. For this to be achieved, different GWs have to ideally exchange CSI because each GW has access to only the feedback of the users it serves. This information exchange creates a large signaling overhead and requires a perfect interconnection between the GWs, which can be hard to achieve in practical deployments. \textcolor{black}{In this context, the concept of cooperation between GWs emerges and can be addressed from different angles. For instance, some works in the literature refer to cooperative GWs when users are not assigned fixed serving GWs. In fact, each user can be served by a different GW in each time slot, and the multiple GWs collaborate to serve all the users in an optimal strategy \cite{Zheng2012multi}. In this same context, non-cooperative GWs are defined such that each GW is required only to serve the users within its cluster. However, cooperation in this sense will require CSI and data exchange between the GWs to show a performance gain with respect to single GW systems\cite{Zheng2012multi}. Another way to define cooperative GWs, which is the one considered in this paper, relates cooperation to signaling exchange between the GWs. In fact, we differentiate between cooperation and non-cooperation between GWs in multi-GWs systems as follows.}
 \paragraph{\textcolor{black}{Cooperation in Multi-GW Sytems}} \textcolor{black}{ \textcolor{black}{It} is when the GWs exchange information, which can be in the form of CSI or the symbols DI. This allows each GW to exploit the CSI or DI of the users in its neighboring clusters in addition to the CSI and DI of the users in its cluster to design the precoding scheme, which targets both intra and inter-cluster interference. }

 \paragraph{\textcolor{black}{Non-cooperation in Multi-GW Sytems}} \textcolor{black}{ \textcolor{black}{It} is when the GWs do not exchange CSI or the symbols DI. This generally results in uncoordinated processing among GWs during the precoding design, and while it targets inter-cluster interference, it leads to intra-cluster interference. }

 \textcolor{black}{Precoding for multi-GW satellite systems was first introduced in \cite{DevillersBertrand2011}. Therein, the authors first demonstrate the benefits of using on-ground precoding compared to hybrid on-board on-ground precoding in a single gateway scenario. Then, to account for realistic feeder link limitations, they extend the analysis to a multiple gateways scenario where each gateway serves a cluster of adjacent beams. In this case, they design a ZF precoder to address inter and intra-cluster interference. Further, ZF precoding was investigated, assuming that each GW has perfect CSI of the users in its served cluster of beams and users in other adjacent clusters, therefore, assuming cooperative GWs. However, data exchange between the GWs was not considered. This work shows the benefits of on-ground precoding on spectral efficiency and the trade-off between canceling intra and inter-cluster interference in single and multiple gateway scenarios. On the other hand, \cite{Zheng2012multi} studied multiple GW cooperation. More specifically, linear mean square error precoding is studied in two cooperation scenarios. GWs share partial CSI in the first scenario and both partial CSI and data in the second scenario. Then, these schemes were compared to full and partial frequency reuse schemes for a single GW. Simulation results demonstrate that adding data sharing on top of CSI sharing in multiple cooperative GWs allows them to outperform non-cooperative single GW solutions in terms of average throughput per beam.} 
 
 The work in \cite{Joroughi2014} extended the precoding technique in \cite{DevillersBertrand2011} by considering individual per-feed power constraints in the ZF precoding formulation to enhance power efficiency at the satellite payload. \textcolor{black}{\cite{Joroughi2016} investigates various unexplored elements multi-GW precoding in the aforementioned works. Firstly, the problem of communication overhead between the GWs is addressed, and a matrix compression technique is proposed to limit cooperation. Secondly, the impact of feeder link interference is considered, which can result from a lack of payload calibration. Thirdly, the authors analyze the imperfect CSI scenario considering the DVB-S2X standard and propose a robust scheme. Finally, the proposed precoding scheme is extended to the multicast context. The feeder link interference impact was, later, addressed in \cite{JoroughiMosquera2016}. Specifically, authors in \cite{JoroughiMosquera2016} design an MMSE and ZF based precoders to jointly address the feeder and user link interference. They show that feeder link interference can majorly decrease the average SINR of the coverage area. However, it can be significantly improved by the increased exchange of CSI among the GWs.}
 
 Authors in  \cite{ChristopoulosPennanen2016} considered the optimal multigroup multicast methods for multi-GW precoding instead of the heuristic approach used in \cite{Joroughi2016}. In particular, they address the problem of system fairness with coordinated optimization algorithms exchanging CSI between the GWs to limit inter-cluster interference. In \cite{Joroughi2017}, a two-stage precoding scheme was proposed, where inter-beam interference is minimized in the first stage, and intra-beam SINR is improved in the second stage. Then, the design was extended to a multi-GW scenario. In \cite{Mosquera2018}, GW cooperation is also kept to a minimum. Specifically, authors in \cite{Mosquera2018} demonstrate that cluster interaction can be bypassed by using statistical CSI instead of instantaneous channel knowledge without significant performance degradation. To this end, they propose a distributed precoding design based on MMSE and use a regularization parameter to account for interfering clusters. 

 In \cite{WangJi2019} and  \cite{WangJiZhou2019}, a multi-GW with feeder link interference scenario was considered to develop a frame-based multicast precoding scheme. The precoding scheme design aims at maximizing the sum rate while satisfying QoS constraints for each user, the per-feed power constraints on-board, and a total power constraint at each GW. \textcolor{black}{ Two solutions were developed for the problem. One solution uses the successive convex approximation approach, and the other solution consists of a beamforming algorithm that computes QoS beamforming and geometric programming based on power allocation iteratively. Further, the analysis considers three realistic limitations of multi-GW precoding schemes: feeder link interference, communication overhead between GWs, and frame-based precoding in a multicast context.}  In \cite{Joroughi2019}, dynamic on-board precoding for a multi-GW multi-beam satellite system is proposed. Therein, they propose a feed selection and signal switching mechanism that does not rely on CSI exchange among GWs, handles GW fails, selects the best on-board feeds to serve each user, and switches signals between GWs and users dynamically and flexibly. However, such a mechanism can impose high complexity on the satellite payload. \textcolor{black}{The centralized GW concept is introduced in \cite{KisseleffSteven2021}. Therein, the multiple GWs of a multi-GW system are all connected to a central server via high-speed fibers, where all digital baseband processing (\textit{e.g.} precoding) is carried out.  However, such a concept comes with various challenges related to three main folds: synchronization, precoding, and user scheduling. Specifically, UTs must receive synchronized signals in both time and frequency from the multiple GWs to preserve the quality of CSI. Furthermore, optimal joint precoding and user-scheduling is non-convex problem, and computing the precoding matrix of all the clusters in a centralized manner is highly expensive.}

\subsection{Implementation: On-board and On-ground}

 In the face of the growing demand for broadband services, satellite systems transitioned from a single-beam to multi-beam coverage. These multiple beams can be generated by deploying smart antenna technologies on the satellite (e.g., an array-fed reflector), enabling beam steering with beamforming (or precoding) implemented on-board (or on-ground) \cite{PieroAngeletti2015}. Conventionally, the beam generation is implemented at the satellite (either analog or digital) \cite{Angeletti2007, EURECOM2, ana2008}. In fact, from a satellite communication perspective, beamforming implies the design of a fixed beam pattern using an on-board BFN, based on geographic positions. Analog implementation is usually adopted due to the satellite's limited flexibility and power resources. Therefore, the BFN employs reflectors and phase-shifters to create the beams from the feeder link signal based on the traffic state and the propagation conditions in the coverage region. More specifically, the radiation patterns from the satellite feeds are linearly combined by the on-board network to create directive beam radiations in the coverage area. Nevertheless, this beam pattern is designed to account only for geographical considerations. Consequently, it can have a quasi-static aspect that does not take into consideration the specific positions of the users. 

\paragraph{\textcolor{black}{On-board and On-ground implementations}}
 \textcolor{black}{On-board beamforming techniques have shown satisfactory performance, but the growing demand for broadband services in multi-beam satellite systems has required increased flexibility in conventional beamforming techniques. This has resulted in the adoption of full or partial on-ground signal processing. In fact, one of the first implementations of on-ground processing is NASA's tracking and data relay satellite system \cite{Brandel1990}. A comprehensive overview of beamforming techniques for mobile satellites, along with the advantages and disadvantages of on-ground and on-board beamforming, is presented in \cite{Tronc2014}. As aggressive frequency reuse became popular for increasing the throughput of multi-beam satellite systems, partial and full on-ground processing gained more interest. Performing computations at the GW increases the satellite's coverage flexibility while maintaining a manageable payload complexity \cite{Angeletti2001}. On-ground beamforming implementation is usually digital and can utilize CSI to compute the precoded signal. As a result, unlike user-agnostic on-board beamforming, precoding optimizes the transmission to users \cite{Shankar2021}.}

 In \cite{AngelettiPiero2009}, a number of payload architectures for space/ground beamforming were proposed to balance the computation load between the GW and the satellite while maximizing the system's flexibility and capacity. Inspired by the hybrid architecture proposed in \cite{AngelettiPiero2009}, an ESA study \cite{esaGroundSpace} of a high-capacity multi-beam system design demonstrated the promising performance gains of precoding and multi-user detection based on a joint space/ground implementation. \textcolor{black}{The study results were reported in \cite{Arnau-Yanez2011}, where the channel estimation issues were discussed, and both perfect and imperfect CSI were considered for the regularized channel inversion precoder. The simulations compare precoding in the feed space (\textit{i.e.} on-ground precoding), precoding the in the beam space (\textit{i.e.} hybrid on-ground on-board precoding), and on-board fixed beamforming. The results demonstrate the throughput gain of feed space and beam space precoding compared to fixed on-board beamforming.} Further, feeder link digitization was studied, demonstrating decreased bandwidth while maintaining satisfactory signal quality. In \cite{Devillers2011}, linear precoding in the feed and the beam space were studied. Multiple precoders were considered: UpConst MMSE, ZF, and regularized channel inversion. The results show that precoding in feed space has a superior performance to precoding in the beam space in terms of availability and throughput. This conclusion is also supported by the work in \cite{Arnau2012}, where it is shown that full-on-ground implementations show superior performance compared to hybrid space/ground implementations.
\paragraph{\textcolor{black}{Hybrid On-board\slash On-ground implementations}}
Future generations of satellite communication systems are anticipated to handle a large number of beams necessitating precoding matrices with high dimensions. \textcolor{black}{Therefore, even though fully on-ground implementations are superior in performance to hybrid implementations, they drastically increase the bandwidth requirements at the feeder link when the processing is fully performed on-ground. To address this problem, higher frequencies like Q/V \cite{Gharanjik2013, GharanjikAhmad2013, Gayrard2009} and optical communications \cite{Saathof2017, Poulenard2015}, or multi-GW architectures \cite{DevillersBertrand2011} can be adopted. However, the current satellite systems lack technological maturity for higher bands, and multi-GW architectures are highly costly to deploy. In this context, hybrid precoding schemes become of great interest as they provide a trade-off between performance and bandwidth requirements.} In principle, this is done by decomposing the original precoding matrix $W$ into two matrices $W_1$ and $W_2$ such that\textcolor{black}{\cite{30}}

\begin{equation}
    W = W_2 W_1,
\end{equation}
 where $W_1$ is a $N_b\times N_b$ complex matrix corresponding to the precoding operation at the GW whereas $W_2$ is an $N_f\times N_b$ complex matrix corresponding to the precoding at the satellite. 
 
 This decomposition is especially useful in multiple-feed-per-beam architectures. In fact, instead of an $N_f\times N_b$ precoding matrix at the GW, the dimensions of the matrix are reduced to $N_b\times N_b$ and the feeder link bandwidth requirement is lowered by a ratio of $\frac{N_f}{N_b}$ \cite{30}. In other words, the benefit of such an approach consists of distributing the processing between the GW and the on-board payload to decrease the number of transmitted signals by the feeder link. Specifically, on-ground precoding has a feeder link of bandwidth 
 \begin{equation}
     \mathrm{B}_{\text{FL on-ground}} = N_f \mathrm{B}_{\text{beam}},
 \end{equation}
where $ \mathrm{B}_{\text{beam}}$ is the available bandwidth per beam. On the other hand, hybrid on-board on-ground precoding has a feeder link of bandwidth

\begin{equation}
     \mathrm{B}_{\text{FL hybrid}} = N_b \mathrm{B}_{\text{beam}}.
 \end{equation}
 Therefore, this approach decreases the processing complexity, the power and the hardware cost at the GW, and the bandwidth requirement at the feeder link. Nevertheless, these gains come with challenges. First, the decomposition lowers the degrees of freedom at the GW. In fact, instead of using all the $N_f$ feeds to mitigate interference between the beams, the $N_b\times N_b$ on-ground precoding matrix will act directly on the $N_b$ beams. This can decline the final throughput by a percentage reaching 20\% \cite{Arnau2012}. Additionally, the on-board analog processing reduces the overall system capacity compared to the fully digital on-ground processing \cite{33}. This is mainly due to the lack of precision of analog processing on the one hand and the quasi-static ergodic channel information used on-board on the other hand. However, the trade-off between complexity and performance of the hybrid on-board/on-ground architectures has been attracting increasing attention during the last two decades, especially with the recent trends pushing towards more flexibility and reconfigurability in the on-board payload \cite{Wansch2018}.

 \textcolor{black}{The paper \cite{Thibault2014} addresses the problem of limited feeder link bandwidth by introducing a hybrid precoding technique. This technique involves using fixed, non-channel-adaptive beamforming for the satellite segment. Additionally, the feed signals are compressed by projecting them onto a lower dimensional subspace, and then preprocessed using linear MMSE with regularized inversion at the GW.} In \cite{JoroughiDevillers2013}, a fixed on-board beam generation pattern design was proposed for a hybrid on-board/on-ground precoding system with low payload complexity. Further, the linear MMSE with a regularized inversion is also considered as in \cite{Thibault2014}. Then, the fixed beam generation process matrix is obtained as an approximated solution of the sum mean squared error minimization problem. Later on, the work in \cite{JoroughiDevillers2013} has been extended to study the return link in \cite{33} and provide a more robust precoding design accounting for the channel perturbations. In \cite{Song2017}, reduced-rank on-ground precoding adopting feed selection is proposed jointly with coarse on-board beamforming. Further, the benefits of the proposed hybrid precoding scheme on the feeder link requirement are shown compared to on-ground precoding.

 In \cite{23}, the feasibility of massive MIMO for SatCom has been studied, and efficient implementation via a hybrid analog/digital precoder was considered. The design of the precoder is based on a two-stage approach. In the first stage, the fully digital precoder is obtained using conventional precoding techniques such as ZF. In the second stage, the optimal hybrid precoder is derived as the minimizer of the Frobenius norm of the difference between the hybrid and the fully digital precoders. \textcolor{black}{In \cite{Joroughi2019}, a joint feed selection, precoding, and signal switching algorithm for the on-board segment is proposed to tackle the feeder link bandwidth scarcity while maintaining an affordable complexity at the payload. Specifically,  this work combines a multi-GW architecture with a hybrid on-ground on-board precoding scheme to get both their benefits on bandwidth requirements.} In \cite{9}, constant-envelope precoding is considered for hybrid on-board/on-ground architecture to offer more robustness against non-linearities. Two designs are proposed considering adaptive and fixed on-board beamforming. Designing the adaptive beamforming is done jointly with the GW signals and results in non-convex optimization problems, which are tacked using algorithmic solutions based on the saddle point method.

The growing demand for broadband services has caused satellite systems to shift from a single-beam to multi-beam coverage. Smart antenna technologies are deployed on satellites to generate multiple beams for beamforming or precoding. Conventionally, beam generation is implemented at the satellite using analog or digital techniques. However, the beam pattern generated is designed based on geographical considerations only, and does not account for specific user positions. On-ground beamforming implementations are digital and utilize Channel State Information (CSI) to compute the precoded signal. Precoding optimizes the transmission to users and increases the system's flexibility and capacity. Future satellite communication systems are anticipated to handle a large number of beams, necessitating precoding matrices with high dimensions. Hybrid precoding schemes provide a trade-off between performance and bandwidth requirements by decomposing the original precoding matrix into two matrices. This approach decreases the number of transmitted signals by the feeder link, but higher frequency bands or multi-GW architectures can also be adopted to address bandwidth requirements.

\subsection{Satellite Service: Mobile and Fixed} 

 An essential aspect of satellite communication is the spectrum of operation, according to which satellite services can be divided into two main categories: fixed satellite services (FSS) and mobile satellite services (MSS). FSS consists of communication between one or multiple space stations with fixed earth stations. It also includes satellite-to-satellite communication links in some cases \cite{ITUfss}. On the other hand, MSS consists of radio-communication links of one or multiple space stations with mobile earth stations \cite{ITUmss}. FSS operates on the Ku and Ka bands. Thus, it benefits from a larger available bandwidth than MSS operating on the L and S bands. However, they suffer from higher path loss and rain fading. 
 
 From a hardware perspective, FSS requires larger and more expensive antennas offering more focused beams. On the other hand, MSS relies on cheaper and less directive antennas to handle the non-stationary nature of its UTs. In addition, FSS includes earth-exploration satellite services, such as meteorological-satellite services, while MSS includes land, aeronautical, and maritime mobile-satellite services. In the face of the growing demand for higher data rates, up-to-date satellite systems rely on multi-beam coverage with a frequency reuse strategy among the beams. This increases the overall throughput with the frequency reuse factor at the cost of higher interference. To push the throughput to its extreme, full frequency reuse with precoding techniques has been proven very effective for FSS \cite{Arapoglou2016, Vazquez2016, Joroughi2017, Christopoulos2015paper}, which have been covered in the previous sections. However, few precoding studies have been carried out for MSS. 

 Recently, with the increasing congestion of the L and S-bands and the need for higher capacity in MSS, higher frequency reuse among beams has become an interesting alternative \cite{Tronc2014}. Further, driven by the high gains of precoding for FSS \cite{Arapoglou2016, Vazquez2016, Joroughi2017, Christopoulos2015paper}, and considering the limited gains of multi-user detection in mobile systems \cite{Cocco2014}, precoding for MSS have been receiving growing attention. Indeed, the mobility of the UT reduces the precoding gain compared to the fixed scenario, as it decreases the accuracy of the CSI \cite{Love2008}. \textcolor{black}{However, MSS can rely on a single GW to handle the total traffic, while FSS needs multiple GWs to handle the high bandwidth requirement in practice. Therefore, it is worth investigating the precoding gain in MSS \cite{Joroughi2016, Zheng2012multi, Tervo2018}.}

  \textcolor{black}{When considering precoding for MSS, a crucial aspect is the time-varying nature of the channel. In fact, the GW can only obtain delayed CSI, which can constrain the precoding performance.} A live precoding test for a mobile satellite system has been reported in \cite{Bastien2016}. Later on, one of the first few studies of precoding for MSS was carried out in \cite{Icolari2017}. A hot spot scenario is considered where the problem of unbalanced service demand in the L-band is addressed. In particular, Icolari \textit{et al.} propose a channel assignment approach that combines FDMA and SDMA, with both orthogonal channels and precoded channels reusing the same frequency in multiple areas as in \cite{Christopoulos2013, Taricco2014}. In this case, the total capacity is the sum of the two capacities resulting from the two access schemes, and it is shown to be higher than a lower-frequency reuse scenario. In \cite{VázquezAngel2018}, the impact of user mobility on the precoding performance of satellite systems is investigated. In fact, channel models such as slow nomadic, maritime and maritime low elevation are considered. In this context, the time-varying nature of the MSS channel results in a delayed CSI at the GW, which is proven of limited impact on the average throughput. Furthermore, both the unicast and multicast scenarios are considered. A large throughput gain is achieved for unicasting, and a sufficiently high gain is also attained for multicasting if a proper scheduling strategy is employed. Another recent work in the multi-beam mobile satellite context is done in \cite{Vahid2018, JoroughiS2019}. Therein, various CSI feedback strategies aiming to maintain a lower CSI variation at the GW are presented. In particular, two CSI feedback mechanisms are proposed, one based only on the latest CSI estimation and the other based on all the previous CSI estimations. Then, adapted precoding schemes for the proposed CSI feedback mechanisms are developed and tested for a maritime communication scenario to maintain a lower complexity of the proposed precoding schemes compared to land-mobile satellite services.

  \subsection{\textcolor{black}{Summary \& Lessons Learnt}}  
  
  \textcolor{black}{This section presents the main contribution of this survey, which is a classification of precoding techniques for HTS from two main levels: a problem formulation level and a system design level. The references discussed in this section are compared and summarized according to the various subcategories in Table \ref{tab:my-table}.}
  
  \textcolor{black}{The problem formulation level is a classification considering the mathematical aspects when designing the precoding scheme. The first aspect under this level involves the scheduling strategy incorporated in the precoding scheme, which we refer to as the precoding group. It can be unicast when one user per beam is served in each time slot or multicast when more than one user is served in each time slot. 
  In the forward link of multi-beam satellites with full frequency reuse, it is more practical to formulate the transmission as a multicast problem. This raises the problem of user clustering, which addresses the strategy of grouping users in each beam. 
  The second aspect concerns the optimization problem the precoding scheme is designed to solve. Therefore, precoding schemes can be closed-form or numerical solutions to an optimization problem. We classify the different optimization problems according to their objective functions: spectral efficiency maximization, power minimization, energy efficiency maximization, secrecy maximization, and in some cases, mixed objectives where the precoding scheme is designed jointly with a MAC layer protocol such as scheduling. The last aspect of the problem formulation level is related to the precoding switching rate, which can be on a block level or a symbol level. Specifically, precoding techniques relying only on CSI to mitigate interference are updated at every coherence time of the channel. Therefore, a block of several symbols can use the same precoding matrix, which explains calling it block-level precoding. On the other hand, precoding schemes that use both CSI and data information to mitigate interference are updated with every new symbol, and therefore its called symbol-level precoding.}
  
  \textcolor{black}{ The problem design level is a classification based on design aspects of the satellite communication system that impact precoding. The first aspect under this level concerns the number of GWs in the HTS system, and we refer to it as system configuration. In fact, the feeder link bandwidth requirement becomes too large when the number of users, beams, and polarizations increases. Multiple GWs help to reduce the feeder link bandwidth requirements, processing complexity, and manage gateway failures. However, multi-gateway systems face challenges such as decreasing degrees of freedom for interference mitigation and collective processing of precoding among the gateways. Different cooperative and non-cooperative GWs strategies are proposed in this context, depending on whether the CSI and DI are shared among the GWs or not, and more robust designs are proposed to account for feeder link interference and imperfect CSI. The second aspect is the precoding implementation, which can be on-board the satellite, on-ground at the GW, or hybrid on-board on-ground. Conventionally, beam generation is implemented at the satellite using analog techniques. However, the beam pattern generated is designed based on geographical considerations only and does not account for specific user positions. On-ground precoding implementations are digital and utilize CSI to compute the precoded signal. Therefore, they optimize the transmission to users and increase the system's flexibility and capacity at the cost of a high feeder link bandwidth requirement compared to fixed on-board implementations. Hybrid precoding schemes provide a trade-off between performance and bandwidth requirements by decomposing the original precoding matrix into two matrices. The last aspect under system design level is
  Satellite communication services are categorized into FSS and MSS, which differ in their spectrum of operation, hardware requirements, and service types. FSS operates on the Ku and Ka bands with larger bandwidth but higher path loss and rain fading, while MSS operates on the L and S bands with cheaper and less directive antennas. To improve throughput, satellite systems use multi-beam coverage, full frequency reuse, and precoding techniques for FSS. However, few precoding studies have been done for MSS, where channel time variability limits the precoding performance due to delayed CSI. Recent works have investigated some precoding aspects of MSS, including a hot spot scenario, user mobility impact, and user scheduling strategies, with proposed CSI feedback mechanisms and precoding schemes to maintain lower complexity.}

\begin{table*}[]
\centering
\caption{\textcolor{blue}{Comparison of precoding schemes for HTS.}}
\label{tab:my-table}
\resizebox{\textwidth}{!}{%
\begin{tabular}{|c|cccc|ccc|c|c|}
\hline
\rowcolor[HTML]{EFEFEF} 
\cellcolor[HTML]{EFEFEF} &
  \multicolumn{4}{c|}{\cellcolor[HTML]{EFEFEF}Problem formulation} &
  \multicolumn{3}{c|}{\cellcolor[HTML]{EFEFEF}System design} &
  \cellcolor[HTML]{EFEFEF} &
  \cellcolor[HTML]{EFEFEF} \\ \cline{2-8}
\rowcolor[HTML]{EFEFEF} 
\cellcolor[HTML]{EFEFEF} &
  \multicolumn{1}{c|}{\cellcolor[HTML]{EFEFEF}} &
  \multicolumn{2}{c|}{\cellcolor[HTML]{EFEFEF}Objective} &
  \cellcolor[HTML]{EFEFEF} &
  \multicolumn{1}{c|}{\cellcolor[HTML]{EFEFEF}} &
  \multicolumn{1}{c|}{\cellcolor[HTML]{EFEFEF}} &
  \cellcolor[HTML]{EFEFEF} &
  \cellcolor[HTML]{EFEFEF} &
  \cellcolor[HTML]{EFEFEF} \\ \cline{3-4}
\rowcolor[HTML]{EFEFEF} 
\multirow{-3}{*}{\cellcolor[HTML]{EFEFEF}Paper} &
  \multicolumn{1}{c|}{\multirow{-2}{*}{\cellcolor[HTML]{EFEFEF}Group}} &
  \multicolumn{1}{c|}{\cellcolor[HTML]{EFEFEF}Closed form} &
  \multicolumn{1}{c|}{\cellcolor[HTML]{EFEFEF}Opt. Problem} &
  \multirow{-2}{*}{\cellcolor[HTML]{EFEFEF}Level} &
  \multicolumn{1}{c|}{\multirow{-2}{*}{\cellcolor[HTML]{EFEFEF}Configuration}} &
  \multicolumn{1}{c|}{\multirow{-2}{*}{\cellcolor[HTML]{EFEFEF}Implementation}} &
  \multirow{-2}{*}{\cellcolor[HTML]{EFEFEF}Service} &
  \multirow{-3}{*}{\cellcolor[HTML]{EFEFEF}Robustness} &
  \multirow{-3}{*}{\cellcolor[HTML]{EFEFEF}Further remarks} \\ \hline
\cite{Joroughi2017} &
  \multicolumn{1}{c|}{Multicast} &
  \multicolumn{1}{c|}{RZF/MMSE} &
  \multicolumn{1}{c|}{\begin{tabular}[c]{@{}c@{}}SE (sum rate\\ maximisation)\end{tabular}} &
  Block &
  \multicolumn{1}{c|}{Single/Multiple GW} &
  \multicolumn{1}{c|}{On-ground} &
  Fixed &
  \begin{tabular}[c]{@{}c@{}}Outdated CSI.\\ No data sharing.\end{tabular} &
  Cooperative \\ \hline
\cite{Icolari2017} &
  \multicolumn{1}{c|}{Unicast} &
  \multicolumn{1}{c|}{LMMSE} &
  \multicolumn{1}{c|}{\texttimes} &
  Block &
  \multicolumn{1}{c|}{Single GW} &
  \multicolumn{1}{c|}{Hybrid} &
  Mobile &
  Perfect CSI. &
  - \\ \hline
\cite{WangJi2019} &
  \multicolumn{1}{c|}{Multicast} &
  \multicolumn{1}{c|}{\texttimes} &
  \multicolumn{1}{c|}{\begin{tabular}[c]{@{}c@{}}SE (sum rate\\ maximisation)\end{tabular}} &
  Block &
  \multicolumn{1}{c|}{Multiple GW} &
  \multicolumn{1}{c|}{On-ground} &
  Fixed &
  \begin{tabular}[c]{@{}c@{}}Perfect CSI.\\ Feeder link interference.\end{tabular} &
  Cooperative \\ \hline
\cite{8510717} &
  \multicolumn{1}{c|}{Multicast} &
  \multicolumn{1}{c|}{MMSE} &
  \multicolumn{1}{c|}{\texttimes} &
  Block &
  \multicolumn{1}{c|}{Single GW} &
  \multicolumn{1}{c|}{On-ground} &
  Fixed &
  Perfect CSI. &
  - \\ \hline
\cite{Arnau2012} &
  \multicolumn{1}{c|}{Unicast} &
  \multicolumn{1}{c|}{RCI/ZF} &
  \multicolumn{1}{c|}{\texttimes} &
  Block &
  \multicolumn{1}{c|}{Single GW} &
  \multicolumn{1}{c|}{\begin{tabular}[c]{@{}c@{}}On-ground/On-board/\\ Hybrid\end{tabular}} &
  Fixed &
  Imperfect CSI. &
  - \\ \hline
\cite{JoroughiS2019} &
  \multicolumn{1}{c|}{Unicast} &
  \multicolumn{1}{c|}{MMSE/ZF} &
  \multicolumn{1}{c|}{\checkmark} &
  Block &
  \multicolumn{1}{c|}{Single GW} &
  \multicolumn{1}{c|}{On-ground} &
  Mobile &
  Outdated CSI. &
  - \\ \hline
\cite{6190179} &
  \multicolumn{1}{c|}{Unicast} &
  \multicolumn{1}{c|}{\texttimes} &
  \multicolumn{1}{c|}{\begin{tabular}[c]{@{}c@{}}EE/SE (sum rate\\ maximisation)\end{tabular}} &
  Block &
  \multicolumn{1}{c|}{Single GW} &
  \multicolumn{1}{c|}{On-ground} &
  Fixed &
  Perfect CSI. &
  - \\ \hline
\cite{JoroughiDevillers2013} &
  \multicolumn{1}{c|}{Unicast} &
  \multicolumn{1}{c|}{\begin{tabular}[c]{@{}c@{}}Worst case MMSE/\\ LMMSE\end{tabular}} &
  \multicolumn{1}{c|}{\texttimes} &
  Block &
  \multicolumn{1}{c|}{Single GW} &
  \multicolumn{1}{c|}{Hybrid} &
  Fixed &
  \begin{tabular}[c]{@{}c@{}}Perfect/\\ Imperfect CSI.\end{tabular} &
  - \\ \hline
\cite{DevillersBertrand2011} &
  \multicolumn{1}{c|}{Unicast} &
  \multicolumn{1}{c|}{ZF} &
  \multicolumn{1}{c|}{\texttimes} &
  Block &
  \multicolumn{1}{c|}{Single/Multiple GW} &
  \multicolumn{1}{c|}{On-ground/Hybrid} &
  Fixed &
  Perfect CSI. &
  Cooperative \\ \hline
\cite{Lu2019} &
  \multicolumn{1}{c|}{Unicast} &
  \multicolumn{1}{c|}{\texttimes} &
  \multicolumn{1}{c|}{\begin{tabular}[c]{@{}c@{}}Secrecy (sum secrecy \\ rate maximisation)\end{tabular}} &
  Block &
  \multicolumn{1}{c|}{Single GW} &
  \multicolumn{1}{c|}{On-ground} &
  Fixed &
  Imperfect CSI. &
  - \\ \hline
\cite{EURECOM2} &
  \multicolumn{1}{c|}{Unicast} &
  \multicolumn{1}{c|}{MMSE/ZF} &
  \multicolumn{1}{c|}{SE} &
  Block &
  \multicolumn{1}{c|}{Multiple GW} &
  \multicolumn{1}{c|}{On-ground} &
  Fixed &
  \begin{tabular}[c]{@{}c@{}}Imperfect CSI.\\ Non-linearity.\end{tabular} &
  - \\ \hline
\cite{Lei2011} &
  \multicolumn{1}{c|}{Unicast} &
  \multicolumn{1}{c|}{ZF} &
  \multicolumn{1}{c|}{\begin{tabular}[c]{@{}c@{}}Mixed (secrecy+\\ Precoding)\end{tabular}} &
  Block &
  \multicolumn{1}{c|}{Single GW} &
  \multicolumn{1}{c|}{On-board/Hybrid} &
  Fixed &
  Imperfect CSI. &
  - \\ \hline
\cite{SpanoDanilo2019} &
  \multicolumn{1}{c|}{Unicast} &
  \multicolumn{1}{c|}{\texttimes} &
  \multicolumn{1}{c|}{Peak power minimsation} &
  Symbol &
  \multicolumn{1}{c|}{Single GW} &
  \multicolumn{1}{c|}{On-ground} &
  Fixed &
  \begin{tabular}[c]{@{}c@{}}Perfect CSI.\\ Non-linear.\\ distortions.\end{tabular} &
  \begin{tabular}[c]{@{}c@{}}M-PSK/\\ Multi-level\\  constellation\end{tabular} \\ \hline
\cite{Christopoulos2014} &
  \multicolumn{1}{c|}{Multicast} &
  \multicolumn{1}{c|}{\texttimes} &
  \multicolumn{1}{c|}{SE (max-min fair)} &
  Block &
  \multicolumn{1}{c|}{Single GW} &
  \multicolumn{1}{c|}{On-ground} &
  Fixed &
  Perfect CSI. &
  - \\ \hline
\cite{ChristopoulosPennanen2016} &
  \multicolumn{1}{c|}{Multicast} &
  \multicolumn{1}{c|}{\texttimes} &
  \multicolumn{1}{c|}{SE (max-min fair)} &
  Block &
  \multicolumn{1}{c|}{Multiple GW} &
  \multicolumn{1}{c|}{On-ground} &
  Fixed &
  \begin{tabular}[c]{@{}c@{}}Perfect CSI.\\ Data sharing.\end{tabular} &
  Cooperative \\ \hline
\cite{Zheng2012} &
  \multicolumn{1}{c|}{Unicast} &
  \multicolumn{1}{c|}{\texttimes} &
  \multicolumn{1}{c|}{\begin{tabular}[c]{@{}c@{}}SE (rate matching,\\ sum rate maximisation, \\  rate balancing)\end{tabular}} &
  Block &
  \multicolumn{1}{c|}{Single GW} &
  \multicolumn{1}{c|}{On-ground} &
  Fixed &
  Perfect CSI. &
  - \\ \hline
\cite{9} &
  \multicolumn{1}{c|}{Unicast} &
  \multicolumn{1}{c|}{\texttimes} &
  \multicolumn{1}{c|}{\checkmark} &
  Symbol &
  \multicolumn{1}{c|}{Single GW} &
  \multicolumn{1}{c|}{Hybrid} &
  Fixed &
  Perfect CSI. &
  - \\ \hline
\cite{Christopoulos2014Aug} &
  \multicolumn{1}{c|}{Multicast} &
  \multicolumn{1}{c|}{\texttimes} &
  \multicolumn{1}{c|}{SE (max-min fair)} &
  Block &
  \multicolumn{1}{c|}{Single GW} &
  \multicolumn{1}{c|}{On-ground} &
  Fixed &
  Perfect CSI. &
  - \\ \hline
\cite{5198782} &
  \multicolumn{1}{c|}{Unicast} &
  \multicolumn{1}{c|}{\texttimes} &
  \multicolumn{1}{c|}{THP} &
  - &
  \multicolumn{1}{c|}{Single GW} &
  \multicolumn{1}{c|}{On-ground} &
  Fixed &
  Perfect CSI. &
  - \\ \hline
\cite{Lin2019} &
  \multicolumn{1}{c|}{Unicast} &
  \multicolumn{1}{c|}{\texttimes} &
  \multicolumn{1}{c|}{Secrecy (EE)} &
  Block &
  \multicolumn{1}{c|}{Single GW} &
  \multicolumn{1}{c|}{On-ground} &
  Fixed &
  \begin{tabular}[c]{@{}c@{}}Imperfect CSI of\\ eavesdroppers.\end{tabular} &
  - \\ \hline
\cite{Zheng2012sec} &
  \multicolumn{1}{c|}{Unicast} &
  \multicolumn{1}{c|}{\texttimes} &
  \multicolumn{1}{c|}{\begin{tabular}[c]{@{}c@{}}Secrecy (sum power\\ minimsation)\end{tabular}} &
  Block &
  \multicolumn{1}{c|}{Single GW} &
  \multicolumn{1}{c|}{On-ground} &
  Fixed &
  \begin{tabular}[c]{@{}c@{}}Imperfect CSI \\ of eavesdropper.\end{tabular} &
  - \\ \hline
\cite{Guidotti2018} &
  \multicolumn{1}{c|}{Multicast} &
  \multicolumn{1}{c|}{MMSE} &
  \multicolumn{1}{c|}{\texttimes} &
  Block &
  \multicolumn{1}{c|}{Single GW} &
  \multicolumn{1}{c|}{On-ground} &
  Fixed &
  Perfect CSI. &
  - \\ \hline
\cite{Zheng2012multi} &
  \multicolumn{1}{c|}{Unicast} &
  \multicolumn{1}{c|}{RZF} &
  \multicolumn{1}{c|}{\begin{tabular}[c]{@{}c@{}}SE (sum rate\\ maximisation)\end{tabular}} &
  Block &
  \multicolumn{1}{c|}{Multiple GW} &
  \multicolumn{1}{c|}{On-ground} &
  Fixed &
  \begin{tabular}[c]{@{}c@{}}Partial CSI.\\ Partial data sharing.\end{tabular} &
  \begin{tabular}[c]{@{}c@{}}Cooperative/\\ Non-cooperative\end{tabular} \\ \hline
\cite{7} &
  \multicolumn{1}{c|}{Unicast} &
  \multicolumn{1}{c|}{ZF} &
  \multicolumn{1}{c|}{\texttimes} &
  Block &
  \multicolumn{1}{c|}{Single GW} &
  \multicolumn{1}{c|}{On-ground} &
  Fixed &
  Partial CSI. &
  - \\ \hline
\cite{23} &
  \multicolumn{1}{c|}{Unicast} &
  \multicolumn{1}{c|}{ZF} &
  \multicolumn{1}{c|}{\checkmark} &
  Block &
  \multicolumn{1}{c|}{Single GW} &
  \multicolumn{1}{c|}{Hybrid} &
  Fixed &
  Perfect CSI. &
  - \\ \hline
\cite{33} &
  \multicolumn{1}{c|}{Unicast} &
  \multicolumn{1}{c|}{Worst case MMSE} &
  \multicolumn{1}{c|}{\texttimes} &
  Block &
  \multicolumn{1}{c|}{Single GW} &
  \multicolumn{1}{c|}{On-board} &
  Fixed &
  Perfect CSI. &
  - \\ \hline
\cite{Joroughi2014} &
  \multicolumn{1}{c|}{Unicast} &
  \multicolumn{1}{c|}{ZF} &
  \multicolumn{1}{c|}{SE (max-min fair)} &
  Block &
  \multicolumn{1}{c|}{Multiple GW} &
  \multicolumn{1}{c|}{On-ground} &
  Fixed &
  \begin{tabular}[c]{@{}c@{}}Perfect CSI.\\ No data sharing.\end{tabular} &
  Cooperative \\ \hline
\cite{JoroughiMosquera2016} &
  \multicolumn{1}{c|}{Unicast} &
  \multicolumn{1}{c|}{ZF/MMSE} &
  \multicolumn{1}{c|}{\texttimes} &
  Block &
  \multicolumn{1}{c|}{Multiple GW} &
  \multicolumn{1}{c|}{On-ground} &
  Fixed &
  \begin{tabular}[c]{@{}c@{}}Perfect CSI.\\ Feeder link interference.\end{tabular} &
  Cooperative \\ \hline
\cite{Zorba2008} &
  \multicolumn{1}{c|}{Unicast} &
  \multicolumn{1}{c|}{\texttimes} &
  \multicolumn{1}{c|}{\checkmark} &
  Block &
  \multicolumn{1}{c|}{Single GW} &
  \multicolumn{1}{c|}{On-ground} &
  Fixed &
  Partial CSI. &
  - \\ \hline
\cite{SpanoDanilo2016} &
  \multicolumn{1}{c|}{Unicast} &
  \multicolumn{1}{c|}{\texttimes} &
  \multicolumn{1}{c|}{Min-max fair} &
  Symbol &
  \multicolumn{1}{c|}{Single GW} &
  \multicolumn{1}{c|}{On-ground} &
  Fixed &
  Perfect CSI. &
  M-PSK \\ \hline
\cite{Zhang2021sec} &
  \multicolumn{1}{c|}{Unicast} &
  \multicolumn{1}{c|}{\texttimes} &
  \multicolumn{1}{c|}{\begin{tabular}[c]{@{}c@{}}Secrecy (worst secrecy \\ rate maximisation)\end{tabular}} &
  Block &
  \multicolumn{1}{c|}{Single GW} &
  \multicolumn{1}{c|}{On-ground} &
  Fixed &
  \begin{tabular}[c]{@{}c@{}}Imperfect CSI of\\ eavesdroppers.\end{tabular} &
  - \\ \hline
\cite{Wang2018} &
  \multicolumn{1}{c|}{Multicast} &
  \multicolumn{1}{c|}{\texttimes} &
  \multicolumn{1}{c|}{SE (max-min fair)} &
  Block &
  \multicolumn{1}{c|}{Single GW} &
  \multicolumn{1}{c|}{On-ground} &
  Fixed &
  Outdated CSI. &
  - \\ \hline
\cite{Chatzinotas2011} &
  \multicolumn{1}{c|}{Unicast} &
  \multicolumn{1}{c|}{\texttimes} &
  \multicolumn{1}{c|}{SE (rate balancing)} &
  Block &
  \multicolumn{1}{c|}{Single GW} &
  \multicolumn{1}{c|}{On-ground} &
  Fixed &
  Perfect CSI. &
  - \\ \hline
\cite{Lin2019Apr} &
  \multicolumn{1}{c|}{Unicast} &
  \multicolumn{1}{c|}{\texttimes} &
  \multicolumn{1}{c|}{\begin{tabular}[c]{@{}c@{}}Secrecy (worst secrecy \\ rate maximisation)\end{tabular}} &
  Block &
  \multicolumn{1}{c|}{Single GW} &
  \multicolumn{1}{c|}{On-ground} &
  Fixed &
  \begin{tabular}[c]{@{}c@{}}Imperfect CSI of\\ eavesdroppers.\end{tabular} &
  - \\ \hline
\cite{Vahid2018} &
  \multicolumn{1}{c|}{Unicast} &
  \multicolumn{1}{c|}{\checkmark} &
  \multicolumn{1}{c|}{\texttimes} &
  Block &
  \multicolumn{1}{c|}{Single GW} &
  \multicolumn{1}{c|}{On-ground} &
  Mobile &
  Outdated CSI. &
  - \\ \hline
\cite{Thibault2014} &
  \multicolumn{1}{c|}{Unicast} &
  \multicolumn{1}{c|}{RZF} &
  \multicolumn{1}{c|}{\texttimes} &
  Block &
  \multicolumn{1}{c|}{Single GW} &
  \multicolumn{1}{c|}{Hybrid} &
  Fixed &
  Perfect CSI. &
  - \\ \hline
\cite{Arnau-Yanez2011} &
  \multicolumn{1}{c|}{Unicast} &
  \multicolumn{1}{c|}{RCI} &
  \multicolumn{1}{c|}{\texttimes} &
  Block &
  \multicolumn{1}{c|}{Single GW} &
  \multicolumn{1}{c|}{On-board/Hybrid} &
  Fixed &
  Imperfect CSI. &
  - \\ \hline
\cite{Qi2018} &
  \multicolumn{1}{c|}{Unicast} &
  \multicolumn{1}{c|}{\texttimes} &
  \multicolumn{1}{c|}{EE} &
  Block &
  \multicolumn{1}{c|}{Single GW} &
  \multicolumn{1}{c|}{On-ground} &
  Fixed &
  Perfect CSI. &
  - \\ \hline
\cite{Christopoulos2015paper} &
  \multicolumn{1}{c|}{Multicast} &
  \multicolumn{1}{c|}{\texttimes} &
  \multicolumn{1}{c|}{\begin{tabular}[c]{@{}c@{}}SE (sum rate\\ maximisation)\end{tabular}} &
  Block &
  \multicolumn{1}{c|}{Single GW} &
  \multicolumn{1}{c|}{On-ground} &
  Fixed &
  Perfect CSI. &
  - \\ \hline
\cite{35} &
  \multicolumn{1}{c|}{Multicast} &
  \multicolumn{1}{c|}{\texttimes} &
  \multicolumn{1}{c|}{EE} &
  Block &
  \multicolumn{1}{c|}{Single GW} &
  \multicolumn{1}{c|}{On-ground} &
  Fixed &
  Perfect CSI. &
  - \\ \hline
\cite{4409391} &
  \multicolumn{1}{c|}{Unicast} &
  \multicolumn{1}{c|}{\texttimes} &
  \multicolumn{1}{c|}{THP} &
  - &
  \multicolumn{1}{c|}{Multiple GW} &
  \multicolumn{1}{c|}{On-ground} &
  Fixed &
  Perfect CSI. &
  - \\ \hline
\cite{VázquezAngel2018} &
  \multicolumn{1}{c|}{Multicast} &
  \multicolumn{1}{c|}{MMSE} &
  \multicolumn{1}{c|}{\texttimes} &
  Block &
  \multicolumn{1}{c|}{Single GW} &
  \multicolumn{1}{c|}{On-ground} &
  Mobile &
  Outdated CSI. &
  - \\ \hline
\cite{Devillers2011} &
  \multicolumn{1}{c|}{Unicast} &
  \multicolumn{1}{c|}{RCI/ZF/MMSE} &
  \multicolumn{1}{c|}{\texttimes} &
  Block &
  \multicolumn{1}{c|}{Single GW} &
  \multicolumn{1}{c|}{On-ground/Hybrid} &
  Fixed &
  Imperfect CSI. &
  - \\ \hline
\cite{22} &
  \multicolumn{1}{c|}{Unicast} &
  \multicolumn{1}{c|}{\texttimes} &
  \multicolumn{1}{c|}{\begin{tabular}[c]{@{}c@{}}Sum power \\ minimisation\end{tabular}} &
  Block &
  \multicolumn{1}{c|}{Single GW} &
  \multicolumn{1}{c|}{On-ground} &
  Fixed &
  \begin{tabular}[c]{@{}c@{}}Imperfect CSI.\\ Phase uncertainty.\end{tabular} &
  - \\ \hline
\cite{WangJiZhou2019} &
  \multicolumn{1}{c|}{Multicast} &
  \multicolumn{1}{c|}{MMSE} &
  \multicolumn{1}{c|}{\begin{tabular}[c]{@{}c@{}}SE (sum rate\\ maximisation)\end{tabular}} &
  Block &
  \multicolumn{1}{c|}{Multiple GW} &
  \multicolumn{1}{c|}{On-ground/On-board} &
  Fixed &
  \begin{tabular}[c]{@{}c@{}}Perfect CSI.\\ Feeder link interference.\end{tabular} &
  \begin{tabular}[c]{@{}c@{}}Cooperative/\\ Limited cooperation\end{tabular} \\ \hline
\cite{Taricco2014} &
  \multicolumn{1}{c|}{Multicast} &
  \multicolumn{1}{c|}{\texttimes} &
  \multicolumn{1}{c|}{\checkmark} &
  Block &
  \multicolumn{1}{c|}{Single GW} &
  \multicolumn{1}{c|}{On-ground} &
  Fixed &
  \begin{tabular}[c]{@{}c@{}}Imperfect CSI.\\ Outdated CSI.\\ Phase offset.\end{tabular} &
  - \\ \hline
\cite{Spano2017} &
  \multicolumn{1}{c|}{Unicast} &
  \multicolumn{1}{c|}{\texttimes} &
  \multicolumn{1}{c|}{\begin{tabular}[c]{@{}c@{}}Spatial peak-to-average \\ power ratio minimization\end{tabular}} &
  Symbol &
  \multicolumn{1}{c|}{Single GW} &
  \multicolumn{1}{c|}{On-ground} &
  Fixed &
  \begin{tabular}[c]{@{}c@{}}Perfect CSI.\\ Non-linear distortions.\end{tabular} &
  M-PSK \\ \hline
\cite{28} &
  \multicolumn{1}{c|}{Multicast} &
  \multicolumn{1}{c|}{MMSE} &
  \multicolumn{1}{c|}{\texttimes} &
  Block &
  \multicolumn{1}{c|}{Single GW} &
  \multicolumn{1}{c|}{On-ground} &
  Fixed &
  Perfect CSI. &
  - \\ \hline
\cite{Song2017} &
  \multicolumn{1}{c|}{Unicast} &
  \multicolumn{1}{c|}{MMSE} &
  \multicolumn{1}{c|}{\checkmark} &
  Block &
  \multicolumn{1}{c|}{Single GW} &
  \multicolumn{1}{c|}{Hybrid} &
  Mobile &
  Statistical CSI. &
  - \\ \hline
\cite{Christopoulos2014Oct} &
  \multicolumn{1}{c|}{Multicast} &
  \multicolumn{1}{c|}{\texttimes} &
  \multicolumn{1}{c|}{SE (max-min fair)} &
  Block &
  \multicolumn{1}{c|}{Single GW} &
  \multicolumn{1}{c|}{On-ground} &
  Fixed &
  Imperfect CSI. &
  - \\ \hline
\cite{Joroughi2016} &
  \multicolumn{1}{c|}{Unicast} &
  \multicolumn{1}{c|}{ZF/LMMSE} &
  \multicolumn{1}{c|}{\texttimes} &
  Block &
  \multicolumn{1}{c|}{Multiple GW} &
  \multicolumn{1}{c|}{On-ground} &
  Fixed &
  \begin{tabular}[c]{@{}c@{}}Imperfect CSI.\\ Reduced data sharing.\\ Feeder link interference.\end{tabular} &
  Cooperative \\ \hline
\cite{Schraml2020} &
  \multicolumn{1}{c|}{Unicast} &
  \multicolumn{1}{c|}{\texttimes} &
  \multicolumn{1}{c|}{\begin{tabular}[c]{@{}c@{}}Secrecy (worst secrecy \\ capacity maximisation)\end{tabular}} &
  Block &
  \multicolumn{1}{c|}{Single GW} &
  \multicolumn{1}{c|}{On-ground} &
  Fixed &
  Perfect CSI. &
  - \\ \hline
\cite{21} &
  \multicolumn{1}{c|}{Multicast} &
  \multicolumn{1}{c|}{\texttimes} &
  \multicolumn{1}{c|}{\begin{tabular}[c]{@{}c@{}}Mixed (scheduling+\\ Precoding)\end{tabular}} &
  Block &
  \multicolumn{1}{c|}{Single GW} &
  \multicolumn{1}{c|}{On-ground} &
  Fixed &
  Perfect CSI. &
  - \\ \hline
\cite{Joroughi2019} &
  \multicolumn{1}{c|}{Unicast} &
  \multicolumn{1}{c|}{RZF} &
  \multicolumn{1}{c|}{\texttimes} &
  Block &
  \multicolumn{1}{c|}{Multiple GW} &
  \multicolumn{1}{c|}{On-Board} &
  Fixed &
  \begin{tabular}[c]{@{}c@{}}Imperfect CSI.\\ No data sharing.\end{tabular} &
  Non-cooperative \\ \hline
\cite{Christopoulos2012} &
  \multicolumn{1}{c|}{Unicast} &
  \multicolumn{1}{c|}{\texttimes} &
  \multicolumn{1}{c|}{\begin{tabular}[c]{@{}c@{}}SE (sum rate\\ maximisation)\end{tabular}} &
  Block &
  \multicolumn{1}{c|}{Single GW} &
  \multicolumn{1}{c|}{On-ground} &
  Fixed &
  Perfect CSI. &
  - \\ \hline
\cite{Mosquera2018} &
  \multicolumn{1}{c|}{Unicast} &
  \multicolumn{1}{c|}{MMSE} &
  \multicolumn{1}{c|}{\texttimes} &
  Block &
  \multicolumn{1}{c|}{Multiple GW} &
  \multicolumn{1}{c|}{On-board/Hybrid} &
  Fixed &
  \begin{tabular}[c]{@{}c@{}}Channel statistics.\\ No data sharing.\end{tabular} &
  Limited cooperation \\ \hline
\end{tabular}%
}
\end{table*}

\section{Precoding in Practice}

 The majority of the precoding literature for multi-beam satellites assumes ideal transmission conditions that are not achievable in practical systems. In fact, Satcoms suffer from various practical impairments that can heavily impact the overall system performance. These impairments can be classified into channel-induced impairments and transceiver and payload-related impairments. The latter includes non-linearity effects of high power amplifiers (HPA), multi-carrier distortion, phase noise, sampling clock offset, and I/Q imbalance \cite{Sharma2021}. On the other hand, channel-induced impairments are related to channel estimation errors, and propagation delays \cite{Sharma2021}. 

\subsection{Transceiver and Payload Impairments}

\subsubsection{Non-linearity }
 High power amplifiers (HPA) are an integral part of satellites. However, they are inherently non-linear. A simple solution to avoid the non-linear effects of HPAs is to reduce the input power, ensuring a more linear operation regime. However, this will cause a loss of a large portion of power in the form of heat. Therefore, HPAs are generally operated close to their saturation point to address the trade-off between energy efficiency and non-linear effects. Nevertheless, the latter operation mode still can introduce non-linearity effects related to the amplitude and phase modulation responses because most of HPAs for satellites utilize the traveling-wave tube amplifier (TWTA) technology \cite{Saleh1981}. In this context, pre-distortion \cite{Deleu2014, Beidas2015, Piazza2014} and/or equalization \cite{Beidas2011} solutions are employed to address distortions while operating the HPA near its saturation. This is also of paramount relevance in the context of inter-modulation distortion. In fact, the satellite transponder jointly amplifies multiple signal carriers simultaneously using a single HPA for cost-effectiveness. However, this generates non-linear inter-modulation distortions in the amplified signals. In this context, the channel becomes non-linear, and Volterra series represent a common mathematical framework to model such non-linearity. \cite{Beidas2011, Delamotte2016}.

\subsubsection{Phase Perturbations}

 The satellite payload introduces phase noise that rotates the phase of the channel. This can be caused by the satellite's motion within its station and thermal noise. Another dominant form of phase noise originates from the instabilities of oscillators on-board the satellite and low noise block down converters (LNBs) at the UTs. In fact, frequency offsets appear during down-conversion, when the frequency of the oscillator at the receiver side is not identical to the carrier frequency of the received signal. To model this effect, the received signal after down-conversion can be written as \cite{Sharma2021}  
\begin{equation}
y(t) = x(t) e^{-j 2 \pi f_n t + \phi_n}, 
\end{equation}
where $y(t)$ and $x(t)$ are the base-band received and transmitted signals respectively, and $\phi_n$ and $f_n$ are the phase and the frequency offsets respectively. 

 Oscillator-induced frequency offsets always occur in practice because the transmitter and the receiver are independent and cannot have oscillators at the exact same frequency. However, this phenomenon is further accentuated when the transmitted signal experiences frequency shifts during propagation. For instance, when there is a relative velocity between the satellite and the UTs, Doppler frequency shifts occur. The Doppler effect is not apparent in GEO satellite systems and fixed UTs. Nevertheless, this effect has to be accounted for in LEO satellite systems and mobile ground terminals. In fact, there have been a lot of recent works accounting for Doppler effects in precoding for LEO satellites \cite{Li2022, Gao2021, You2020j}.

\subsection{Channel Impairments}
 
 Unlike its terrestrial counterpart, SatCom channels can have long round trip delays (RTD) (0.5 s for GSO satellites), time-varying characteristics, and significant Doppler effects depending on the constellation and type of service. Therefore, the common hypothesis of perfect CSI is highly unrealistic because of channel impairments such as latency, and estimation errors. This can be accounted for by writing the estimated channel matrix $\hat{\textbf{H}}$ as\textcolor{black}{\cite{Sharma2021}}
\begin{equation}
    \hat{\textbf{H}} = \textbf{H} + \textbf{E},
\end{equation}
where $\textbf{H}$ is the true channel matrix and $\textbf{E}$ is the error channel error matrix, whose elements can be modeled as independent and identically distributed Gaussian following on the classical Clarke’s isotropic scattering model \cite{Onggosanusi2001, Sharma2021}.

The long RTD effect should also be considered for a more accurate channel model. In fact, the CSI received at the GW is not instantaneous but somewhat outdated. In general, this can be modeled as\textcolor{black}{\cite{Sharma2021}}
\begin{equation}
    \hat{\textbf{H}}[n] = \textbf{H}[n-D] + \textbf{E}[n],
\end{equation}
where $D$ denotes the delay and $\hat{\textbf{H}}[n]$, $\textbf{H}[n]$ and $\textbf{E}[n]$ denote the estimated channel, true channel and error matrices at the $n$-th time instance, respectively. 

 The long RTD is generally assumed to affect the channel's phase, while its effect on the magnitude of the channel can be negligible during the feedback interval \cite{42}. The main component influencing the magnitude of the channel is rain attenuation, which induces slow variations and can be easily estimated \cite{Amr2009}. Therefore, the effects of outdated CSI can be considered phase offsets in a static environment due to the propagation path as in \cite{Taricco2014}. 

\subsection{Designs Accounting for Phase Perturbations}

 Phase uncertainties are the dominant factor influencing the accuracy of the CSI, and their presence increases are the center frequency increases, as for beyond Ka-band frequencies with fixed users \cite{Shankar2021}. These perturbations arise from the nature of the satellite channel, as well as the satellite transceiver and payload. However, the quality of the CSI has a significant influence on precoding performance. Therefore, a realistic assessment of the precoding performance necessitates accounting for imperfect and outdated CSI. This is done by adding randomness to the estimated channel to represent the phase noise process. Various distributions considering apriori knowledge are assigned to this process, such as Tikhonov, uniform, and normal \cite{Martínez2019}. In this context, the SINR is a random variable, and common robust precoding problems\textcolor{black}{\cite{Shankar2021}} can be formulated under power constraints as

\paragraph{Outage \textcolor{black}{Minimization}}
\begin{equation}
    \min_{\textbf{w}_1, \ldots, \textbf{w}_{N_b}} \max_{k} ~\mathrm{P}(\mathrm{SINR_k} \leq \epsilon),
    \label{modelprob}
\end{equation}
where $\epsilon$ is the outage threshold. 
\paragraph{Average SINR}
\begin{equation}
    \max_{\textbf{w}_1, \ldots, \textbf{w}_{N_b}, k} ~\mathrm{E}(\mathrm{SINR_k} ).
    \label{modelexpec}
\end{equation}

 Unlike the deterministic formulations that assume a perfect and instantaneous CSI, the above formulations are based on stochastic uncertainty models that consider the impact of phase noise. The model in (\ref{modelprob}) is probabilistic, and the precoding scheme is designed to guarantee a certain outage level, while the model in (\ref{modelexpec}) is expectation-based and aims to optimize the average performance. This consideration is often neglected in the literature. However, there have been several works proposing robust precoding schemes in this context. 

 In \cite{Taricco2014}, three sources of phase noise are considered. Namely, the phase offsets originating from the signal propagation path, the LNBs, and the satellite on-board oscillators. This is modeled by adding the sum of three phase perturbations to each element of the channel matrix. Each phase is normally distributed with standard deviations reported by the ESA in \cite{esaStandardDeviation}. The effect of imperfect CSI estimation is also considered, and outdated CSI is accounted for as a signal propagation phase offset. Then, the performance of a linear precoding scheme based on \cite{Devillers2011} is assessed considering user clustering under the aforementioned practical impairments, and system trade-offs are discussed. The work in \cite{42} proposes a robust precoding design in the face of long RTD in multi-beam satellite systems. The amplitude of the channel is assumed to be constant during the feedback interval, while the phase is assumed to be rapidly varying, causing an outdated CSI at the GW. Therefore, the channel phase used for precoding at an instant $t_1$ is modeled as the sum of the channel phase at $t_0 = t_1 - 250 ms$, and a normally distributed error. Then, the precoding weights are derived as the solution of the minimization of the transmit power under per antenna power constraints and the maximum probability of outage constraints. The numerical results demonstrate the considerable gain of the robust precoding scheme in terms of availability and transmit power. 

 However, the probability of outage constraints arising from the randomness of SINR complicates the computation of precoding weights. This issue has been tackled in \cite{22}, where the problem of transmit power minimization under outage probability constraints is further investigated for a more flexible representation. Therein, the authors derived a conservative approximation of the probability of outage based on Taylor series expansion and large deviation inequality to provide a tractable solution to the power minimization problem in the face of phase perturbation. The numerical results demonstrate a superior power performance of the proposed algorithm compared to \cite{42}. However, the power gain comes at the cost of a higher probability of outage compared to \cite{42}. 

 In \cite{Joroughi2017}, a robust multicast multigroup precoding is proposed, where the CSI corruption at the GW is modeled as a general perturbation matrix using the first order perturbation theory of eigenvectors and eigenvalues \cite{Liu2007}. Robustness in the context of multigroup multicast precoding is also studied in \cite{Wang2018}. Therein, a channel model considering phase noise is presented. Then, two optimization problems are investigated, and the relationship between them is established: the maximization of the minimum average SINR and the power minimization with partial CSI. Further, authors in \cite{Wang2018} propose a low complexity and robust clustering strategy that only necessitate partial CSI. Their results highlight the performance gain of the robust precoding in terms of SINR distribution and per-user rate, especially when the variance of the phase noise increases. 

 Nevertheless, all the aforementioned works address robust precoding in the context of fixed satellite services. In \cite{JoroughiS2019}, robust precoding is studied for mobile satellite services. In fact, MSS also suffer from degraded CSI that is not only due to the long RTD and payload impairments but also to the mobility of users. Authors in \cite{JoroughiS2019} assume a perfectly calibrated payload and investigate the impact of users' mobility and long RTD on precoding. They propose a robust precoding scheme based on a reliable CSI feedback mechanism that exploits the past CSI measurements to provide the current CSI estimate. Then, they compare the sum-rate performance considering both outdated and instantaneous CSI to show that their proposed robust precoding scheme provides a large gain in throughput in spite of the degradation with respect to the real-time CSI case. 

 Most of the aforementioned contributions deal with phase noise in precoding by representing the phase uncertainty as a Gaussian process \cite{Taricco2014}\cite{42}. However, additional practical phenomena affecting oscillators should be taken into account. Namely, near-carrier power spectral density \cite{Chorti2006, McNeill2017}. In fact, the output signal of an oscillator can not be a unique spectral tone but rather dispersed over a small band, which makes the synchronization between multiple geographically distant antennas a challenging task. In this context, the impact of synchronization among transmitted waveforms is addressed in \cite{Martínez2019}. Therein, the authors model the phase noise of the satellite oscillator and show the degradation of precoding performance due to the independence of oscillators in a realistic setting. 

 Although various theoretical works in the literature investigate the practical impairments in precoding for satellites, testbeds are still limited. In this context, a field trial of precoding for two co-located satellites has been carried out in \cite{Storek2020}. The effects of carrier offset and oscillator phase noise are accurately modeled and compensated at GW. Then, ZF precoding is implemented based on the proposed robust CSI estimation strategy. The numerical results reported by the testbed show rate and modulation error ratio enhancements. 

\subsection{Designs Accounting for Non-linearity }

Most precoding literature for satellites assumes a linear channel and assesses precoding performance without accounting for non-linearity effects introduced by HPAs. These effects are even more dominant when peak to average power ratios (PAPR) are high for signals, namely, those with spectral-efficient modulation such as 19/32 point multi-ring constellations. Therefore, the impact of HPAs has to be considered to guarantee an accurate evaluation of precoding gain, especially since precoding increases the PAPR of the input signal by modifying its distribution \cite{45, Andrenacci2015}. The first step towards a more robust precoding design in the face of non-linearity consists of including these effects in the received signal model. Various models in the literature capture non-linearity effect \cite{Delamotte2016}, commonly, Volterra series-based models as in \cite{45}. Therein, the signal received by the $i$-th user can be written as
\begin{equation}
    y_i = g_{1,i} \textbf{h}_i^H \textbf{W} \textbf{s} + g_{3,i} \textbf{h}_i^H ((\textbf{W} \textbf{s}) \odot (\textbf{W} \textbf{s}) \odot (\textbf{W} \textbf{s})^*) + n_i, 
    \label{eq:nonlinear}
\end{equation}
where $\odot$ is the Hadamard product, $g_{1,i}$ and $g_{3,i}$ are the first and third Volterra coefficients respectively. 

 From (\ref{eq:nonlinear}), we can see that the received signal is a non-linear mapping of the precoding matrix, which calls for different signal processing frameworks. One strategy imposes constraints on the maximum signal amplitude, forcing HAPs to operate in their linear regime \cite{Spano2018}. In this case, the received signal in (\ref{eq:nonlinear}) can be approximated as (\ref{eq:linear}), and the previously covered precoding techniques can be applied. The other strategy solves the precoding problem in the general framework of non-linear signal pre-distortion (SPD) \cite{30}. In fact, there has been an intensified attention, in the SatCom community, toward SPD as a technique to mitigate non-linearity impairments \cite{Piazza2014, Beidas2010, 30}. This technique consists of compensating the non-linear effects of HPAs by a non-linear transformation of the waveform. Traditionally, the processing related to such techniques happens in the GW to exploit the powerful computational capacity on the ground. However, the on-board implementation has also been attracting growing attention, as it circumvents the limitations of tight emission masks \cite{30}. 

 There have been several works proposing joint designs of precoding and pre-distortion. For instance, authors in \cite{Alvarez2005} combine THP with a fractional pre-distorter based on a gain-based Look-Up Table. In \cite{Cheng2013}, codebook-based precoding is combined with a linear adaptive pre-distortion technique. The power minimization problem is addressed, where the codebook-based precoding and power assignment weights are jointly optimized with the pre-distortion matrix. In \cite{45}, multistream Crest factor reduction (CFR) and SPD are designed in a cascade to mitigate non-linearities of HPAs as well as interference. However, unlike the two aforementioned works, where either single channel systems \cite{Alvarez2005} or linear processing of the channel \cite{Cheng2013} are considered, the work in \cite{45} proposes a non-linear processing at the GW, with multistream CFR and SPD. This approach provides various performance gains and opens the door for a joint pre-distortion and precoding framework. 

 Constant-envelope precoding has also been proposed as a robust precoding technique in the face of non-linear distortions of amplifiers. In this technique, the precoded signals are designed to have a constant amplitude, and it has been well studied in the literature \cite{Mollén2016, Mohammed2013, Mohammed2012, MohammedSaif2013, LiuFan2017, Domouchtsidis2019}. Non-linear distortions on the transmitted signals depend on their instantaneous transmitted power, which worsens with high variations \cite{Spano2018, Spano2017}. Therefore, maintaining a constant amplitude of the transmitted signal across the satellite antennas increases the robustness against TWTAs' non-linearities. Constant-envelope precoding for multi-beam satellites has been proposed in \cite{9}. Therein, an on-ground precoding scheme is designed considering a fixed on-board beamformer. Then an on-ground precoding scheme is designed jointly with an adaptive on-board beamformer. The proposed constant-envelope precoding schemes are compared with linear MMSE precoding, and numerical results show their effectiveness and robustness. 

 In \cite{Maturo2019}, a real-time demonstration of precoding for a multi-beam satellite system is reported, where a proper collection of CSI is considered, and HPAs non-linearity is modeled. The results using the hardware demonstrator prove the feasibility of precoding using the super-framing structure of DVB-S2X. Another hardware precoding demonstration is reported in \cite{Duncan2019} and experimental verification of Volterra-based distortion models for multicarrier satellite systems is reported in \cite{Delamotte2016}.

\subsection{\textcolor{black}{Summary \& Lessons Learnt}}
\textcolor{black}{Satellite communication systems face several impairments that affect the quality of communication. These impairments result mainly from payloads, transceivers, and the nature of the satellite communication channel. One major impairment is the non-linearity caused by HPAs. These non-linear distortions can be modeled using the Volterra series framework, and solutions in this context include pre-distortion and equalization techniques. Another important impairment to take into account while designing precoding for satellites is phase perturbations. Specifically, phase perturbations are caused by oscillators on-board the satellite and LNB converters at the UTs. Frequency offsets appear during down-conversion, and Doppler frequency shifts occur when there is a relative velocity between the satellite and the UTs.  In addition to the transceiver and payload impairments, SatCom channels can have long round trip delays (RTD), time-varying characteristics, and significant Doppler effects depending on the constellation and type of service. Therefore, the common hypothesis of perfect CSI is unrealistic because of channel impairments such as latency and estimation errors.
To account for phase noise, various stochastic uncertainty models, such as Tikhonov, uniform, and normal distributions, can be considered. In this context, the SINR becomes a random variable. Therefore, the precoding scheme can be designed to minimize the probability of a certain outage level or to maximize the expected SINR.  }

\section{Trends in Precoding for HTS}

\begin{figure*}[t!]
    \centering
    \includegraphics[width=\linewidth]{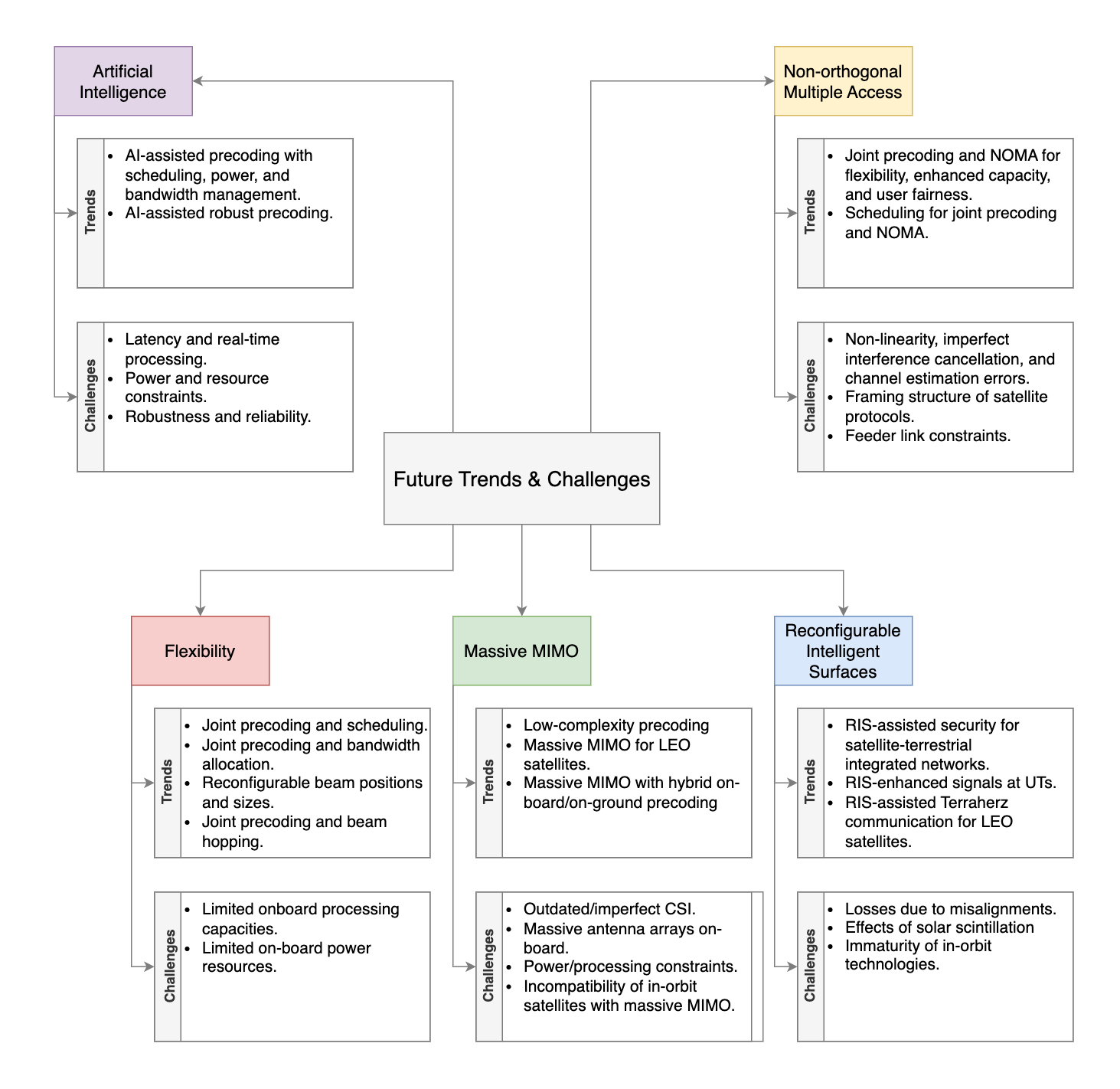}
    \caption{\textcolor{black}{ \textcolor{black}{Trends and challenges in precoding for HTS.}}}
    \label{fig:trends}
\end{figure*} 

\textcolor{black}{In this section we discuss the trends in precoding for HTS to inspire further research in this field. A summary of the discussed research trends is illustrated in Fig. \ref{fig:trends}.}

\subsection{Flexibility}

Most in-orbit satellite communication systems are based on a regular beam layout plan, where a four-frequency reuse pattern is employed to mitigate inter-beam interference, and power is allocated uniformly across all the beams. This leads to a quasi-static beam pattern that does not necessarily adapt to the heterogeneous nature of traffic. In fact, demand generally varies across the beams during the day and along the different seasons. Therefore, a quasi-static beam layout plan can cause a mismatch between the demand and the offered capacity. Some beams can be experiencing a hot-spot scenario and exhausting the offered bandwidth without properly satisfying the demand. On the other hand, some other beams can be under-used and waste spectral resources. This calls for more traffic-aware radio resource management and transmission strategies. 

Demand-based optimization in the forward link of multi-beam satellite communication systems has been receiving intensified attention thanks to the advances in on-board processing and large-scale active antennas \cite{Kisseleff2021}. Dynamic beamforming is investigated in \cite{Chaker2021} to provide an adaptive beam layout plan using directive and irregular beams. Further, precoding is employed to combat the inter-beam interference from overlapping beams. Numerical results show that the proposed adaptive beam layout plan considerably reduces capacity mismatches across the beams compared to a regular beam layout plan. The hot-spot problem has been addressed using flexible precoding for mobile satellite systems in \cite{Icolari2017} and for fixed satellite systems in \cite{18}. Overall, demand-aware optimization can be introduced on various levels. It can be applied for scheduling as in \cite{Honnaiah2022}, where a demand-based scheduling strategy is investigated for precoded HTS. It also can be applied for bandwidth and/or power allocation as in \cite{Salih2021, Abdu2022}. In particular, \cite{Salih2021} shows that the combination of flexible bandwidth allocation with linear precoding guarantees higher efficiency in demand satisfaction and bandwidth than benchmark schemes. Another strategy to design demand-aware transmission relies on re-configurable beam positions and sizes to spatially adapt the demand \cite{Chaker2021, Puneeth2021, Jubba2021}. Finally, beam hopping allows for a demand-aware coverage in the time domain \cite{Chen2021}, and its combination with precoding has been investigated in \cite{28}.

\subsection{Massive MIMO}

Massive MIMO technology has been widely applied for terrestrial systems thanks to its capability to enhance data rates and spectral efficiency \cite{Geoffrey2014}. \textcolor{black}{ Therefore, it is natural to investigate the extension of its performance benefits to satellite communication systems.} However, deploying this technology for satellite communication systems faces various challenges. 
First, massive MIMO performance greatly depends on accurate instantaneous CSI. Nonetheless, as it is described in section V, satellite communication systems suffer from various practical impairments that hinder the acquisition of accurate instantaneous CSI. In fact, the CSI obtained at the satellite is outdated and imperfect due to the long round trip delay and channel estimation error. \textcolor{black}{Therefore, integrating massive MIMO into satellite communication systems requires the development of appropriate CSI feedback mechanisms that bypass the performance degradation introduced by inaccurate channel estimation.} Second, massive MIMO communication requires massive antenna arrays whose deployment can be quite 
challenging considering the satellite's payload limitations.
\textcolor{black}{In fact, satellites have space restrictions that can prevent deploying the many antennas required for massive MIMO. This means that the design of the satellite and its antennas must be carefully optimized to maximize the benefits of massive MIMO while minimizing the space it occupies on-board the satellite. Further, satellite communication systems typically have strict power constraints, making it challenging to provide the power needed for a large number of antennas. Therefore, power-efficient antenna designs and signal processing are needed to mitigate this issue. Third, the widely adopted four frequency reuse schemes for in-orbit satellites are incompatible with massive MIMO technology, which calls for new considerations in the design of frequency reuse strategies. Finally, the absence of multi-path fading in the satellite channel significantly impacts the performance of massive MIMO. Thus it should be taken into consideration to analyze the potential benefits of this technology. These challenges represent key problems to be addressed in order to integrate massive MIMO technology into satellite communication systems.}

Recently, motivated by the developments in multi-beam payloads incorporating active antennas with a large number of radiating elements, \textcolor{black}{massive MIMO in satellites has been receiving increased attention, and precoding emerges as a prolific technique to exploit the benefits of this technology fully}. In particular, there have been several recent works proposing massive MIMO for LEO satellite communication systems \cite{You2020, You2020j, Angeletti2020}. This has also been motivated by the attention that LEO satellites are receiving thanks to their lower deployment and launching costs, as well as their reduced transmission delays. Massive MIMO is proposed to be implemented for LEO satellites equipped with uniform planar arrays and using statistical CSI instead of instantaneous CSI. In particular, \cite{You2020} shows that the performance of a statistical CSI precoder approaches that of an instantaneous CSI precoder asymptotically. Authors in \cite{Angeletti2020} investigate a pragmatic approach for massive MIMO implementation at a system and satellite payload level, which they extended in \cite{AngelettiPiero2021} to a heuristic and computationally efficient solution. In \cite{Guo2021}, broad coverage transmission is investigated for massive MIMO satellite systems, and a precoder design is proposed based on the min-max criterion. Further, Hybrid beamforming is investigated in \cite{Yang2022} for LEO massive MIMO satellite systems to design an energy-efficient scheme that is robust CSI errors.

\subsection{Reconfigurable Intelligent Surfaces}

Reconfigurable intelligent surfaces (RIS) are emerging as one of the most potential technologies for beyond 5G and 6G communication systems, promising higher spectral and energy efficiency at a low cost and low hardware complexity. The main idea behind RIS consists of using thin materials equipped with low-cost passive antennas to reflect signals and shift their phases toward desired users. RIS can be deployed on walls or ceilings and controlled on an electromagnetic level using a software controller. This technology has been receiving intensified attention over the last decade. It has been widely studied for terrestrial systems in contexts like MIMO \cite{Shicong2020}, millimeter wave communication \cite{Yongxu2020}, and cognitive radio \cite{Yuan2021}. Studies of RIS for terrestrial systems highlight the capability of this technology to enhance data security and user fairness in a cost-effective manner. 
\textcolor{black}{ Thus, integrating this technology into satellite communication systems can have attractive benefits. In fact, satellite communication systems can suffer performance degradation caused by shadowing in the UT environment. In this case, RIS mounted on a high building can be exploited to enhance the effective gain of the signal transmitted from the satellite to the ground terminal by shifting the signal's phase toward the UT. RIS can also be used to enhance security in the downlink of satellite communication networks. In fact, the nature of wireless channels and the large coverage areas of satellites increase the vulnerability of these systems to eavesdropping. A RIS-assisted link can enhance the signal at the UT while degrading it at the eavesdropper. In the context of full frequency reuse, problems that jointly optimize precoding weights at the satellite and phase shifts are the RIS can be investigated. Furthermore, RIS can be exploited to enhance inter-satellite links. In fact, These links suffer from misalignment fading and path loss, which can be compensated for using RIS mounted on the satellites to enhance the signals' propagation. }

There have been several recent studies investigating RIS-assisted networks for satellite communication systems. \textcolor{black}{In particular, cognitive integrated satellite-terrestrial networks have been investigated in \cite{Sai2021}, where the issue of downlink security was addressed using a RIS-assisted link.} In \cite{Matthiesen2021}, RIS has been considered to enhance the link budget between the user terminals and a LEO satellite. It has also been considered to enhance LEO inter-satellite links in the terahertz band in \cite{04281}. Further, RIS-assisted networks have been studied for GEO satellites in \cite{00497}, where the RIS adjusts the phase of the signal from the satellite toward the ground user to boost the effective gain of the system. Therein, the transmit power of the satellite is jointly optimized with the phase shifts applied at the RIS to maximize the channel capacity. Another recent work in \cite{9726800} investigates RIS for hybrid satellite-terrestrial relay networks where the beamforming weights' optimization for the satellite and the base station is performed jointly with the optimization of the RIS's phase shifts to minimize the total transmit power. \textcolor{black}{Thus, this opens the door to investigating coupled problems where RIS phase shifts are optimized jointly with precoding/beamforming weights at the satellite.}

\subsection{\textcolor{black}{Non-orthogonal Multiple Access}}

\textcolor{black}{Protocols such as TDMA, FDMA, CDMA, and their variants fall under orthogonal multiple access (OMA) protocols, in contrast to non-orthogonal multiple access (NOMA) transmission techniques \cite{7676258}. NOMA is primarily a physical layer technique that allows different users to use the same radio resources to transmit their signals. Then, multi-user detection and successive interference cancellation are employed at the receiver side to recover the individual data streams. However, some variants of NOMA incorporate MAC layer techniques such as scheduling to optimize the performance and allocate resources dynamically. Over the last few years, NOMA has been actively researched for various wireless communication systems, and it has demonstrated outstanding results in terms of capacity, user fairness, and latency for terrestrial communication systems\cite{7676258}. Therefore, it is only natural to investigate its application to satellite communication systems.}

\textcolor{black}{ In multi-beam satellite communication systems, NOMA can be applied in the return or the forward links. Although NOMA has been widely studied for the return link of multi-beam satellites, more research needs to be dedicated to its application to the forward link \cite{8930827}. In fact, the forward link of a multi-beam satellite has different characteristics than that of terrestrial systems, which makes applying NOMA, in this case, less of straight forward extension of its terrestrial counterpart. Multi-beam satellite systems enable frequency reuse strategies across the different beams. In non-aggressive reuse schemes such as four or three colors, UTs can use single user-decoding thanks to the spectrum isolation between the different beams. However, in more aggressive frequency reuse strategies, precoding becomes a more interesting way to cope with increased inter-beam interference. Therefore, the application of NOMA to the forward link of multi-beam satellites depends on the implemented frequency reuse scheme. In fact, in non-aggressive frequency reuse, NOMA can be used within a single beam to mitigate intra-beam interference, thanks to the low level of inter-beam interference, as long as there is a considerable SNR asymmetry among users. Nevertheless, when considering more aggressive frequency reuse strategies, inter-beam interference becomes significant, which deteriorates the performance of NOMA in single beams. In this scenario, strategies to integrate NOMA in the forward link transmission have to be investigated. In \cite{8930827}, beam cooperation is discussed as a solution to bypass the performance degradation of NOMA due to inter-beam interference. In other words, frames corresponding to users from different beams should be jointly processed, which is far from the standard NOMA and more compatible with precoding.}

\textcolor{black}{The combination of NOMA with precoding techniques in the forward link of multi-beam satellites further pushes the spectral efficiency of full-frequency reuse strategies to its extreme. In a simple full-frequency reuse unicast scenario,  $N_b$ users (with one user per beam) are scheduled in each time slot to be served simultaneously. Then, the precoding matrix is designed to mitigate inter-beam interference between the scheduled users. Incorporating NOMA into this setting allows the satellite to serve $L N_b$ users simultaneously instead of $N_b$, with $L$ being the number of non-orthogonal frames per beam. In other words, the scheduler selects a total of $L N_b$ users, with $L$ users per beam. Then, the satellite applies precoding to mitigate inter-beam interference, while NOMA is applied in each beam to mitigate intra-beam interference. However, to guarantee the benefits of NOMA, the selected users within the same beam should have a significant SNR imbalance. In this context, authors in \cite{7842127} show that MMSE precoding in the case of L=2 reduces inter-beam interference. It is also worth mentioning that in realistic multi-beam satellite systems, more than one user is served by the same physical frame. Therefore, the scheduler has to group users with similar channel conditions in each beam as one multicast group to guarantee a good precoding performance, which contradicts the scheduling required by NOMA based on SNR imbalance. Therefore, combining NOMA with precoding in a multicast scenario calls for a scheduling strategy that provides a trade-off between the two criteria \cite{9473484}. }

\textcolor{black}{NOMA with precoding for the forward link of multi-beam satellites has been receiving increased attention during recent years \cite{8930827}. For instance, authors in \cite{Caus2016} investigate sub-optimal user-scheduling strategies to incorporate in NOMA-based multi-beam satellites. Furthermore, \cite{Wang2019Anyue} addresses the challenge of aligning the offered capacity with traffic demands and proposes a solution incorporating NOMA to alleviate intra-beam interference and allow for aggressive frequency reuse. Nevertheless, numerous aspects remain to be further investigated for NOMA in multi-beam satellites. For instance, it is imperative to study the performance of NOMA under practical satellite impairments such as non-linearity, channel estimation errors, and imperfect interference cancellation. Moreover, NOMA has to be studied in conjunction with flexible resource allocation to adapt to dynamic traffic changes. Finally, physical layer security in NOMA-based multi-beam satellites is another essential aspect to be studied by future research. }

\subsection{\textcolor{black}{Artificial Intelligence}}

\textcolor{black}{Artificial intelligence techniques such as machine learning (ML), deep learning (DL), and reinforcement learning (RL) have shown outstanding results in automating processes in various fields. Although their extension to satellite communication and wireless communication, in general, is still at its beginning, there are various applications where these techniques can be highly beneficial.
ML techniques can assist SatCom systems in tasks such as interference detection, flexible payload configuration, and congestion prediction. Current interference detection strategies in satellite control centers rely on human expertise to analyze the spectrum and identify an abnormal error rate rise or a QoS deterioration. However, this task can be automated using unsupervised ML as discussed in \cite{VázquezML}. Flexible payload configuration is another interesting use case of ML models. In fact, the current satellite communication systems require human assistance to determine payload configuration, such as frequency and power allocation, as well as the beam pattern. However, future satellite communication systems should be able to adapt quickly and flexibly to changes in traffic and interference levels. Therefore, flexible resource allocation can be an interesting application of ML models in the payload configuration of satellites. In fact, as proposed in \cite{VázquezML}, sub-carrier allocation can be determined by an ML model exploiting data about the CSI, the interference levels, and the capacity requested by UTs. Moreover, ML models can be highly beneficial in predicting congestion in traffic. In fact, in satellite communication, traffic can be quite irregular due to the geographical distribution of users, which induces some areas to have high demand while others have low demand. Generally, these congestion events are predicted based on traffic patterns from the preceding days. Nevertheless, incorporating ML into this process can increase the accuracy and reactivity time to enhance the system's adaptation and therefore improve performance \cite{VázquezML}. }

\textcolor{black}{In precoding-related applications, ML mechanisms can be used for resource management, particularly for user scheduling. In fact, as discussed in section \ref{mixed}, the precoding performance significantly depends on the underlying scheduling strategy. Therefore, optimal performance is obtained by jointly optimizing scheduling and precoding parameters, which leads to a highly complex optimization problem. One way to enhance the performance of the scheduling strategy while maintaining a relatively lower complexity can be to apply a learning technique for clustering. For instance, authors in \cite{9815569} proposed a scheduling strategy based on unsupervised learning techniques, namely, K-means, hierarchical clustering, and self organization. The clustering exploits both the CSI of users and their geographical positions to determine the multicast groups to be served simultaneously, and it is done in conjunction with RZF precoding. The results are encouraging and demonstrate an increased average performance per beam compared to geographical clustering. This opens the door for further research to investigate the benefits of unsupervised learning on scheduling and precoding performance. }

\textcolor{black}{In addition to unsupervised learning, supervised learning can be exploited for resource management of satellite communication systems. For instance, recent work in \cite{Ortiz} shows the potential benefit of using neural networks in bandwidth and power management of HTS. In fact, a neural network classifier is considered to choose an optimal resource allocation configuration from a required traffic per beam. Another recent work in \cite{8302493} utilizes deep reinforcement learning to perform dynamic channel allocation for efficient use of the radio resources. Moreover, ML techniques can be used to generate more robust precoding schemes. These models can be trained to compensate for the non-linearity and estimation errors of the satellite channel.}

\subsection{\textcolor{black}{Summary \& Lessons Learnt}}

\textcolor{black}{This section covers some catalyzing wireless communication technologies driving the research trends in precoding for HTS. These technologies include flexibility, massive MIMO, RIS, NOMA, and ML. }

\textcolor{black}{ The trend towards flexibility promotes adaptive demand-based optimization in multi-beam satellite systems to reduce mismatches between the traffic demand and the offered capacity. Flexible precoding can be used to address hot-spot scenarios, and demand-aware optimization can be applied to scheduling, bandwidth, and power allocation. On the other hand, the extension of various successful terrestrial technologies, such as massive MIMO, RIS, and NOMA, is at the center of the current research trends. The application of massive MIMO to HTS faces various challenges, such as channel estimation errors and satellite payload limitations. However, recent research focuses on LEO satellites with uniform planar arrays and statistical CSI and employs precoding techniques to exploit the benefits of massive MIMO. Recent research also explores RIS-assisted networks in satellite systems, addressing issues such as downlink security and link budget enhancement, which can be investigated using joint optimization of RIS phase shifts and satellite beamforming/precoding weights. }

\textcolor{black}{Furthermore, the application of NOMA in the forward link of multi-beam satellite communication systems presents unique challenges and opportunities. In a full-frequency reuse scenario, combining NOMA with precoding allows serving a larger number of users simultaneously, with precoding mitigating inter-beam interference and NOMA addressing intra-beam interference. Research in this area can further investigate practical impairments, flexible resource allocation, and physical layer security. In addition to NOMA, ML is another technology that holds great potential for revolutionizing satellite communication systems. ML can automate tasks like interference detection, flexible payload configuration, and congestion prediction. In precoding-related applications, ML mechanisms can be used for resource management, particularly user scheduling. It can also enhance the performance of precoding schemes by compensating for non-linearity and estimation errors in the satellite channel. The application of these techniques in satellite communication systems is still in its early stages but shows promising benefits for improving performance and automation.
}

\section{Conclusion}
Satellite communication systems are witnessing various technological breakthroughs making them an attractive solution for future communication systems. Specifically, the next generation of satellite communications is anticipated to handle high data rates by exploiting multi-beam and full frequency reuse technologies. Therefore, precoding emerges as a prolific interference management technique, enabling high spectral and energy efficiency. Given the versatile nature of precoding design, this survey presents a classification of the recent precoding techniques for HTS communication. 

The survey starts by providing an overview of the physical and modeling aspects of the satellite communication system. Then, a brief overview of the satellite PHY and MAC layers techniques is presented to serve as a reference in the precoding analysis. However, the main contribution of this work is a comprehensive classification of the precoding literature for multi-beam satellites considering two different perspectives. The first perspective considers the precoding problem formulation to classify the schemes based on the scheduling strategy (\textit{i.e.} precoding group), the \textcolor{black}{the precoding optimization problem} , and the switching rate (\textit{i.e.} precoding level). The second perspective categorizes precoding techniques based on the satellite system design, which includes precoding for single-GW or multi-GW systems (\textit{i.e.} system architecture), precoding for on-board, on-ground, or hybrid on-board/on-ground implementations (\textit{i.e.} precoding implementation), and precoding for fixed or mobile satellites (\textit{i.e.} satellite services). This classification provides a canvas of the recent precoding schemes and covers the trending direction in the literature. However, while the majority of the works in the literature consider perfect channel and transmission conditions, real satellite communication systems are subject to various channel and transceiver impairments. Therefore, this survey highlights HTS systems' practical impairments and presents robust precoding techniques in the literature. \textcolor{black}{Finally, research trends in precoding for HTS are presented before concluding}.

\bibliographystyle{IEEEtran}
\bibliography{ref.bib}

\end{document}